\documentclass[aps,prb,graphicx,twocolumn,superscriptaddress,longbibliography]{revtex4-2}

\usepackage{graphicx} 
\usepackage{float}

\graphicspath{{plots/}}

\usepackage{amsmath}
\usepackage{amsfonts}
\usepackage{amssymb}
\usepackage{amsthm} 
\usepackage{mathrsfs}
\usepackage{braket}

\usepackage[pdftex]{hyperref}
\hypersetup{ 
    colorlinks,
    citecolor=blue,
    filecolor=blue,
    linkcolor=red,
    urlcolor=blue 
}

\usepackage{times}
\usepackage{algorithm}

\usepackage{algpseudocode}

\newcommand{\p}[1]{\left(#1\right)}
\newcommand{\sqarep}[1]{\left[ #1 \right] }

\newcommand{\abs}[1]{\left|#1\right|}

\newcommand{\commentout}[1]{}

\usepackage{soul}

\begin{document}

\title{A Divide-and-Conquer Algorithm to Solve Disordered Interacting Few-Particle Systems in One Dimension}

\title{A Divide-and-Conquer Algorithm for Disordered and Interacting Few-Particle Systems in One Dimension}

\author{Llu\'{i}s Hern\'{a}ndez--Mul\`a}
\email{lhernandezmula@gmail.com}
\affiliation{Institut f\"ur Theoretische Physik, Universit\"at Innsbruck, A-6020 Innsbruck, Austria}
\author{Andreas M. L\"auchli}
\affiliation{Laboratory for Theoretical and Computational Physics, Paul Scherrer Institute, 5232 Villigen, Switzerland}
\affiliation{Institute of Physics, \'{E}cole Polytechnique F\'{e}d\'{e}rale de Lausanne (EPFL), 1015 Lausanne, Switzerland}

\date{\today}

\begin{abstract}
We present an algorithm to solve very large one-dimensional disordered and interacting few-particle systems. Our approach exploits the localized nature of the eigenfunctions in real space to achieve a linear scaling with the total system size $L$. This allows us to solve for all eigenfunctions of single-particle systems with different types of disorder up to one billion sites. Based on this technology we collect very detailed histograms of properties of eigenfunctions, such as the localization length or the participation ratio as a function of their energy. These histograms reveal surprisingly rich fine structures, whose origins we discuss. We also apply the algorithm to single particle problems where not all eigenfunctions are localized and show how this is diagnosed. Finally we extend the algorithm to interacting two-particle problems in the presence of disorder and demonstrate that our algorithm is well suited to analyze the effect of interactions on wavefunctions.

\end{abstract}

\maketitle 

\section{Introduction}
Since the work of Basko, Aleiner and Altshuler~\cite{BASKO20061126} showing that localization phenomena can exist in weakly interacting systems for strong enough disorder, ongoing research tried to understand the properties of such systems, also known as Many-Body Localized (MBL). The defining property of MBL systems is that they do not fulfill the Eigenstate Thermalization Hypotheses (ETH, \cite{PhysRevE.50.888,PhysRevA.43.2046}). Despite the lack of thermalization is by itself an interesting fundamental problem,  there are several potential  applications for such systems that makes them interesting for quantum computation purposes, in particular such systems are possible candidates for realizing quantum memories \cite{PhysRevB.89.144201, Bauer_2013, PhysRevB.88.014206, PhysRevLett.107.030503}. Experimental realizations of systems showing the expected properties of the MBL phase are available \cite{Kondov2015, Schreiber2015, Bordia2016, Smith2016, Choi2016, Bordia2017, Bordia2017_2, Lueschen2017, Xu2018, Rispoli2019, Lukin2019, Kohlert2019}. Despite all the effort done, there is an ongoing debate about the majority of the properties of MBL systems. It is unclear if the knowledge obtained from numerical simulations of small systems in  one dimension (1D) would survive the thermodynamic limit. For instance, the existence of mobility edge first shown in Ref.~\cite{PhysRevB.91.081103} for systems up to $L=22$ sites was challenged by the delocalization mechanism described in Ref.~\cite{PhysRevB.93.014203}, where the presence of ergodic bubbles within the localized phase would make the mobility edge unstable at large $L$. It is also an open discussion whether the regime previously considered deep in the MBL is truly MBL \cite{PhysRevE.102.062144,ABANIN2021168415}. One of the reasons for this large uncertainty is the lack of numerical methods for solving large interacting systems, where enough statistics needs to be obtained to distinguish between finite size effects and real physical properties. The MBL phase does not affect exclusively the ground state. Therefore, methods to target any eigenfunctions are needed, and the basic DMRG \cite{PhysRevLett.69.2863} or quantum Monte Carlo are not enough. There is promising progress towards an efficient algorithm combining Matrix Product States (MPS) and more sophisticate DMRG~\cite{PhysRevB.94.041116, PhysRevLett.116.247204, PhysRevLett.118.017201, PhysRevB.99.104201, PhysRevX.7.021018}, but such methods still face some real challenges when solving large enough systems.

\begin{figure*}[ht]
\centering{\includegraphics[width=0.6\linewidth]{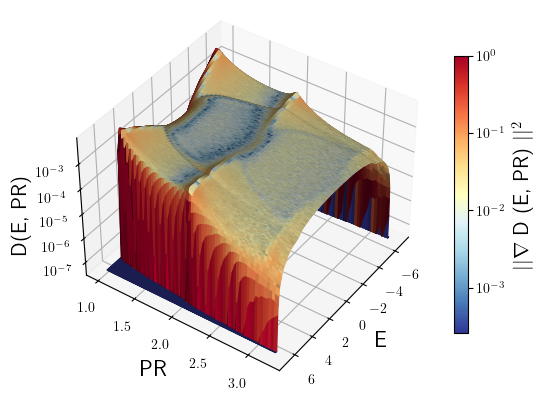}}\\
\centering{\includegraphics[width=0.45\linewidth]{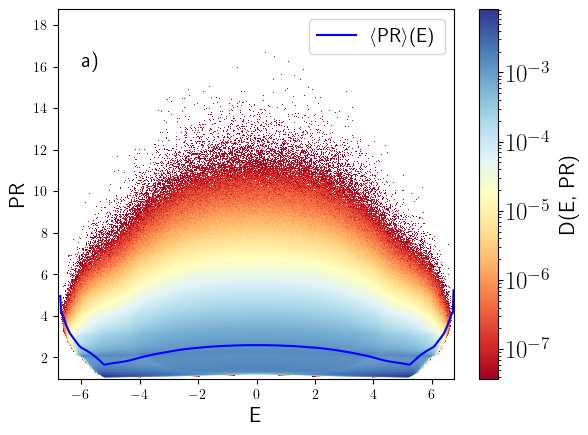}
\includegraphics[width=0.45\linewidth]{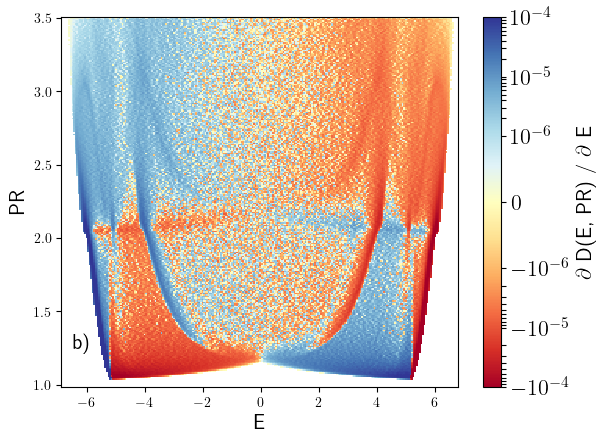}}
\caption{ Histogram of energy ($E$), Participation Ratio (PR) and the density of states, $\rho$, obtained from the full set of eigenstates in a system with $L=10^9$ sites and a on-site random disorder which follows a box distribution with disorder strength $W_{\text{And}}=10$ ($W_{\text{MBL}} = 2.5$ in the notation of the MBL community). In the 4D-histogram, the colorscale is given by the norm of the gradient of the density of states. The colorscale in Plots a) and b) is given by the density of states and the derivative of the density of states over the energy, respectively. Note the different $y$-axis ranges in a) and b).}
\label{example_intro}
\end{figure*}

We have taken a bottom-up approach to develop an algorithm that aims at solving large systems as long as they are in the localized phase, starting with the few-particles scenario. In the first part of this paper, we introduce a new algorithm, based on the Divide-and-Conquer (DaC) scheme in order to tackle the 1D Anderson problem. In the second part of this paper, we adapt our algorithm to deal with the two-interacting particles (TIP) problem, where we focus on the effect of the interaction.

The Anderson problem consists of a tight-binding model with on-site disorder, and was introduced in 1958 as the first instance of localization phenomena in disordered media ~\cite{Anderson_1958}. The localized nature of the problem was a novel feature compared to the classical counterpart of the Brownian motion, which predicts a diffusive growth of the space explored by the particles. The difference between the classical and quantum predictions drew the interest towards this topic~\cite{Kramer_1993, MIRLIN2000259, Markos2006}. In particular in one dimension it is well established that any small amount of disorder on an on-site random potential leads to localized eigenfunctions with exponentially decaying tails, whose envelope is controlled by the localization length, $\xi$. The value of $\xi$ depends on the disorder and the energy of the eigenfunction, but it does not depend on the system size $L$. 

Our method takes advantage of the finite value of the localization length. By disregarding the exponentially small tails, we can fit the eigenfunctions into subsystems. The size of the subsystems, $M$, depends on $\xi$, and is thus $L$-independent. Once we have found all the eigenfunctions that fit on the considered subsystem, we move to another subsystem, until we have considered all sites of the global system. Combining the solutions obtained in each subsystem allows us to find explicitly the full set of the eigenfunctions, their energies and calculate any observable of interest. From the obtained data using the DaC algorithm, we generate 4D-histograms of the Energy, Participation Ratio and the Density of states, as can be seen in Fig.~\ref{example_intro}. The Participation Ratio (PR) is related to the localization length $\xi$ and measures roughly over how many sites eigenfunctions are spread. The color-scale represents the gradient of the density of states $\rho$. To better appreciate some of the underlying structure of the 4D-histogram, we also include histograms of the Energy-PR and the colorscale given by both the density of states and the derivative with respect to the energy of the density of states, in the plots a) and b) of Fig.~\ref{example_intro}.

From our statistical analysis, we can distinguish several patterns, such as the discontinuities in the density or extended plateaus, which can be seen as red and blue color areas of the 4D-histogram respectively, and are sensitive to the microscopic details of the disorder potential. These distinct patterns are specific to the box disorder, and do not appear in any of the other potentials studied (Gaussian, binary or Aubry-Andr\'{e}). Rather, we believe they stem from the continuous but bounded nature of the box distribution. The analysis of these observables shows how much the properties of the systems can be influenced by the microscopic details of the random potential.

The structure of this paper is as follows: We start by introducting the Anderson Hamiltonian and the various types of disorder and the important observables of interest in Sec.~\ref{seq:Hamiltonians}. Then we provide the basic intuition behind the Divide-and-Conquer (DaC) algorithm and describe the algorithm in great detail in Sec.~\ref{DaC_General_idea}. After the DaC is introduced, we apply the method to solve the Anderson problem in two regimes, the strong (where microscopic details matter) and weak (microscopic details do not matter) disorder regimes. The specifics of the implementations of the DaC algorithm for the single-particle are explained in Section~\ref{sec:DaC_single_particle} and the numerical results are in Section~\ref{sec:Num_results}. Then, we study systems that are not fully localized, where the algorithm can only provide the localized eigenfunctions. Once we have studied in detail the single particle physics, we move towards the two-particle physics, where we analyse the role of interaction. In order to deal deal with this problem, some adaptations of the DaC algorithm have to be implemented, as we justify in Section~\ref{sec_alg_TIP}. The numerical results of the TIP problem are shown Section~\ref{sec_num_TIP}.

\section{Hamiltonians and Observables of interest} 

\label{seq:Hamiltonians}

In this section we introduce the relevant Hamiltonians for the single and few-particle particle cases, and discuss some observables of interest for Anderson localization and the interacting few-particle problem in the presence of disorder.

The Anderson model was first introduced in Ref.~\cite{Anderson_1958}. We review some of the key aspects of the one-dimensional Anderson model in this section, but for more in-depth treatments, there is an extensive literature reviewing the properties of the Anderson model:~\cite{Kramer_1993, MIRLIN2000259, Markos2006, Evers2008}.

The interplay between interactions and disorder for the two particle problems was first studied in Ref.~\cite{Shepelyansky1994_4}. Few-particle interacting systems are still actively discussed~\cite{Frahm2016_4, Frahm2016_2D, Schmidtke2017_4, Stellin2019}.

\subsection{Single particle problem: Model and conventions}

The standard model to study the Anderson problem in 1D is the disordered tight-binding Hamiltonian. In the spinless fermionic language, in 1D and with open-boundary conditions (OBC), the Hamiltonian reads:
\begin{equation}
H = \sum_{i}^{L-1} t_{i,i+1} \p{c_i^\dag c_{i+1}^{\phantom{\dag}}+c_{i+1}^\dag c_{i}^{\phantom{\dag}}}-\sum_{i=1}^{L} \epsilon_i n_i,
\label{Anderson}
\end{equation}
where $\epsilon_i$ is the chemical potential at site $i$, $t_{i,i+1}$ is the hopping amplitude between nearest neighbor sites $i$ and $i+1$, $c_j$ ($c_j^\dag$) annihilates (creates) a spinless fermion at site $j$, and $n_j = c_j^\dag c_j$ is the number operator~\footnote{Note for the single-particle problem the fermionic or bosonic nature of the second quantized operators plays no role.}.
\\
We allow for disorder both in the nearest-neighbor hopping $t_{i,i+1}$ as well as in the local potential $\epsilon_i$. However we focus mostly on constant $t_{i,i+1}=t=1$ with different on-site disorder distributions for $\epsilon_i$:
\\
\begin{center}
\begin{tabular}{|c|c|c|}
\hline
	Disorder type & Distribution of $\epsilon$ & Variance $V$\\
\hline
\hline
	Box & $\epsilon_i$ uniform in $[-W/2, W/2]$ & $V = W^2/12$\\
\hline
	Gaussian & $\epsilon_i$ gaussian centered around $0$ & $V = W^2/4$\\
\hline
	Binary & $\epsilon_i=\pm W/2$, $p_+=p_-=1/2$ & $V = W^2/4$\\
\hline
\hline
	Aubry-Andr\'{e} & $\epsilon_j = \cos\p{2\pi\beta j} W$ & $V = W^2/2$\\
\hline
\end{tabular}
\end{center}

The Aubry-Andr\'{e} (AA) potential is deterministic~\cite{aubry1980analyticity}, while the others are genuinely random. Considering irrational $\beta$ ensures the formation of a quasiperiodic potential, that localizes the eigenfunctions if $W>2t$, while for values $W\leq 2t$, all the eigenstates are delocalized ~\cite{aubry1980analyticity}. Our choice for $\beta$ is the inverse of the golden ratio, $\beta = (\sqrt{5}-1)/2$. Systems subject to this quasi-random potential configuration exhibit no Griffith effects, because there is a complete spatial correlation~\cite{PhysRevB.87.134202}. This type of disorder is often used in experiments~(\cite{Schreiber2015, Bordia2016, Bordia2017, Bordia2017_2, Rispoli2019}).

We expect that the eigenstates in systems with binary disorder distribution to be more delocalized than for other types of on-site disorder, in the regime of large disorder. There are two reasons for that and both are due to the fact that the potential can only take two discrete values.
First, there are intervals with all the sites with exactly the same potential. Within this region, the physics is like the clean tight-binding model, in particular, eigenfunctions are delocalized. Second, the potential can lead to the creation of periodic clusters with the same distribution of the disorder among the sites which  partially restores the translational invariance in sections of the chain. This translation invariance allows for the eigenfunctions to be extended in those regions. Note that the formation of these patterns is independent on the value of $W$. Both possibilities are exponentially suppressed with the number of involved sites in the pattern, allowing the survival of localization in the thermodynamic limit. The effect of these configurations can be clearly observed in our numerical data, leading to much more delocalized eigenfunctions than in the other distributions.

At last, we also consider disorder in the hopping coefficient (bond disorder), for which $t_{i,i+1} \in [1-\Delta t, 1 + \Delta t]$, following a uniform, "box" distribution. When the bond disorder is considered, the local potential $\epsilon_i$ is taken constant and its value does not matter. In this scenario, the Hamiltonian has a chiral symmetry~\cite{PhysRevB.18.569}. One of the physical consequence of the chiral  symmetry is that eigenfunctions come in pairs for each disorder configuration, in the sense that if an eigenfunction energy $E$ is present, then there is another eigenfunction with energy $-E$. Both the localization length and the density of states diverge at energy $|E| \to 0$~\cite{PhysRevB.18.569}. 

Regarding the convention of the disorder strength $W$, we follow the choice in the Anderson localization community and denote it by $W_{\text{And}}$ throughout the paper. For an easier comparison with the standard convention in the MBL community, we will give also the corresponding value, $W_{\text{MBL}}$, which is given by $W_{\text{MBL}} = W_{\text{And}}/4$.

\subsection{Single particle problem: Quantities of interest}

In the numerical study of single particle physics and Anderson localization, we focus on the following quantities:

\begin{enumerate}
\item 
\textbf{The density of states,} $\rho(E)$ and $\text{D}(E, PR)$.

The density of states is the number of states found at a certain energy $E$:

\begin{equation}
    \rho(E) = \frac{1}{L} \sum_{i=1}^L \delta (E - E_i), 
\end{equation}
where $E_i$ is the energy of the $i$-th eigenstate of the system. $\rho(E)$ is normalized: $\int dE\; \rho(E)=1$.

If we consider a discretization of the energy given by discrete energy intervals $I^E_i$ with width $\Delta E_i$, we can write the density of states as:

\begin{equation}
\label{eqn_density_def}
    \rho(E\in I^E_i) = \frac{1}{L \; \Delta E_i} n(I^E_i), 
\end{equation}
where $n(I^E_i)$ counts the number of eigenstates with an energy in the interval $I^E_i$. With the DaC algorithm, we have access to all the energies of the system and we will use Eqn.~\eqref{eqn_density_def} to obtain the density of states from the complete spectrum.

We also define a new quantity, D(E, PR), which is related to the density of states $\rho(E)$, but it additionally depends on the value of the Participation Ratio (PR), defined below in Eqn.~\eqref{eqn_D_E_PR_def}:
\begin{equation}
\begin{split}
\text{D}(E\in I^E_i, &\text{PR} \in I^{\text{PR}}_j) = \\
&\frac{1}{L \; \Delta E_i \Delta \text{PR}_j } n(I^E_i, I^{\text{PR}}_j), 
\end{split}
\label{eqn_D_E_PR_def}
\end{equation}
where $n(I^E_i, I^{\text{PR}}_j)$ counts the number of eigenstates with an energy in the interval $I^E_i$ and PR in the interval $I^{\text{PR}}_j$.

\item
\textbf{The localization length, $\xi$}.
    
The localization length captures how fast the exponentially suppressed tails of the eigenstates decay. For an eigenstate localized around the site $x_0$, the amplitude at a far-away site $x$ is given by the localization length:
\begin{equation}
    |\psi_{x_0}(x)| \lesssim \exp\p{-\frac{\abs{x-x_0}}{\xi}}.
\end{equation}
In Ref.~\cite{Thouless_1972}, a relation between the localization length and the density of states is derived, for random potentials:
\begin{equation}
\xi^{-1} (E) = \int dx \rho(x)\log(\abs{E-x}).
\label{Eqn:localization_E}
\end{equation}

It is also possible to calculate the localization length of an individual eigenstate, $\xi^{-1}(\ket{\psi_\beta})$, with energy $E_\beta$, given by:
\begin{equation}
\label{eqn_calculate_loc_DaC}
\xi^{-1}(\ket{\psi_\beta}) = \lim_{L\to\infty} \frac{1}{L-1} \sum_{\alpha\neq\beta} \ln\p{ E_\beta - E_\alpha }.
\end{equation}
We use Eqn.~\eqref{eqn_calculate_loc_DaC} to calculate the localization length of the eigenstates from the results obtained using the DaC algorithm.

In the limit of weak disorder, there is an analytical solution, shown in~\cite{Kramer_1993}, for the localization length as a function of the energy and the variance of the random potential, $V$:
\begin{equation}
\xi(E, V) = \frac{96}{12V}\sqarep{1-\p{\frac{E}{2t}}^2}.
\label{Eqn:Thouless}
\end{equation}
By inspection of Eqn.~\ref{Eqn:Thouless}, it is clear that this expression only holds for values of the energy $\abs{E} < 2t$. In~\cite{Czycholl1981}, a correction for the localization length at energy zero was derived:
\begin{equation}
\xi(E=0, V) \sim \frac{105t^2}{12V}.
\label{Eqn:correction_localization}
\end{equation}
The physical origin of this correction is an anomaly in the density of states at $E\sim 0$~\cite{Kappus1981}. We will confirm these results using very high statistics later in this paper. 
    
\item
\textbf{The Participation Ratio,} PR.
    
The Participation Ratio of a wavefunction $\ket{\psi}$ is defined as:
\begin{equation}
    \text{PR}(\ket{\psi}) = \frac{1}{\sum_{i} \abs{\braket{i|\psi}}^4 },
\end{equation}
where $\braket{i|\psi}$ is the amplitude of the wavefunction at site $i$. The value of the PR is an indicator of how much (de)localized the wavefunction is. For a plane wave in a system of $L$ sites, the value of the associated PR is $L$. 

\end{enumerate}

\subsection{Two particle problem: Hamiltonian and Observables}

The Hamiltonian $H$ used to study the Two-Interacting Particles (TIP) problem is the following:
\begin{equation}
\label{eqn_H_TIP}
\begin{split}
&H = H_0 + H_I, \\
&H_0 = t \sum_{\braket{i,j}} \p{c_i^\dag c_{j}+c_j^\dag c_{i}}-\sum_{i=1}^{L} \epsilon_i n_i,  \\
&H_I = U \sum_{i=1}^{L-1} n_i n_{i+1} .
\end{split}
\end{equation}
$H_0$ is the same as in the Anderson model and the term $H_I$ adds a nearest-neighbor interaction, where $U$ is referred as the interaction strength. The value of the hopping, $t$, is always homogeneous. The Hamiltonian of the TIP problem has the same form as the standard Hamiltonian to study the MBL phase, namely the XXZ model, when written in terms of fermionic operators. The only difference is an overall factor 1/2, $t = J/2$ and $U = J_z/2$, where $J$ and $J_z$ are the commonly used parameters for the hopping and interaction, in the XXZ model. 

In the regime of strong disorder, we consider three random distributions, namely the box, Gaussian and binary, and the Aubry-Andr\'{e} potential. In the weak disorder regime, only the previous random potentials are considered, not the Aubry-Andr\'{e} potential. The reason is the formation of metallic states at certain values interaction strength, which depend on the disorder strength, when the disorder follows the Aubry-Andr\'{e} potential. The existence of metallic states were first shown in~\cite{Flach2012_4} and studied in more detail in~\cite{Frahm2015_4, Frahm2016_2D}. Such metallic states cannot be obtained with the DaC algorithm. Moreover, since the largest size for the subsystems that we can deal with is $M=200$ sites for the TIP, we cannot distinguish with certainty the appearance of metallic states from localized eigenstates whose localization length is too large to be fitted in a subsystems of size $M=200$ sites.

The observables studied are the energies of the eigenstates and their Participation Ratio (PR) in real space, which is obtained via the normalized probability to find one particle in each site and it can be calculated using the following equation:
\begin{equation}
\text{PR}(\ket{\psi}) = \p{\sum_{x=1}^L n_x^2}^{-1},
\end{equation}
with
\begin{equation}
n_x = \frac{1}{2} \sum_{y=1}^L \abs{ \braket{x,y|\psi} }^2, \quad \sum_{x=1}^L  n_x = 1.
\end{equation}

There are other quantities which are also considered in the community of few-particle physics, like the PR in energy representation and the fluctuations of the center of mass~\cite{Frahm2016_4}. In our study of the TIP problem, we have also calculated all the mentioned observables, but the conclusions are always the same. Therefore, we only present the results regarding the PR in real space.

\section{The Divide-and-Conquer algorithm: General Idea}
\label{DaC_General_idea}

Our algorithm is based on the simple intuition that in a system, in which all eigenfunctions are spatially localized, it should be possible to obtain and represent each  eigenfunctions in an interval centered on that wavefunction and large enough to host the wave function up to some precision. Since the size of these intervals are dictated by the localization length of the wave functions and not by the total system size, there is the potential for a method to scale linearly with systems size $L$, instead of the standard expectation of $L^2$ or $L^3$ for eigenvalue solvers. In the one-dimensional Anderson problem~\cite{Anderson_1958} all eigenfunctions are expected to be localized, so this model is a natural playground to explore these ideas. 

The basic idea of the algorithm we call "Divide-and-Conquer" (DaC) is to solve several subsystems of a large total system and combine their results, instead of solving at once the entire system. Solving one subsystem allows us to find those eigenfunctions of the full system that are localized in the considered subsystem. In order to solve one subsystem, $S$, we need the write down the Hamiltonian $H^S$ which describes the physics only inside $S$. The number of Hamiltonians that we need to solve with the DaC method scales linearly with the system size $L$, but their dimensions scale with on the localization length $\xi$. In the localized regime, where $\xi$ does not depend on $L$, this leads to a much more efficient algorithm, allowing us to obtain all eigenfunctions of the Anderson model in systems up to $L=10^9$, i.e. one billion sites.

The basics steps of the algorithm are the following:

\begin{algorithm}[H]

\caption{Divide-and-Conquer method}
\begin{algorithmic}[1]

\State Generate and solve the Hamiltonian of a subsystem.

\State Discriminate between real and spurious eigenfunctions which arise due to artificial boundary conditions.

\State Eliminate already obtained eigenfunctions (equilibrium) or do the time evolution of a wave function (dynamics). In both cases, calculate the observables of interest.

\end{algorithmic}

\end{algorithm}

\subsection{Splitting the Hamiltonian}
\label{DaC_split}

Given an Hamiltonian, $H$, it can be separated in the two terms, $H^S$, which describes the physics exclusively inside a chosen subsystem $S$, and $H^\text{Env}$, which encodes all the other terms. The original Hamiltonian is the sum of both terms, $H = H^S + H^\text{Env}$. As an example, let us consider a local Hamiltonian $H$, consisting in on-site operators, $h_i^{(1)}$, and two consecutive sites operators, $h_{i, i+1}^{(2)}$:
\begin{equation}
\label{eqn_general_H}
H =\sum_{i=1}^L h_i^{(1)} + \sum_{i=1}^{L-1} h_{i,i+1}^{(2)}.
\end{equation}
If $S$ is a subsystem extending over a number of sites included in the interval $[\alpha, \Omega]$, then we define $H^S$ as
\begin{equation}
H^S = \sum_{i=\alpha}^\Omega h_i^{(1)} + \sum_{i=\alpha}^{\Omega-1} h_{i,i+1}^{(2)}.
\end{equation}

Let us assume we can obtain the eigenfunctions of $H^S$. Due to the artificial boundaries created from detaching the subsystem $S$ from the rest of the system, not all the eigenfunctions of $H^S$ can be embedded to form an eigenfunction of $H$. As it will be explained in step~\ref{variance_criteria}, there is an efficient way to relate the variance of the $\ket{\Phi}$, one of the eigenfunctions of $H^S$, with $|| H^\text{Env}\ket{\Phi} ||^2 $, allowing for the fast discriminate between the real and spurious eigenfunctions.

We want to emphasize that depending on both $H$ and $H^S$, it might happen that none of the eigenfunctions of $H^S$ can be used as a good approximations of the eigenfunctions of $H$.

\subsection{ Dividing the system into subsystems }
\label{DaC_subsystems}

\begin{figure}
\begin{center}
\includegraphics[width=\linewidth]{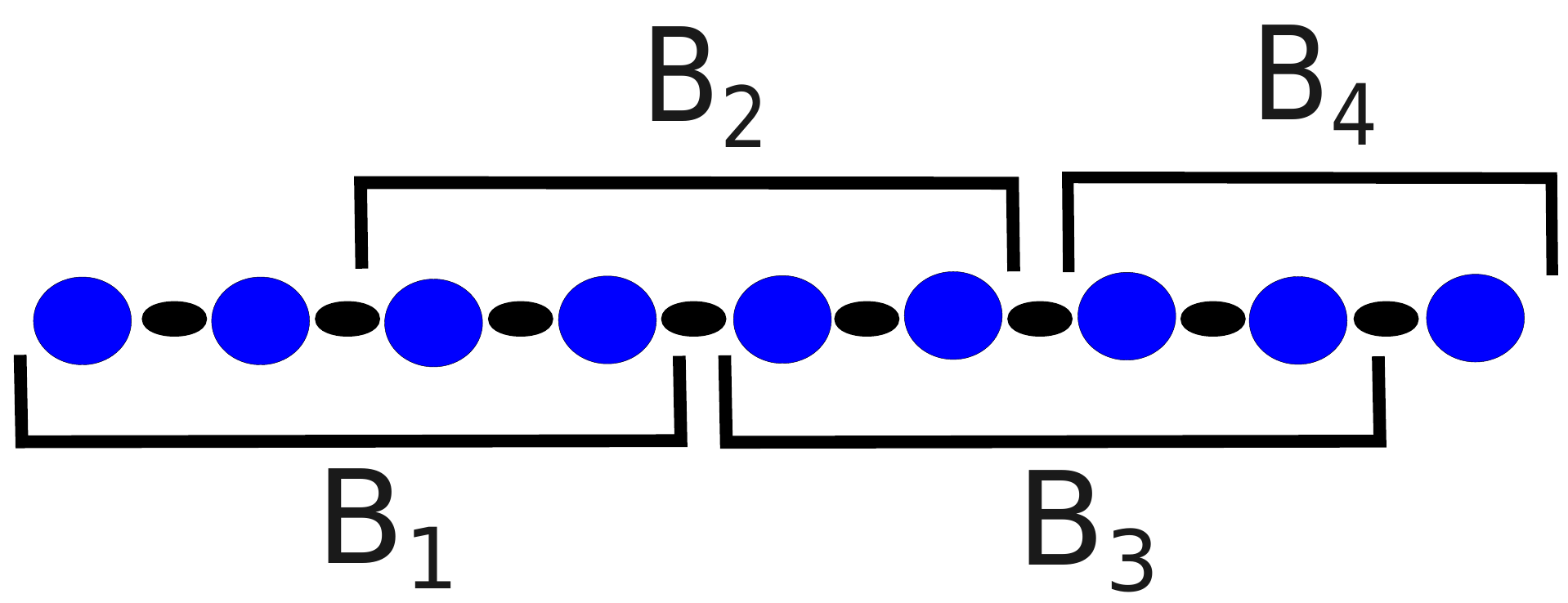}
\end{center}
\caption{ Covering the system using the set of subsystems using the Half-shifted partition, $\{B_i\}$, as explained in the main text.}
\label{fig:cartoon_subsystem}

\end{figure}

Once a method to calculate accurate approximations of the eigenfunctions in a given interval is available, it can be applied into different subsystems of $M$ sites which, once they are combined, cover the full system of $L$ sites. There are several choices for the set of subsystems and we explain three of them:

\begin{enumerate}
\item
\textbf{ Site-by-site partition}

One possible covering of the system is obtained by shifting the subsystems one site at each time. The set of subsystems is given by the intervals $A_i$:
\begin{equation}
A_i = (i, i+M], \quad i\in (0,L-M].
\end{equation}
There are two problems with this choice. First, several eigenfunctions are obtained numerous times, especially if the subsystem size is much larger than the support of some of the eigenfunctions. As a result, it is necessary to eliminate many eigenfunctions, which have been obtained in multiple subsystems. The second issue is the number of subsystems needed to cover the full system, which with this choice is $(L-M)$.

\item

\textbf{Half-shifted partition}

Another choice for the set of subsystems is to consider the intervals $B_i$ defined as 
\begin{equation}
    B_i = \left( i\times M/2, i\times M/2+M \right], \; 0\leq i \leq \lceil{\frac{2L}{M}}\rceil.
\end{equation}
A representation of this covering is displayed in Fig.~\ref{fig:cartoon_subsystem}. Using this division, an eigenfunction can only be found twice, in two consecutive subsystems. The number of subsystems is now $N=\lceil{2L/M}\rceil$ instead of $(L-M)$. A problem related with this choice is that the subsystem size $M$ needed to obtain the same number of eigenfunctions as in the site-by-site option is twice as large. This partition is the one we have implemented for most of the results shown in this paper.
Instead of shifting the subsystem by half its size, we can move the subsystems a factor $\Delta \cdot M$, with $\Delta \leq 0.5$. 

\item

\textbf{ Self-adjusting partition }

It is also possible to determine the minimum size of the subsystems while solving the system. In this approach, we need to decide locally if we have found all the eigenstates with finite overlap with a given site $x$. 

If we solve a subsystem with size $M$ centered on site $x$ and obtain $N$ eigenfunctions, $\{\ket{\psi_{\alpha}}\}_\alpha$, then the single-particle reduced density matrix, RDM$_{1}$, which is obtained from the density matrix tracing out the orthogonal complement of the subspace generated by $\{\ket{\psi_{\alpha}}\}_\alpha$, can be calculated via:
\begin{equation}
\label{def_RDM}
\text{RDM}_{1} = \frac{1}{N} \sum_{\alpha} \ket{\psi_\alpha}\bra{\psi_\alpha}.
\end{equation}
If all eigenstates with a finite overlap at site $x$ have been found, then the corresponding diagonal entry of RDM$_{1} \times N$ is one. If it is not the case, up to numerical precision, then we need to increase the size of the subsystem, keeping the site $x$ at the center, until the corresponding diagonal entry of RDM$_{1} \times N$ is one. Afterwords, we place the following subsystem centered in the first site $y$ where the corresponding entry of RDM$_{1} \times N$ is not one.

\end{enumerate}

\subsection{Efficient discrimination between real and spurious eigenstates}
\label{variance_criteria}
In step~\ref{DaC_split}, we have mentioned that the quantity $|| H^\text{Env}\ket{\Phi} ||^2$ can be used to determine how close $\ket{\Phi}$ is to one of the eigenfunctions of $H$. The reason supporting this statement is the following. If a wavefunction $\ket{\Phi}$ is localized in the interval $S$, then its variance with respect the $H$ is:
\begin{equation}
\sigma^2_{\Phi}(H) = \sigma^2_{\Phi}(H^S) + ||H^\text{Env}\ket{\Phi}||^2,
\label{Eqn:cutoff}
\end{equation} 
where $\sigma^2_{\Phi}(H^S)$ is the variance with respect the Hamiltonian of the subsystem, $H^S$, and $H^\text{Env} = H - H^S$. Moreover, if $\ket{\Phi}$ is an eigenfunction of $H^S$, then $\sigma^2_{\Phi}(H) = ||H^\text{Env}\ket{\Phi}||^2$. For such localized wavefunctions, the calculation of $||H^\text{Env}\ket{\Phi}||^2$ can be done efficiently, at least for local Hamiltonians, since $H^\text{Env}$ acts mainly outside of the interval where $\ket{\Phi}$ is localized. As an example, let us consider the Hamiltonian of Eqn.~\ref{eqn_general_H} and a wavefunction $\ket{\Phi}$ localized in an interval which starts at site $\alpha$ and it finishes at site $\Omega$. In this case, $||H^\text{Env}\ket{\Phi}||^2$ can be calculate efficiently as the following:
\begin{equation}
||H^\text{Env}\ket{\Phi}||^2 = || h_{\alpha-1,\alpha}^{(2)} \ket{\Phi}||^2 + ||h_{\Omega,\Omega+1}^{(2)} \ket{\Phi}||^2.
\label{Eqn:Variance_general}
\end{equation}

\begin{figure}
\begin{center}
\includegraphics[width=\linewidth]{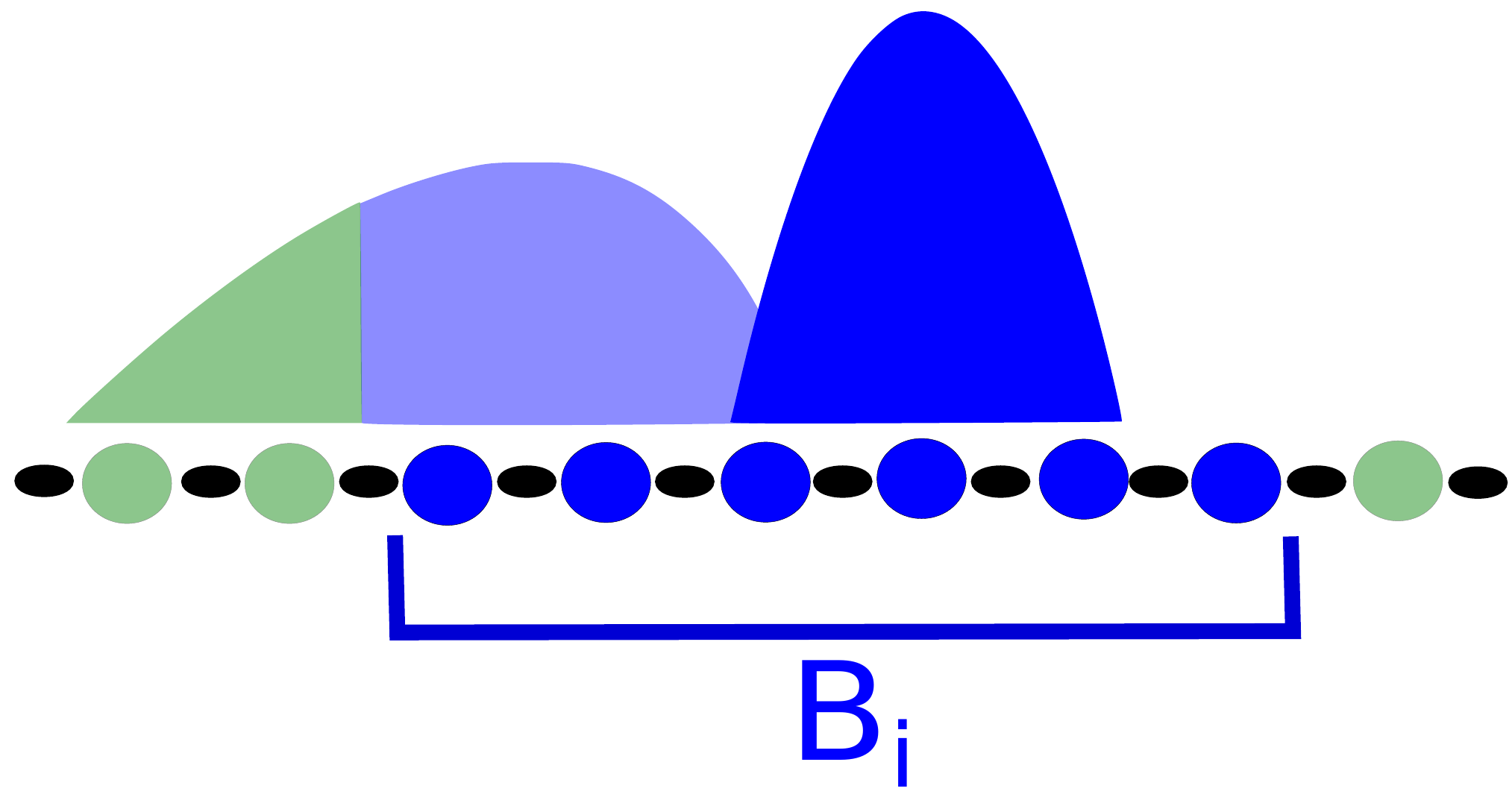}
\end{center}
\caption{ Cartoon representation of the physical intuition behind the efficient discrimination between real and spurious eigenstates. When solving the subsystem $B_i$, consisting on the blue circles, we encounter two types of eigenstates. One of them are the eigenstates that do not have an overlap with the boundaries of $B_i$, like the one represented in opaque blue, and the other type are those who have an overlap with the edges of the subsystem, like the one in translucent blue. The former can be embedded to form a real eigenstate of the system, while the latter is an artifact of the virtual boundaries imposed by the DaC algorithm, the real eigenstate also include contributions from outside the interval $B_i$, represented in translucent green. The spurious eigenstate must be disregard. }
\label{fig:cartoon_variance}

\end{figure}

Physically, the previous equations tells us that, given an eigenfunction of the subsystem with zero amplitude at the edges of the subsystem, then it can be embedded into an eigenfunction of the system. For a cartoon representation of this idea, see Fig.~\ref{fig:cartoon_variance}.

\subsection{ Combination the results of different subsystem}
\label{ordering}

Our algorithm combines the set of obtained eigenfunctions in different subsystems in order to retrieve the eigenfunctions of the full Hamiltonian $H$.  However, the same eigenfunction could be found in different overlapping subsystems. Therefore, it must be checked whether there are repeated eigenfunctions, and if so, store only a single one of them.

Two steps are applied in order to determine whether an eigenfunction has already been found. First, the absolute value of the energy difference among the eigenfuctions is calculated. If for a given eigenfunction, its energy difference with all the others is larger than any possible numerical error, then it is considered as a new eigenfunction. Otherwise, the absolute value of the scalar product between the considered eigenfunction and the eigenfunctions of the previous subsystem is calculated. If the scalar product is smaller than a given cutoff $\theta$, then we have obtained a new eigenfunction. This approach guarantees that the final set of eigenfunctions are orthogonal up to the cutoff $\theta$. In order to determine if the chosen value of $\theta$ is small enough, we calculate the population on each site, via the single-particle reduced density matrix, in order to make sure that it is not larger than one. Another method to discriminate new and repeated eigenfunctions would be to apply the Gram-Schmidt process, where there is no need to add an arbitrary cutoff, but with the caveat that it is much slower method than the scalar product.

\subsection{Parallelization}

It is straightforward to apply the DaC algorithm in parallel, since the set of subsystems to be solved can be split into different nodes. 

Let us assume that the set of subsystems to solve is $\{B_1, B_2, ..., B_M\}$. If we want to obtain the eigenfunctions which are localized on the subsystems $B_j$, there is no need to solve any of the previous subsystems $B_i$, with $i=1,..,j-1$. 

In order to avoid storing several times the same eigenfunction, we must check which of the obtained eigenfunctions on the subsystem $B_j$ can also be found in previous subsystems. Note that if two subsystems do not overlap, then it is not possible to obtain the same eigenfunctions from both subsystems. Therefore, before storing any of the eigenfunctions obtained in a subsystem $B_j$, we must solve the previous overlapping subsystems. The number of overlapping subsystems depends on how one decides to divide the system.

\section{The Divide-and-Conquer algorithm: Single Particles}
\label{sec:DaC_single_particle}

In this section, we adapt the general ideas of the DaC algorithm to the specific case of the disordered tight-binding model, described by the Hamiltonian:
\begin{equation}
H = \sum_{\braket{i,j}} t_{i,j} \p{c_i^\dag c_{j}+c_j^\dag c_{i}}-\sum_{i=1}^{L} \epsilon_i n_i.
\label{H_Anderson}
\end{equation}

\subsection{Efficient discrimination between real and spurious eigenvectors}

Applying Eqn.~\ref{Eqn:Variance_general} for the Hamiltonian in Eqn.~\ref{H_Anderson}, the calculation of the variance of $\ket{\Phi}$ is:
\begin{equation}
||H^\text{Env}\ket{\Phi}||^2 = t_{\alpha-1, \alpha}^2 \abs{\Phi_{\alpha-1}}^2 +t_{\Omega, \Omega+1}^2 \abs{\Phi_{\Omega}}^2,
\label{Eqn:Variance}
\end{equation}
where $\Phi_i$ is the amplitude of the state $\ket{\Phi}$ on the site $i$ and $\alpha, \Omega$ are the first and last site of the subsystem, respectively.

\subsection{Termination criteria}

A termination criterion for the algorithm to stop is also required. An obvious choice is once all the $L$ eigenfunctions are obtained. From the discussion in step~\ref{ordering}, we know that the set of obtained eigenfunctions might not be orthogonal. In order to make sure that the set of obtained eigenfunctions are linearly independent and orthogonal, the trace of the single-particle reduced density matrix, defined in Eqn.~\ref{def_RDM}, is also calculated. If the trace is one up to numerical precision, then the algorithm has solved the problem. Otherwise, we need to increase the size of the subsystems to obtain the missing eigenfunctions. Note that there is no need to start from scratch, we can increase only the subsystems where the missing eigenfunctions have support. In order to determine those sites, we can look which elements of the diagonal entries of the single-particle reduced density are not one. The missing eigenfunctions have support on the corresponding sites.

\subsection{ Time evolution }
\label{time_evolution}

Using our scheme we can also investigate the time evolution of a spatially localized initial wave function. 

Due to the localized nature of the eigenfunctions, only a few of them have a finite overlap with a site localized initial wave-function. Thus, we only need those eigenfunctions in order to calculate the time evolution of the wave-function. Moreover, if only a certain precision $\epsilon$ is required, we can further reduce the number of eigenfunctions needed and the accuracy of the (approximated) eigenfunctions. Regarding the number of eigenfunctions, we can disregard those with an overlap with the wave-function smaller than $\delta$, chosen such that the error in the observable is bounded by $\epsilon$. Similarly, if an error of order $\delta$ in the amplitude of the eigenfunction leads to an error when calculating the value of the observable smaller than $\epsilon$, we can consider as a valid eigenfunctions those whose variance is smaller than $\delta^2$. The exact relation between $\delta$ and $\epsilon$ depend on the observable considered. Since we calculate the time evolution of the wave-function using (a subset of) the eigenfunctions of the system, we can target any time of interest with the same computational cost, in particular, we are not limited to small times.

\begin{figure*}
\centering
\begin{tabular}{c c}
\includegraphics[width=0.5\linewidth]{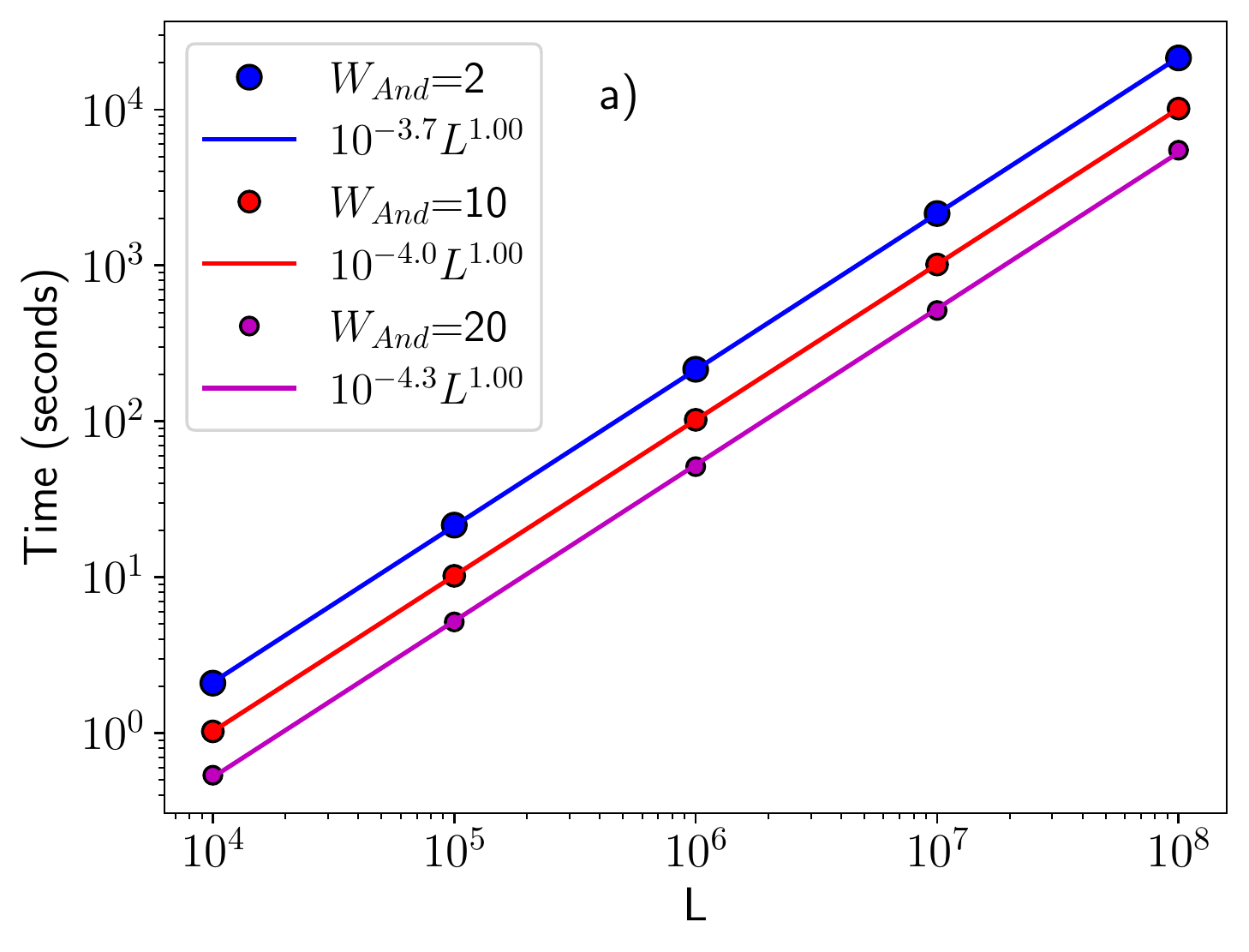} &
\includegraphics[width=0.5\linewidth]{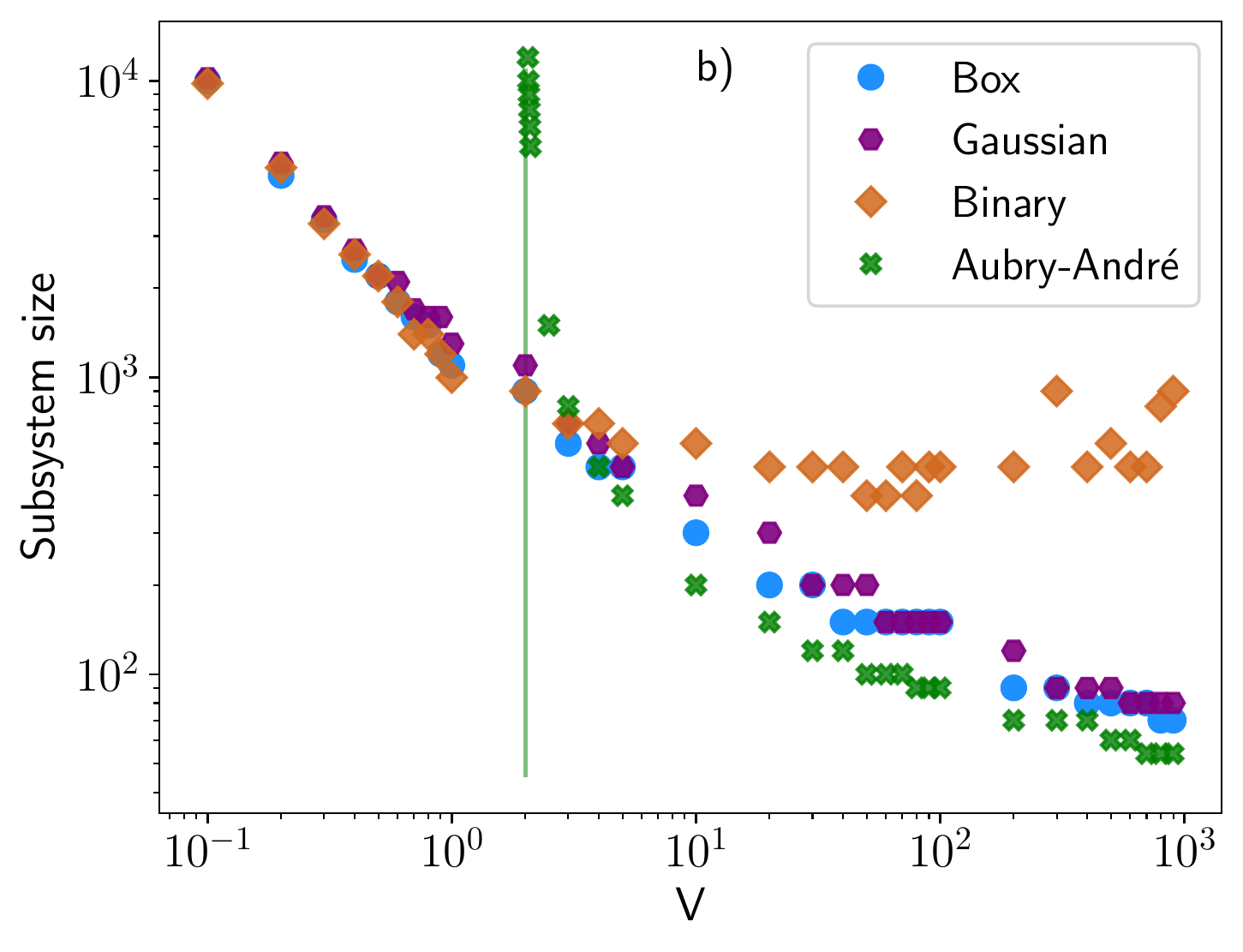}

\end{tabular}
\caption{ Algorithmic features. Left, scaling of the time (using 4 cores of a Intel Core i7-7700 CPU) with system size. In all cases, a linear scaling is achieved. Right, scaling of the subsystem size needed to fully solve a system with $L=10^6$ sites as a function of the variance $V$. The vertical green line at $V=2$ indicates the phase transition of the Aubry-Andr\'{e} model.  }
\label{fig:scaling_alg}

\end{figure*}

\subsection{Performance of the algorithm}

Before applying the DaC approach to study the Anderson problem, we want to point out the expected scaling of the resources with the system size $L$. Let us assume that with subsystems of size $M$ the termination criteria are fulfilled and $M$ is $L$-independent. In this scenario, there is a linear scaling in time with respect to $L$, $O(L\times F(M)/M)$, where $F(M)$ is the cost of solving one subsystem of size $M$. For the single-particle physics, the dimension of the Hilbert space is the same as the length of the system and the cost of diagonalizing a generic Hamiltonian scales as $F(M)=M^3$. For the Anderson model, since we are dealing with a tridiagonal matrix, the diagonalization costs scales as $F(M) = M^2$, implying that the scaling of our algorithm is $O(L\times M)$.

We want to remark that despite the good scaling of our algorithm with the system size $L$, there is also an scaling with $M$, which is related to the localization length $\xi$ of the eigenfunctions. The value of the localization length depends both on the variance $V$ of the potential and on the type of disorder. Note that in the case of $\xi \sim L$, we recover again the expected scaling for tridiagonal matrices, $O(L^2)$. This is the ergodic regime, where our method do not provide any advantage compare with ED.

Even in the localized regime ($\xi \ll L$), the assumption that $M$ does not depend on $L$ is not guaranteed to hold. The reason for is that in larger systems, it is more likely to find larger ergodic regions where the local fluctuations of the potential are smaller than its average fluctuations. This would lead to a larger localization length $\xi$ for the eigenfunctions on that region. Larger values of $\xi$ requires a larger subsystem size, $\overline{M} > M$. We only need to use $\overline{M}$ on that specific region, outside of the region, we can still work with subsystems of size $M$. Therefore, this does not affect the overall performance of the method.

In Fig.~\ref{fig:scaling_alg}, the performance of the algorithm is shown for increasing system size and for decreasing variance of the potential $V$. The data showing the scaling of the computation time with respect the system size $L$, in Fig.~\ref{fig:scaling_alg}a), was obtained using 4 cores of a Intel Core i7-7700 CPU. The disorder follows a box distribution distribution with three different disorder strengths, $W_{\text{And}}=2, 10, 20$. The size of the markers is related with the size of the subsystem, $M$, with values of $M=1000, 500, 250$ respectively. We can see the linear scaling in all the cases in all the cases, but the total time depends on the value of the subsystem size needed to obtain all the eigenstates.

The plot of Fig.~\ref{fig:scaling_alg}b) shows the scaling of the subsystem size, $M$, with the disorder strength of the potential, which is characterized by the its variance, $V$. Three different types of random potentials are considered, namely, the box, Gaussian and binary distribution, and one deterministic and quasiperiodic potential, the Aubry-Andr\'{e} potential. 

For the random potentials, the values of $M$ are quite similar in the regime of small values of the variance, $V < 1$. The difference between the type of random potential is visible at large values of the variance. On one hand, the values of the subsystem size for the binary disorder saturate at a value $M\sim 600$. On the other hand, for the box and Gaussian distributions, the values of $M$ decreases as the variance increases, obtaining all the eigenstates with $M< 100$ for the largest values of $V$. For the Aubry-Andr\'{e} potential, in the regime of strong variance, the eigenstates are more localized than for the random distributions, leading to smaller sizes of the subsystem. 
When the variance of the potential approaches to the value two, there is a divergence in the size of the subsystem, indicating that we are reaching a phase with delocalized eigenstates.

\section{Numerical results for single particles}
\label{sec:Num_results}

Memory and time constraints limit the maximum subsystem size to $M_\text{max} = 26.000$ sites. The cutoff on the maximum subsystem size is responsible for a minimum variance of the potential that can be solved, that is  $V_c=0.05$ ($W_{\text{And}}^{\text{box}, 0}\sim 0.77$, $W_{\text{And}}^{\text{bi}, 0}, W_{\text{And}}^{\text{G}, 0} \sim 0.44$). Therefore, systems with disorder lower than $V_c$ cannot be fully solved by using subsystems of $M_\text{max}$ or less sites. To calculate the variance of the candidates for eigenfunctions  $\ket{\tilde{\Phi}}$, we use Eqn.~\ref{Eqn:Variance}. In the case of strong disorder, the cutoff for the variance is $\sigma_0^2 = 10^{-32}$, ensuring numerical precision of the level of the amplitude of the eigenfunctions. In the weak disorder regime, in order to reduce the subsystem size needed, we reduce the cutoff of the variance to $\sigma_0^2 = 10^{-16}$, leading to a precision of observables of order $O(10^{-8})$, which it is small enough for our purposes. In order to combine the different solutions of the subsystems (see \ref{ordering}), we consider that eigenfunctions with an overlap smaller than $\Theta = 10^{-5}$ to be different. In the dynamics, for weak disorder regime, we calculate the values of the observables with a precision of $ 10^{-1}$, while in the strong disorder regime, the errors are bounded by $10^{-3}$. All energy scales are expressed in units of the hopping parameter, $t=1$.

\begin{figure}
\centering
\begin{center}
\includegraphics[width=\linewidth]{4D_box_W5d0.png} \\
\includegraphics[width=\linewidth]{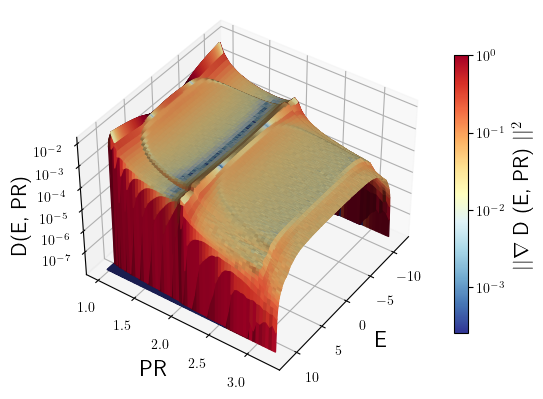} \\
\includegraphics[width=\linewidth]{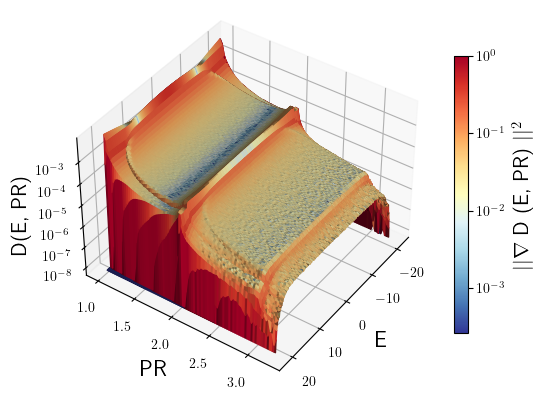}
\end{center}
\caption{ Histograms of energy, PR and density of states, from a system of  $L=10^9$ sites with a potential following a box distribution. From top to bottom, the disorder strength is $W_{\text{And}}=10, 20, 40$ ($W_{\text{MBL}} = 2.5, 5, 10$). The color is obtained from the norm of the gradient of the density of states.}
\label{fig:E_PR_eigen_W10}

\end{figure}

\subsection{Strong disorder physics: Properties of Eigenfunctions}

We start our study of the Anderson problem when the disorder strength is much stronger than the hopping. In this parameter range, the localization length of the eigenfunctions is very small (apart from the discussed challenges for the binary disorder), so we expect to find all the eigenfunctions by applying the DaC algorithm with small subsystems, allowing us to solve systems of $L=10^9$ sites.

From the full set of eigenfunctions obtained for three different values of the disorder strength, $W_{\text{And}}=10, 20, 40$ ($W_{\text{MBL}} = 2.5, 5, 10$), we can generate 4D histograms, displayed in Fig.~\ref{fig:E_PR_eigen_W10}, of the Energy-Participation Ratio-Density of states in the axis and the color-scale from the norm of the gradient of the density of states. 

We proceed to describe the patterns in those three plots, their similarities and differences. In all the cases, there is a pronounced enhancement of the density of states at integer PR. The physical origin of this property are the resonances between an number of nearby sites, where the disorder potential has similar values on a series of consecutive sites, which allows for the delocalization of the eigenstates on those sites, contributing to an integer value of the PR for such eigenstates.  This feature is shared with the other types of distributions, c.f.~Appendix~\ref{appendix_box_G}.

An even more intriguing feature is the appearance of ridges  and approximate plateaus of the density of states. These patterns seem only to appear for the box distribution, and not for the other distributions considered (see Appendix~\ref{appendix_box_G}).

The rather sharp discontinuity at the edge of the spectrum is related to the separation between the regions with finite and zero density of states, and is pronounced here because of the bounded disorder (box) distribution. There is yet another discontinuity of $\rho$, a {\em ridge}, which starts symmetrically at small PR and zero energy, and which continues in a curved way towards larger values of $|E|$ as the PR grows. While this ridge is visible, but not very pronounced in Fig.~\ref{fig:E_PR_eigen_W10}, its effect is amplified in Fig.~\ref{example_intro}(b) (and Fig.~\ref{fig:derivative_E_PR_several_W_strong} in Appendix ~\ref{appendix_box_G}), where only the gradient is displayed. 

In order to visualize better the previously explained structures, in Appendix~\ref{appendix_box_G}, we include two types of histograms of the Energy-PR, one with the color scale given by the density of states and the color scale of the other type is obtained from the derivative of the density of states with respect to the energy. Also, in order to make sure that the previously explained features are not an artifact of imposing a dependency of the PR on the density of states, in Fig.~\ref{fig:density_box_vs_G} and Fig.~\ref{fig:localization_box_vs_G} in Appendix~\ref{appendix_box_G}, we show density of states (only energy dependent), the localization length and their derivatives for the box and Gaussian distributions, emphasizing their differences.

\subsection{Strong disorder physics: Dynamics of localized initial states}

With our method, we can also study the properties of states out-of-equilibrium. In particular, we will look into the dynamics of the set $\{\ket{\psi_i(t)}, \; i\in[1,L]\}$, where $\ket{\psi_i(0)} = a_i^\dag \ket{0}$. The wavefunctions $\ket{\psi_i(0)}$ represent a particle fully localized at site $i$. As previously, the observable of interest is the Participation Ratio (PR) and the system size is $L=10^9$ sites.

\begin{figure}[h]
\begin{center}
\includegraphics[width=1\linewidth]{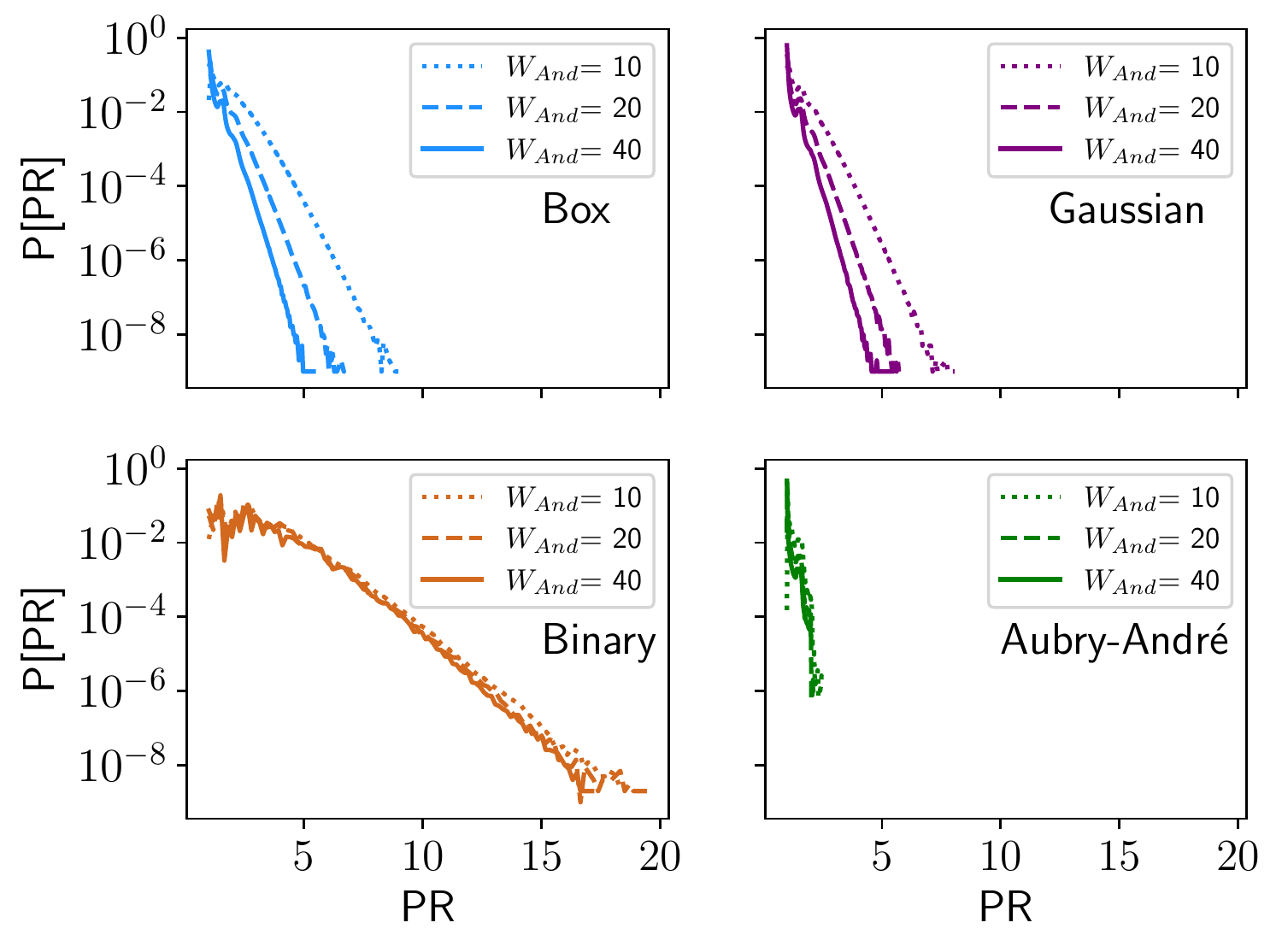}
\end{center}
\caption{ Distribution of the PR in the long-time limit of $L = 10^9$ wavefunctions, each of them initially localized in one site of the system. We consider three disorder strengths, $W_{\text{And}}=10, 20, 40$ ($W_{\text{MBL}} = 2.5, 5, 10$), represented by a dotted, dashed and solid lines, respectively. In each plot, a different type of distribution is used, namely box, Gaussian, binary and Aubry-Andr\'e distributions.}
\label{fig:dyn_W_large}

\end{figure}

In Fig.~\ref{fig:dyn_W_large}, we show the distribution of the $L$ values of the PR in the long-time limit, $\{\text{PR}(\ket{\psi_i(t\to\infty)}), \; i\in[1,L]\}$. The value of each of the $\text{PR}(\ket{\psi_i(t\to\infty)})$ have been obtained from the average over eleven values of the time, evenly distributed in the interval $t\in[9.500, 10.500]$.

We consider the same three values of the disorder strength as before, $W_{\text{And}}=10, 20, 40 \; (W_{\text{MBL}}=2.5, 5, 10)$ and disorder following box, Gaussian, binary and the Aubry-Andr\'e distributions. In all the random potentials, the distribution of the PR exhibits quite extended tails, specially in the binary disorder, leading into a significant difference between the mode and mean value of the PR. Note that in the binary distribution, there is almost no difference in the distributions for the considered values of the disorder strength, since the dynamics for large PR is governed by the number of sites with the same value of potential, which creates translation invariant regions, where the particles can freely propagate. The size of the regions does not depend on the value of $W$, but it is exponentially suppressed. For the Aubry-Andr\'e model, which is deterministic, we can observe much lower values of PR and the distribution is not as broad as in the other cases. The physical reason for the narrow distribution of the PR is the lack of rare-regions, where the fluctuations of the potential are smaller than the average fluctuation. Such regions would lead to large fluctuations in the values of the PR, but they are not present in the Aubry-Andr\'e model.

\subsection{Weak disorder physics: Properties of Eigenfunctions}

Finding the eigenfunctions in systems with a low disorder strength $W$ is more numerically demanding, since the localization length and the required subsystem size are much larger than for system with strong disorder. This regime is of interest for the Anderson problem in one dimension, where the variance $V$ is expected to be the only parameter that determines the physics of the problem and the microscopic details of the potential are irrelevant. Even a small disorder is able to drastically change the physics of the system compared to a clean tight-binding model.

In the following, we will reproduce some of the results of weakly disordered system according to existing literature, which we will use to back-test the results of our algorithm and its accuracy. We consider systems of $L=10^8$ sites and the random potentials $\epsilon_i$ distributed according to a box, Gaussian and binary distribution.

\begin{figure}
\centering
\begin{center}
\includegraphics[width=\linewidth]{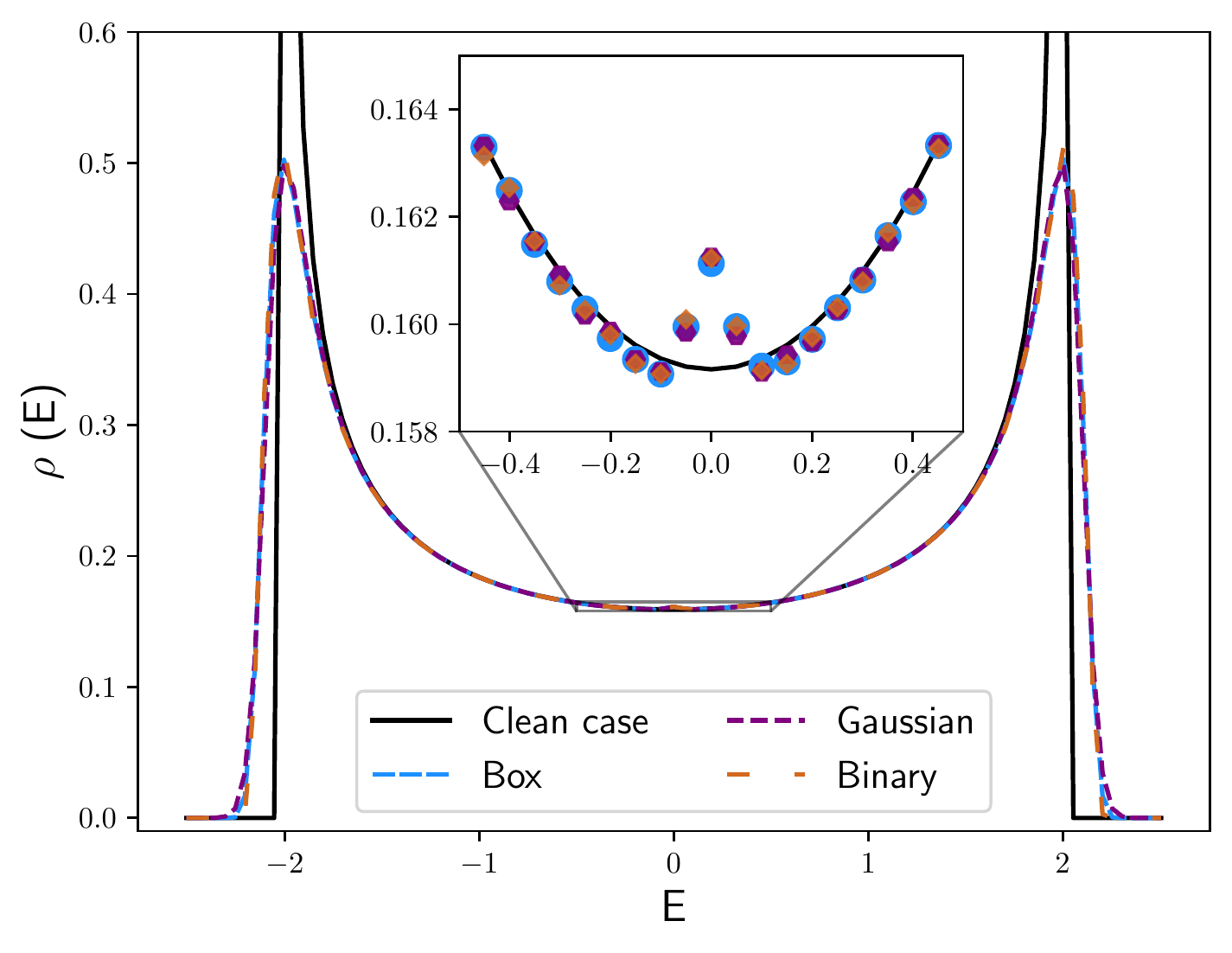}
\end{center}
\caption{ Density of states as function of the energy obtained from the full spectrum of a system of $L=10^8$ sites. The disorder distribution has variance $V=0.05$ and it follows a box, Gaussian and binary distribution. In black, for comparison, it is displayed the density of states for the clean case.}
\label{fig:density_V0d05}

\end{figure}

The values of the density of states for all energies can be seen in Fig.~\ref{fig:density_V0d05}, where we have added the density of states for the clean case as a black dashed line for a better comparison. In the inset plot, we focus on the middle of the spectrum, where a cusp in the density of states at $E=0$, already predicted in~\cite{Kappus1981}, is visible. The values of the density of states are independent of the microscopic details.

\begin{figure}
\centering
\begin{center}
\includegraphics[width=\linewidth]{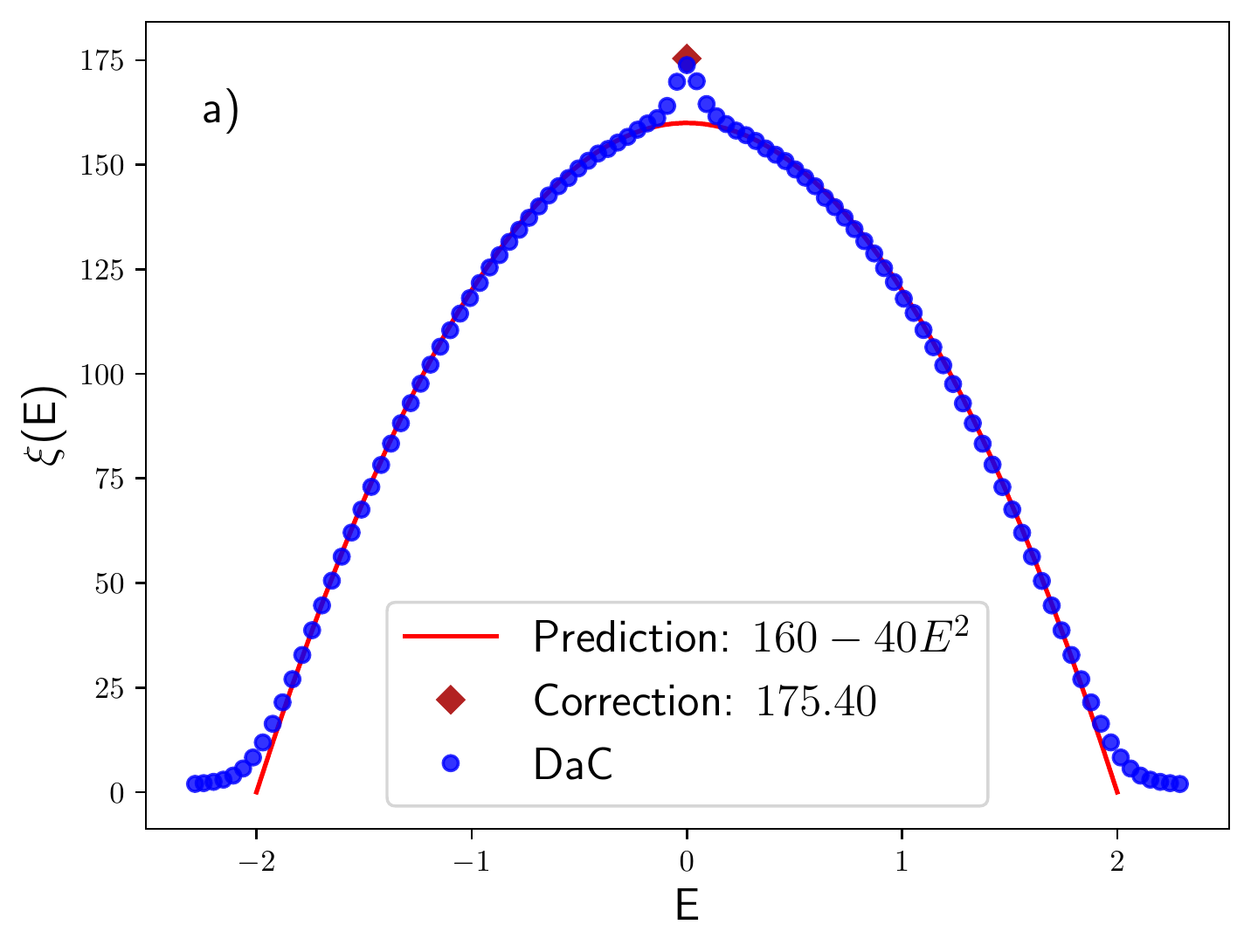}
\includegraphics[width=\linewidth]{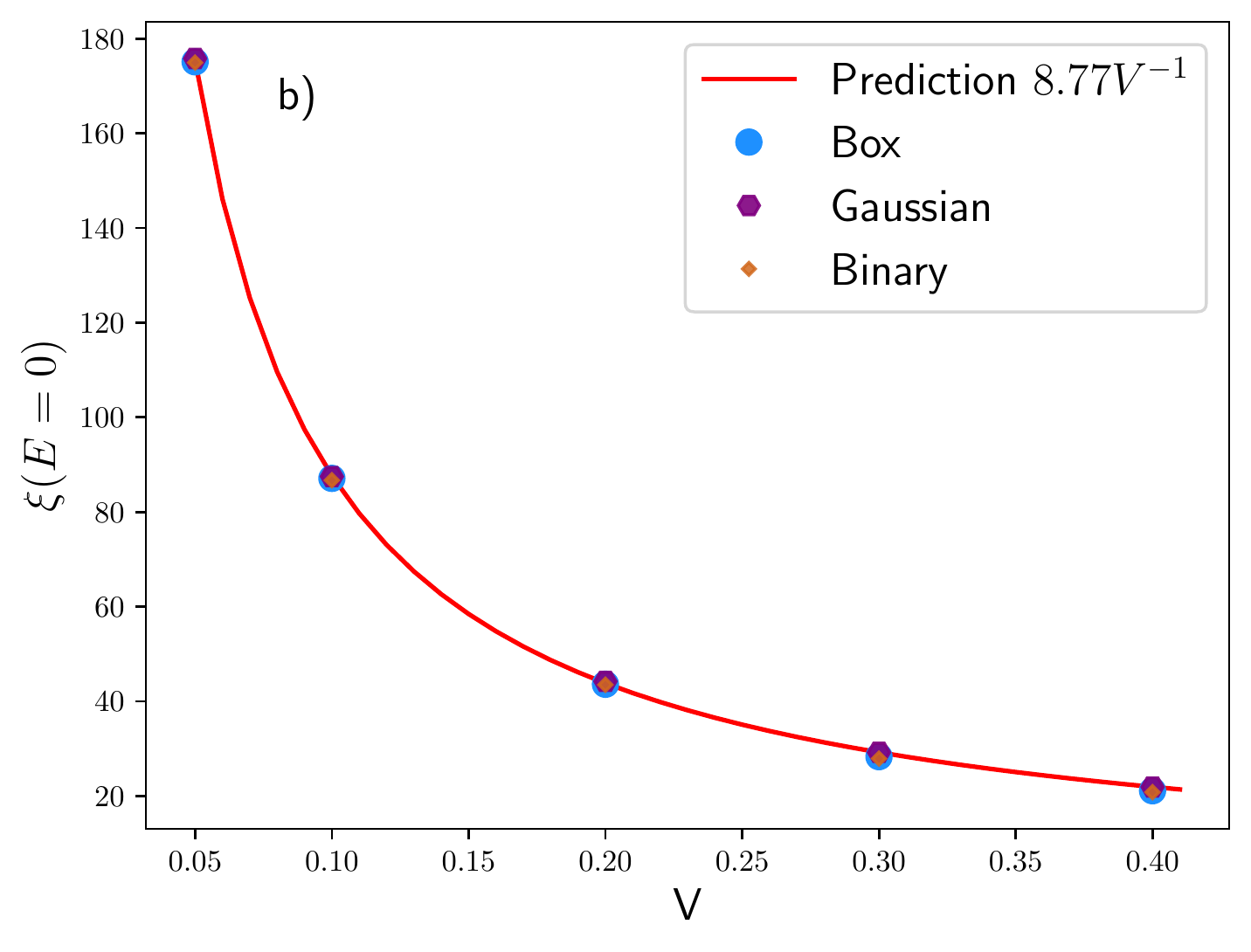}
\end{center}
\caption{ Values of the localization length obtained from solving a system of $L=10^8$ sites compared with the theoretical prediction. In the top plot, it is displayed the dependence of the localization length on the energy, for a potential with variance $V=0.05$. In the bottom plot, the scaling of the localization length at energy zero as a function of the variance of the potential. }
\label{fig:localization_V0d05}

\end{figure}

The DaC algorithm gives us access to the full spectrum, consisting of $L=10^8$ energies. This information combined with Eqn.~\ref{eqn_calculate_loc_DaC}, allows us to calculate the localization length, $\xi$, for some of the eigenstates. In Fig.~\ref{fig:localization_V0d05}, we display the behavior of the localization length, compared with the analytical results, given in Eqn.~s~\ref{Eqn:Thouless} and~\ref{Eqn:correction_localization}. In Fig.~\ref{fig:localization_V0d05}a), we show the energy dependence of the localization length, from the data obtained using the DaC algorithm and a potential following a box distribution with variance $V=0.05$, reproducing the expected results. In Fig.~\ref{fig:localization_V0d05}b), the dependence of the localization length at energy zero on the variance of the potential is shown and it follows the predicted value from perturbation theory. As in Fig.~\ref{fig:density_V0d05}, the results do not depend on the microscopic details, i.e., the type of disorder.

We have also calculated the distribution of energy gap, $r$, and it follows a Poissonian distribution as expected in localized regime~\cite{PhysRevB.75.155111} (see results in Appendix~\ref{appendix}).

\begin{figure}
\centering
\begin{center}
\includegraphics[width=1\linewidth]{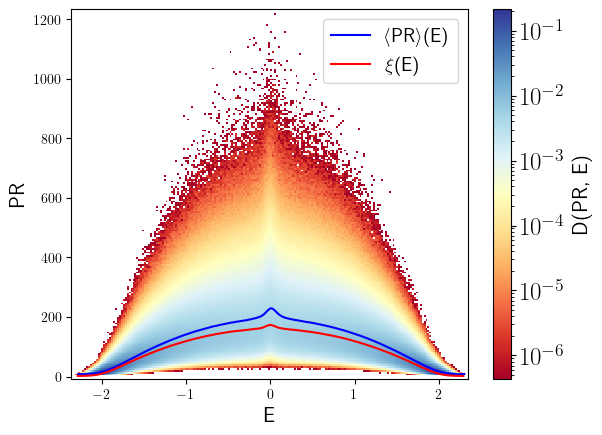} 
\end{center}
\caption{ Histogram of energy-PR of the eigenstates of a system of $L=10^8$ sites. The potential has a variance $V=0.05$ and it follows a box distribution. The solid blue line is the mean PR and the solid red line is the localization length. The color is determined by the value of the density of states $\rho$.}
\label{fig:E_PR_eigen_V0d05}

\end{figure}

It is important to mention that our method transcend the simple recovering of the already known physics for the Anderson model. In addition we can generate full histograms of Energy-PR, and example of which is displayed in Fig.~\ref{fig:E_PR_eigen_V0d05}. We only show data for the box distribution, since there is no visible difference when the other distributions are considered. In Fig.~\ref{fig:E_PR_eigen_V0d05}, the colormap represents the density of states, the blue solid line is the mean value of the PR for the corresponding energy window, and the red solid line is the localization length, $\xi(E)$. The distribution of the PR is quite broad, specially in the middle of the spectrum, and it reaches values up to 1.200 for our system, while the maximum value of localization length is only around 175. The mean value of the PR follows more closely the localization length, exhibiting the same energy dependence and similar values, as we can see in the solid lines of Fig.~\ref{fig:E_PR_eigen_V0d05}. Note also the white region, where the density of states is 0, at the middle of the spectrum for low PR. This is not due to lack of statistics, but it comes from the fact that eigenfunctions in the middle of the spectrum decay slower than the ones with energies at the edge of the spectrum, leading to larger values of the PR.

\subsection{Weak disorder physics: Dynamics of localized initial states}

We have also studied how the distribution of the PR for a set of $L=10^8$ wavefunctions, $\{\ket{\psi_i(t)}, \; i\in[1,L]\}$ evolves in time in the weak disorder regime. As before, the initial state of the wavefunction $\ket{\psi_i(t)}$ is one particle localized at site $i$, $\ket{\psi_i(0)} = a_i^\dag \ket{0}$.

\begin{figure}[h]
\centering
\begin{center}
\includegraphics[width=1\linewidth]{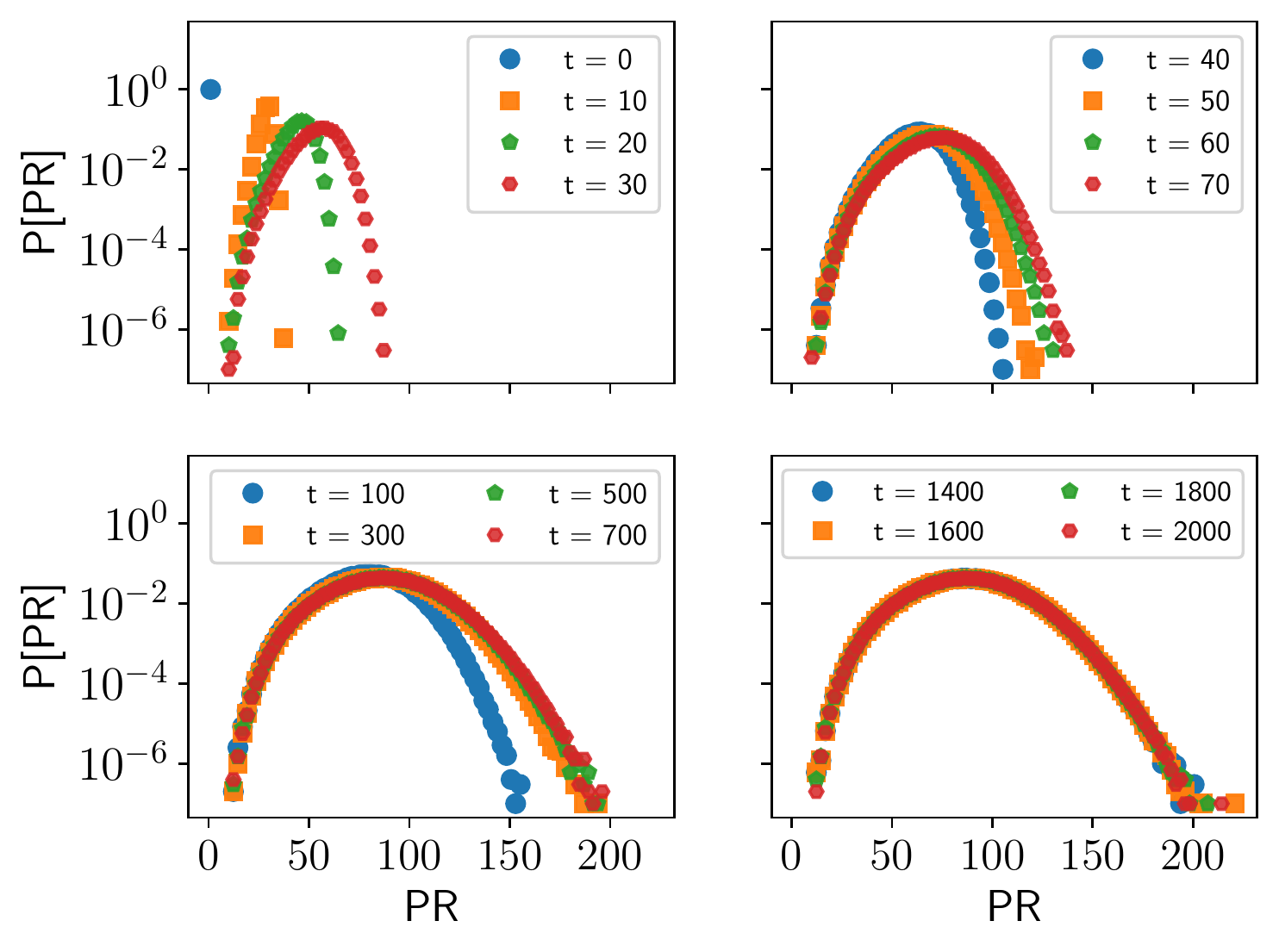}
\end{center}
\caption{ Distribution of PR, for a system of $L=10^8$ sites with a potential from the box distribution with variance $V=0.05$ at different times $\text{t}$. We consider $L$ different wavefunctions, each of them initially localized in one of the $L$ sites of the system. }
\label{fig:dyn_V0d05}

\end{figure}

In Fig.~\ref{fig:dyn_V0d05}, we display the distribution of the PR for several times $t$. We can see how the peak and the width of the distribution are spreading, until converging at $t \geq t_f \sim 500$. 

Due to the choice of the initial state, the distribution of the PR in the dynamics is not as broad as in the case of the eigenfunctions. Since the eigenfunctions that are more delocalized have a smaller overlap with any of the considered initial state, compared with the more localized eigenfunctions on the corresponding site, their contribution to the PR is greatly reduced by the more localized eigenfunctions. The maximum value of the PR in the dynamics is around the mean value of PR from the eigenfunctions.

\subsection{Localized vs. extended eigenfunctions}

In this section, we want to show how the algorithm deals with systems with both localized and delocalized eigenfunctions. Since the DaC algorithm does not impose any a priori physics, only those eigenfunctions which are localized can be found with our method.

Distinguishing with certainty between a localized eigenfunction with a large localization length, which are not obtain via DaC for a subsystem of size $M$, and a truly delocalized eigenfunction is beyond the scope of our DaC algorithm. Therefore our method should not be used for the purpose of identifying the ergodic-localized phase transition. Still, our method can provide hints regarding where the phase transition might be. If for a certain parameter regime we obtain all eigenfunctions for larger and larger systems without increasing the subsystem size $M$, then it can be affirmed that the system is in the localized regime, providing an empirical upper bound for the ergodic-localized phase transition.

\subsubsection{Aubry-Andr\'e model}

The Aubry-Andr\'e model presents a phase transition between ergodic and localized eigenstates at $W = 2t$~\cite{aubry1980analyticity}. We expect that the DaC is able to find all the eigenfunctions for $W > 2t$, and being unable to find any eigenfunctions for $W \leq 2t$.

\begin{figure}[h!]
\begin{center}
\includegraphics[width=\linewidth]{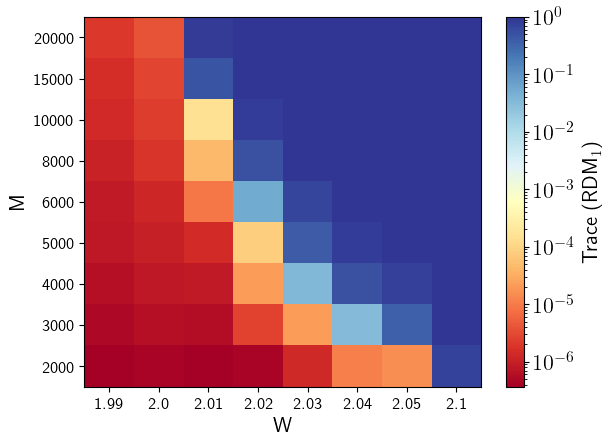}
\end{center}
\caption{ Histogram of the trace of the Reduced Density Matrix for the single particle, RDM$_{1}$, for several values of the subsystem size $M$ and disorder strength $W$ in the Aubry-Andr\'e model, in a system of $L=10^8$ sites. At $W=2.01t$ and $M=20.000$, the trace of the density of states is 0.998, not one. For $W \geq 2.02t$, the maximum value obtained for the trace of the density of states is one, indicating that all eigenfunctions have been obtained. Note the
particular ticks on the $y$-axis.}
\label{fig:phase_AA}

\end{figure}

In Fig.~\ref{fig:phase_AA}, we show the trace of the single-particle reduced density matrix, RDM$_{1}$, obtained from the eigenstates found in a system of $L=10^8$ sites, for several values of $M$. If we find all eigenstates, then the trace of RDM$_{1}$ is one.

From the data obtained via the DaC algorithm, the regime $W > 2t$ is profoundly different from the regime at $W\leq 2t$. For the strongest values of disorder $W$ considered, we find all the eigenfunctions for quite small subsystem sizes $M \sim 2.000$. For parameters closer to the transition point, we see a significant change in the trace of RDM$_{1}$ by increasing $M$. For $W=2.01t$ and $M=20.000$, there are around $0.2\%$ eigenfunctions missing. Despite our algorithm is not able to find all eigenfunctions with the maximum subsystem size considered, the value of $M$ has a considerable effect on the proportion of eigenstates obtained, suggesting that the system is in the localized regime. For $W \leq 2t$, only a very small proportion of the spectrum is obtained, around 300 out of $10^8$ eigenfunctions, for all the considered $M$, which varies from 2.000 to 20.000. Moreover, the obtained eigenfunctions are found on the edges on the system, so they are localized due to the boundaries of the system. 

From our results we can ensure that the system is fully localized when the disorder strength is $W \geq 2.02t$.  It is likely that for $W = 2.01t$ the system is also fully localized, but a subsystem size $M=20.000$ is still too small to capture the eigenfunctions with the largest PR present in the system. For $W \leq 2t$ the system looks like fully delocalized. Without any prior knowledge of the physics of the Aubry-Andr\'e model, our data suggests a phase transitions for $2t \leq W_c < 2.01t$.

\subsubsection{Bond disorder}

The last type of disorder considered in the single-particle physics is the bond disorder, where there is no on-site potential on the system, but there is a random hopping term. Both the density of states and the localization length in these systems diverge at $E \to 0$~\cite{PhysRevB.56.12970}, therefore the DaC algorithm should not be able to find all the eigenfunctions, in particular the ones at the middle of the spectrum with large localization length. 

\begin{figure}
\centering
\begin{center}
\includegraphics[width=1\linewidth]{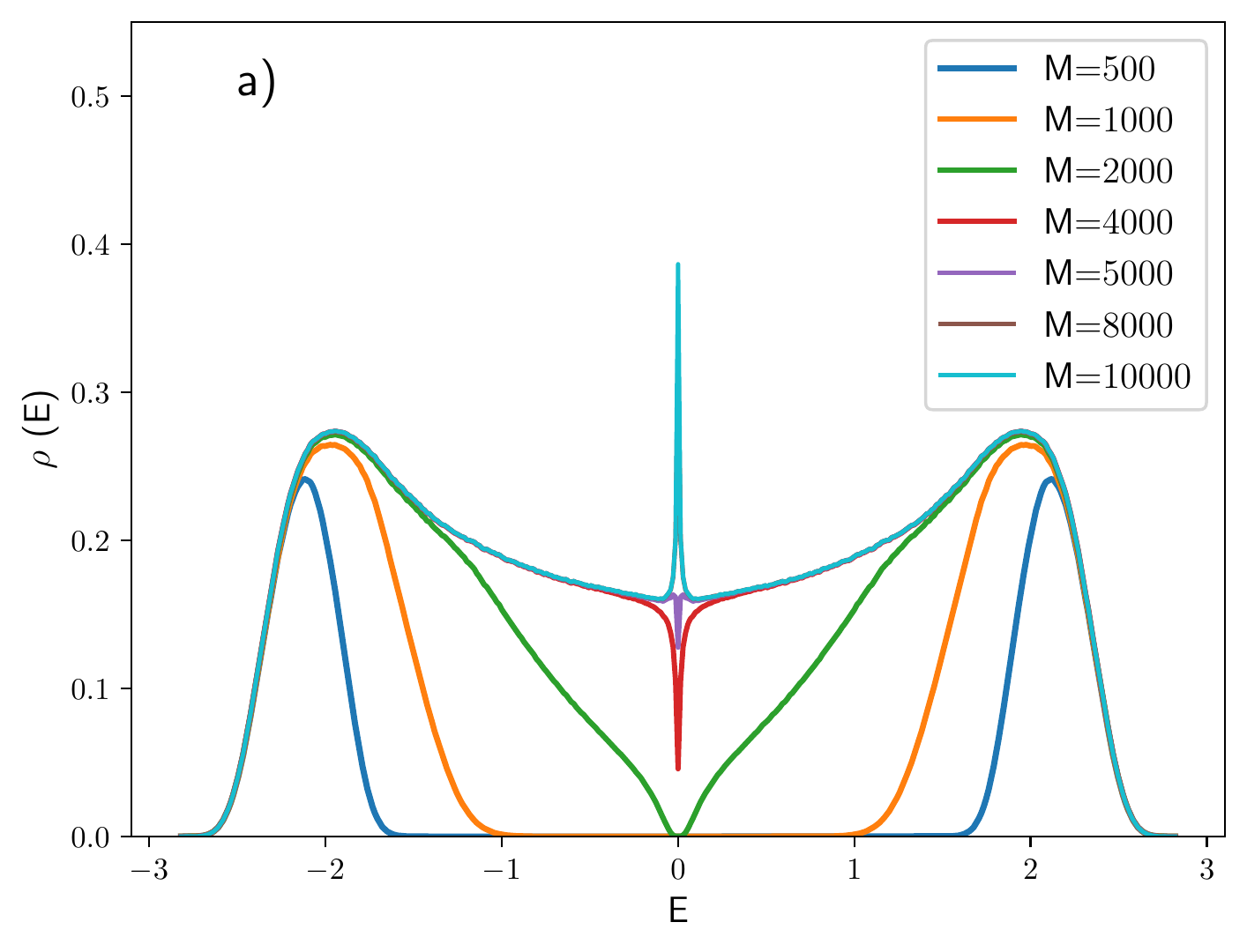} \\
\includegraphics[width=1\linewidth]{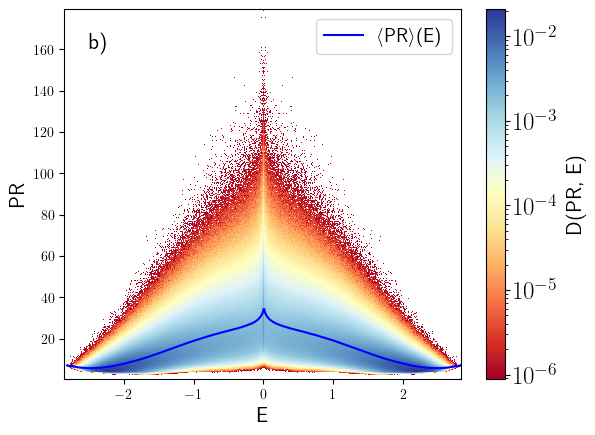}
\end{center}
\caption{ Top, density of states, $\rho$, as a function of energy, $E$, from several subsystem size $M$. The system has $L=10^8$ sites and the bond disorder has strength $\Delta t = 0.5$. Bottom, histogram of the energy-PR from the obtained eigenstates with $M=10.000$. We are able to find more than $99.8\%$ of the eigenfunctions when $M=10.000$. The missing eigenstates are expected to have energy close to zero and be delocalized, thus the mean value of the PR at energy zero obtained via DaC algorithm underestimates the real value of the mean PR at energy zero.}
\label{fig:rho_bond}

\end{figure}

We apply the DaC algorithm with different subsystem sizes, $M$, in order to count the number of eigenfunctions that we are able to obtain for each subsystem size. We consider a system of $L = 10^8$ sites and with the strength of the bond disorder being $\Delta t = 0.5$. In Fig.~\ref{fig:rho_bond}a), the density of states is displayed for several values of $M$. While for small values of $M$ we only obtain eigenfunctions at the edge of the energy-band, enlarging $M$ allows us to find increasingly more eigenfunctions, even at $E\sim0$. For values of the subsystem size $M\geq 4.000$, the density of states outside the central region have converged, indicating that we have found all the eigenfunctions at those energy windows. Most remarkable, the divergence at $E\sim 0$ starts to be visible, for $M\geq 8.000$. 


On the plot of Fig.~\ref{fig:rho_bond}b), we show the histogram of the Energy-PR, from the data obtained using a subsystem of size $M=10^4$ sites. For this value of $M$, we are able to find $99.8\%$ of the eigenstates, therefore the overall structure of the histogram is correct. The eigenstates at the edges of the spectrum are quite localized, with low values of the PR. As the energy of the eigenstates gets closer to zero, both the mean and maximum values of the PR increases. At energy zero, we can see a kink in the mean PR, indicating a divergence in the localization length. There is also an accumulation of data at energy zero, as expected from the divergence in the density of states. 

The actual value of the mean PR at energy zero obtained from the DaC algorithm is not reliable, since all the missing eigenstates are expected to have energy close to zero and they are delocalized. Their delocalization implies that the corresponding value of the PR depends on the system size $L$ ($L = 10^8$), and they have a large influence in the mean PR at energy zero. We want to emphasize that not all the eigenstates with energy close to zero are delocalized, some of them have quite small value of the PR, indicating a strong localization.

\section{The Divide-and-Conquer algorithm: Two Particles}
\label{sec_alg_TIP}

In the second part of the paper, we will deal with the problem of Two Interacting Particles (TIP), where we focus on the effect of the interactions on both the eigenstates and the dynamics. Before we can apply the DaC algorithm to learn about the properties of the TIP, we need to apply some modifications to the algorithm, in order to adapt it to this new application scenario.

In the single particle problem case our goal was to calculate all the individual $L$ eigenfunctions. The TIP however has $L\times(L-1)/2$ eigenfunctions. With our short range (nearest neighbor) interactions, we can clearly see that in our scheme two non-interacting eigenfunctions located in two separate intervals will not be affected by the presence of the interactions. So our goal is {\em not} to calculate all those unaffected eigenfunctions, because this is redundant information already obtained in the single particle case. We therefore focus mainly on those TIP eigenfunctions which are potentially affected by the interactions by finding TIP eigenfunctions in subsystems of size $M$, much smaller than $L$. 

This change of perspective induces a major modification in the algorithm  related to the termination criterion, but other minor technical modifications are also needed. The basic features of the DaC algorithm remain the same, see Section~\ref{DaC_General_idea} for details.

\subsection{Choice of the appropriate basis for the Hamiltonian}

The dimension of the Hilbert space for two spinless fermionic particles in a subsystem of size $M$ is $D = M(M-1)/2$, but the number of non-zero entries scales linearly with $M$, suggesting that we are dealing with a banded matrix.

If we would consider the standard computational basis: $\{\ket{1,2},...\ket{1, M}, \ket{2,3}, ...\ket{M-1, M}\}$, the band-width of the matrix would be $M$. If we apply the Reverse Cuthill-McKee algorithm~\cite{Cuthill1969_4, RCM_4}, we obtain a matrix with a band-width of $M/2$. The maximum size for the subsystem that we can solve is $M=200$, whose Hilbert space dimension is $19.900$.

\subsection{Efficient calculation of the variance}

Applying Eqn.~\ref{Eqn:Variance_general} for the Hamiltonian in Eqn.~\ref{eqn_H_TIP}, the calculation of the variance of $\ket{\Phi}$ is:
\begin{equation}
||H^\text{Env}\ket{\Phi}||^2 = t_{\alpha-1, \alpha}^2 \abs{\Phi_{\alpha-1}}^2 +t_{\Omega, \Omega+1}^2 \abs{\Phi_{\Omega}}^2,
\label{Eqn:Variance_TIP}
\end{equation}
where $\alpha$ is the first site of the subsystem, $\Omega$ is the last site and $\abs{\Phi_i}^2 = \sum_j \abs{\braket{i, j|\Phi}}^2 $ is the probability to find one of the two particles at site $i$.

\subsection{Termination Criterion}
\label{criteria_stop_N2}

Finding a criterion to determine if the subsystem size $M$ is large enough, in the context of the TIP problem, is the major modification that must be implemented in DaC algorithm, since we need to incorporate new physical intuition into the code.

Before we explain the termination criterion, we must mention again that our main goal in the TIP problem is to study the effect of a nearest-neighbor interaction, described by the term $H_I$ in Eqn.~\ref{eqn_H_TIP}, on the full set of eigenstates. Due to our choice of interaction and the localized nature of the eigenstates of single-particle sector, we only need to consider those eigenfunctions of the two-particle sector whose mean distance between particles is either smaller or similar to the sum of their respective single-particle localization length.

We proceed now to explain the termination criterion, for a justification, see Appendix~\ref{appendix_justify_stop_N2}. Given a subset of eigenstates, $\{\phi_\alpha\}$, of the Hamiltonian $H$, defined in Eqn.~\ref{eqn_H_TIP}, we define a quantity which we refer as mean missing population, $\overline{\epsilon}$, as follows:
\begin{equation}
\overline{\epsilon}(\{\phi_\alpha\}) = 1 - \frac{1}{L-1}\sum_{i=1}^{L-1}\sum_\alpha \abs{\braket{\phi_\alpha |i,i+1}}^2.
\end{equation}

The termination criterion is reached when the mean missing population, for a certain subsystem size $M_0$, is smaller than a given cutoff. If we would still increase the value of the subsystem size, we will obtain more eigenstates, but these additional eigenstates will be independent of the value of the interaction.

\begin{figure}
\centering
\begin{tabular}{c c}
\includegraphics[width=0.5\linewidth]{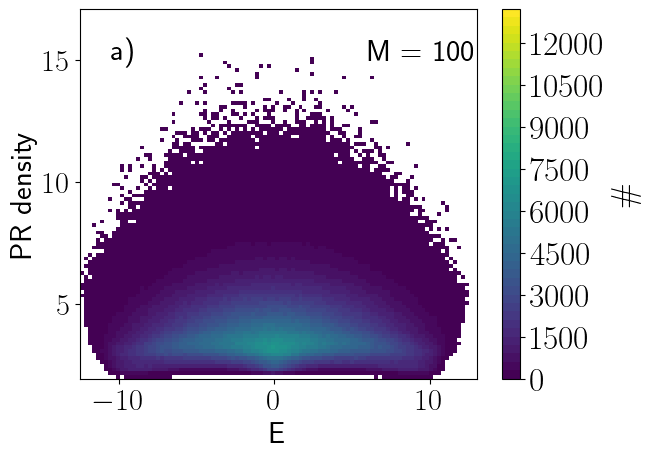} &
\includegraphics[width=0.5\linewidth]{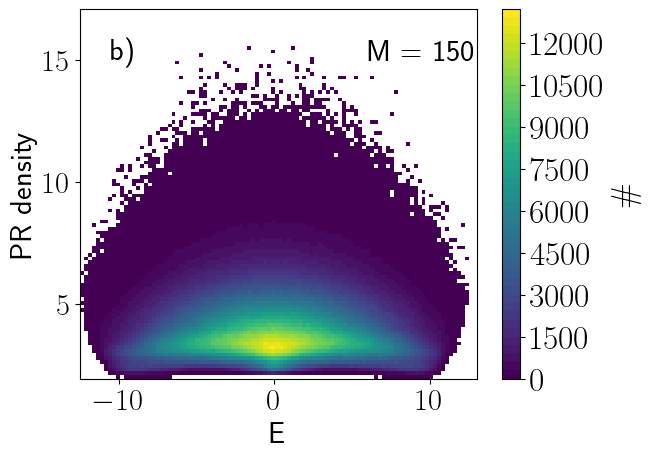} \\
\includegraphics[width=0.5\linewidth]{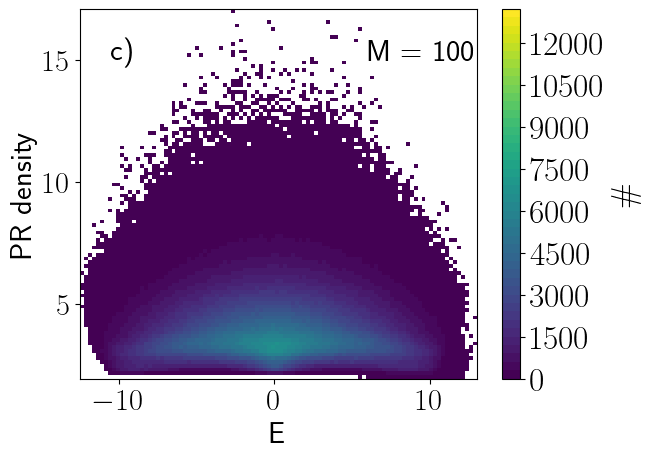} &
\includegraphics[width=0.5\linewidth]{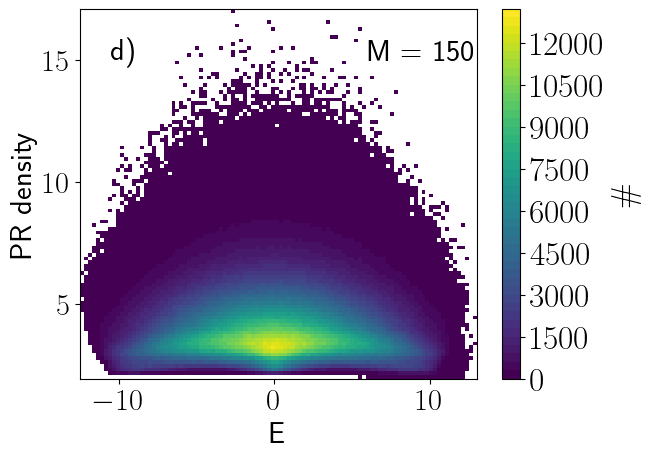} \\
\includegraphics[width=0.5\linewidth]{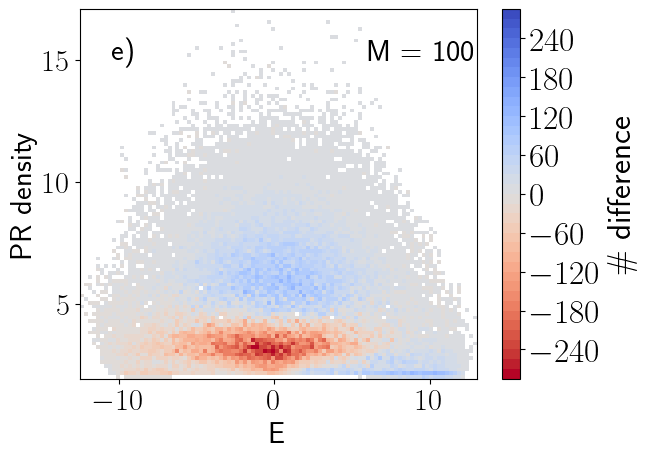} &
\includegraphics[width=0.5\linewidth]{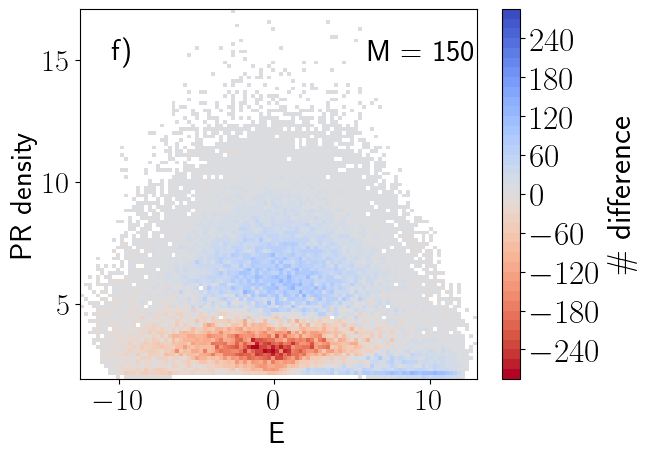}
\end{tabular}
\caption{Histograms generated from the data obtained with the DaC method in a system of $L=10^5$ sites with the random potential following a box distribution, with disorder strength $W_{\text{And}} = 10$ $(W_{\text{MBL}} = 2.5)$, using a subsystem of size $M=100$ sites (left column) and $M=150$ sites (right column). In (a) and (b), the histograms display the number of eigenfunctions obtained when $U = 0$, for the corresponding value of $M$. In (c) and (d), the histograms display the number of eigenfunctions obtained when $U = 2$, for the corresponding value of $M$. 
The histograms (e) and (f) are obtained subtracting the histograms on (c, a) and (d, b), respectively, displaying the effect of the interaction over the energy and PR.
}
\label{fig:effect_M}

\end{figure}

The plots in Fig.~\ref{fig:effect_M}, obtained from a system of $L=10^5$ sites with a potential from a box distribution with a disorder strength $W_{\text{And}} = 10$ $(W_{\text{MBL}} = 2.5)$, provide the numerical justifications of the previous claims. In Fig.~\ref{fig:effect_M}(a,b,c,d), we can see the histograms of the Energy and density PR from the subset of obtained eigenstates, using the DaC algorithm for two sizes for the subsystem, namely $M = 100$ (a,c) and $M = 150$ (b,d). There is a clear effect of subsystem size, leading to an increasing number of eigenstates found for larger values of $M$.

If we consider Fig.~\ref{fig:effect_M}(e,f), where the effect interaction, consisting in moving data from the red to the blue regions, is displayed, then there is no visible difference between the two histograms, indicating that the results have converged in $M$. The corresponding mean missing population for the interactions and subsystems size used is shown in Table~\ref{tab:mean_missing}.

\begin{table}[]
    \centering
\begin{tabular}{|c|c|c|}
\hline
	$M$ & $U$ & Mean missing population \\
\hline
\hline
	100 & 0 & $6.3\cdot 10^{-5}$ \\
\hline
	100 & 2 & $9.4\cdot 10^{-5}$ \\
\hline
	150 & 0 & $1.2\cdot 10^{-14}$ \\
\hline
	150 & 2 & $1.1\cdot 10^{-5}$ \\
\hline
\end{tabular}
\caption{Table showing the mean missing population depending on the subsystem size $M$ and the interaction $U$, for the system considered in Fig.~\ref{fig:effect_M}.}
\label{tab:mean_missing}
\end{table}

\section{Numerical results for two interacting particles}
\label{sec_num_TIP}

Our main goal is to study which is the effect of the interaction strength $U$ on the full set of eigenfunctions for large systems consisting of $L=10^6$ sites, which has a Hilbert space of dimension $D_L = 5\cdot 10^{11}$.

In order to obtain histograms showing the effect of the interaction, we apply the DaC method with exactly the same subsystems to all the considered interaction strengths $U$ and for the non-interacting case, when $U$ is equal to zero. 

In order to calculate the variance of the candidates for eigenfunctions  $\ket{\tilde{\Phi}}$, we use Eqn.~\ref{Eqn:Variance_TIP}, with a cutoff for the variance $\sigma_0^2 = 10^{-16}$, leading to a precision of observables of order $O(10^{-8})$. In order to combine the different solutions of the subsystems (see \ref{ordering}), we consider that eigenfunctions with an overlap smaller than $\Theta = 10^{-7}$ to be different. In the dynamics, for weak disorder regime, we calculate the values of the observables with a precision of $O(10^{-2})$, while in the strong disorder regime, the errors are of the order of $10^{-5}$. All energy scales are expressed in units of the hopping parameter, $t=1$.

\subsection{Strong disorder physics: Effect of the interaction on the Eigenfunctions }

We start by studying of how the eigenstates of the TIP problem are affected by the value of the interaction strength in the regime of strong disorder, where the disorder strength has a value of $W_{\text{And}} = 40$ ($W_{\text{MBL}} = 10$). As before, we consider the potentials box, Gaussian, binary random distributions and the Aubry-Andr\'{e} potential.

\begin{figure}
\centering
\begin{tabular}{c c}
\includegraphics[width=0.5\linewidth]{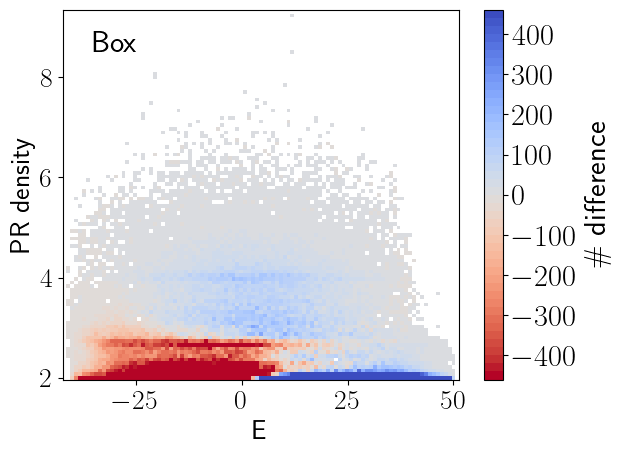} &
\includegraphics[width=0.5\linewidth]{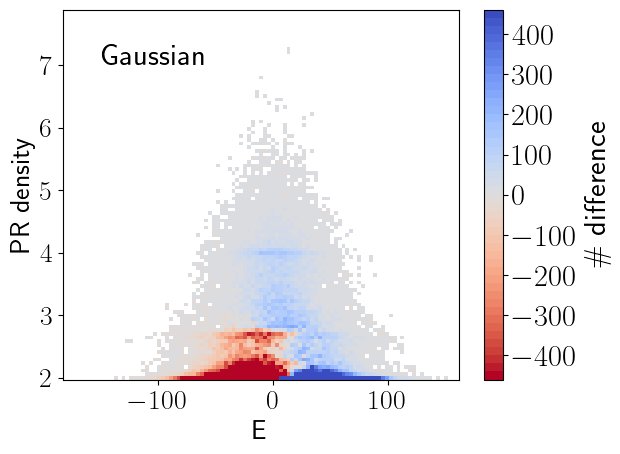} \\
\includegraphics[width=0.5\linewidth]{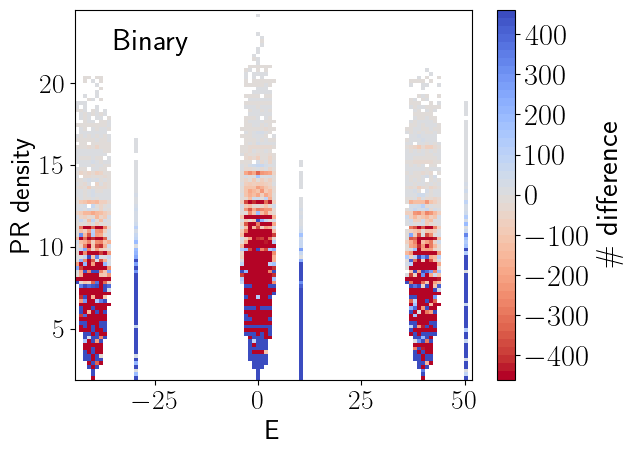} &
\includegraphics[width=0.5\linewidth]{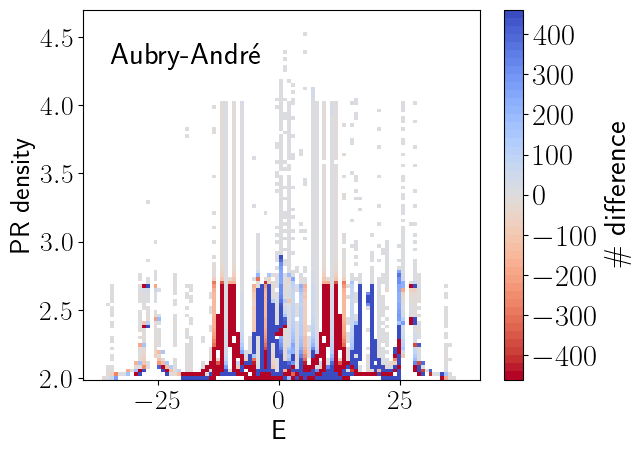}
\end{tabular}
\caption{ Histograms showing the effect of the interaction strength $U = 10$, on the distribution of the energy and PR for systems of $L=10^6$ sites. The disorder follows the usual four potential distributions, indicated in the plots, and the disorder strength is $W_{\text{And}} = 40$ ($W_{\text{MBL}} = 10$). The interaction moves the data from red to blue areas.
}
\label{fig:effect_Jz_histo_E_PR_Wlarge}

\end{figure}

In the histograms of Fig.~\ref{fig:effect_Jz_histo_E_PR_Wlarge}, we can visualize the role of the interaction, with strength $U = 10$, in the distributions of the energy and PR and the impact of the type of distribution for the disorder. Graphically, the interaction move data from the red to the blue regions.

In the strong disorder regime, the major contribution of the interaction is to shift the energy by a term $U$, and this is common in all four distribution. There are also other effects of the interaction. In the box and Gaussian distributions, a faint blue region at the middle of the spectrum and on top of the red region starts to be visible, suggesting the appearance of interaction-induced delocalization effect~\cite{Schmidtke2017_4}, even at this large value of the disorder strength. 

In the binary disorder we see the opposite effect, the blue regions are below the red regions, indicating that the interaction leads towards more localized eigenstates. From the data regarding the Aubry-Andr\'{e} model in the strong disorder regime, it is not clear if the interaction induces delocalization or localization.

\subsection{Strong disorder physics: Effect of the interaction in the dynamics}

\begin{figure}
\centering
\begin{tabular}{c c}
\includegraphics[width=0.5\linewidth]{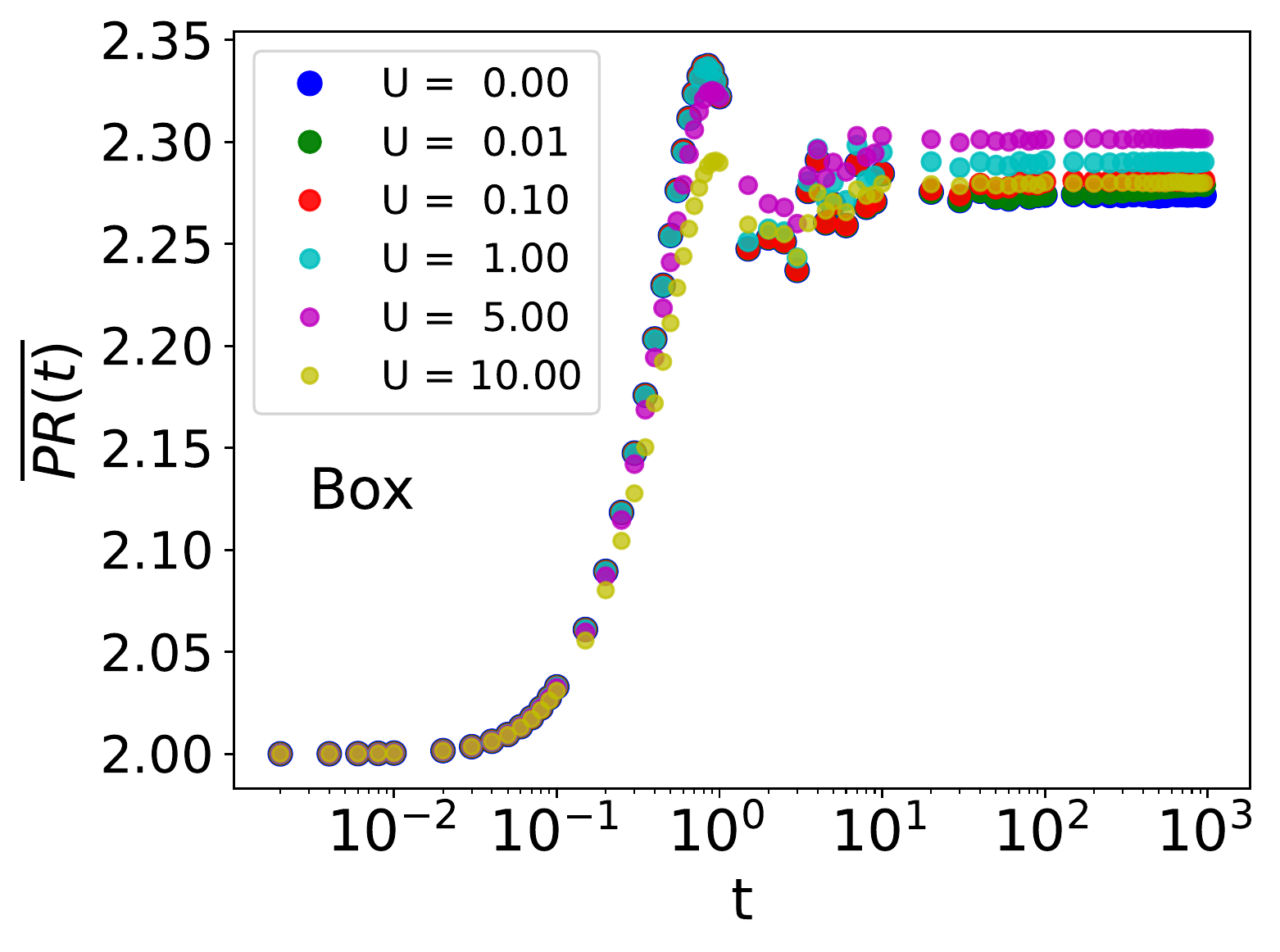} &
\includegraphics[width=0.5\linewidth]{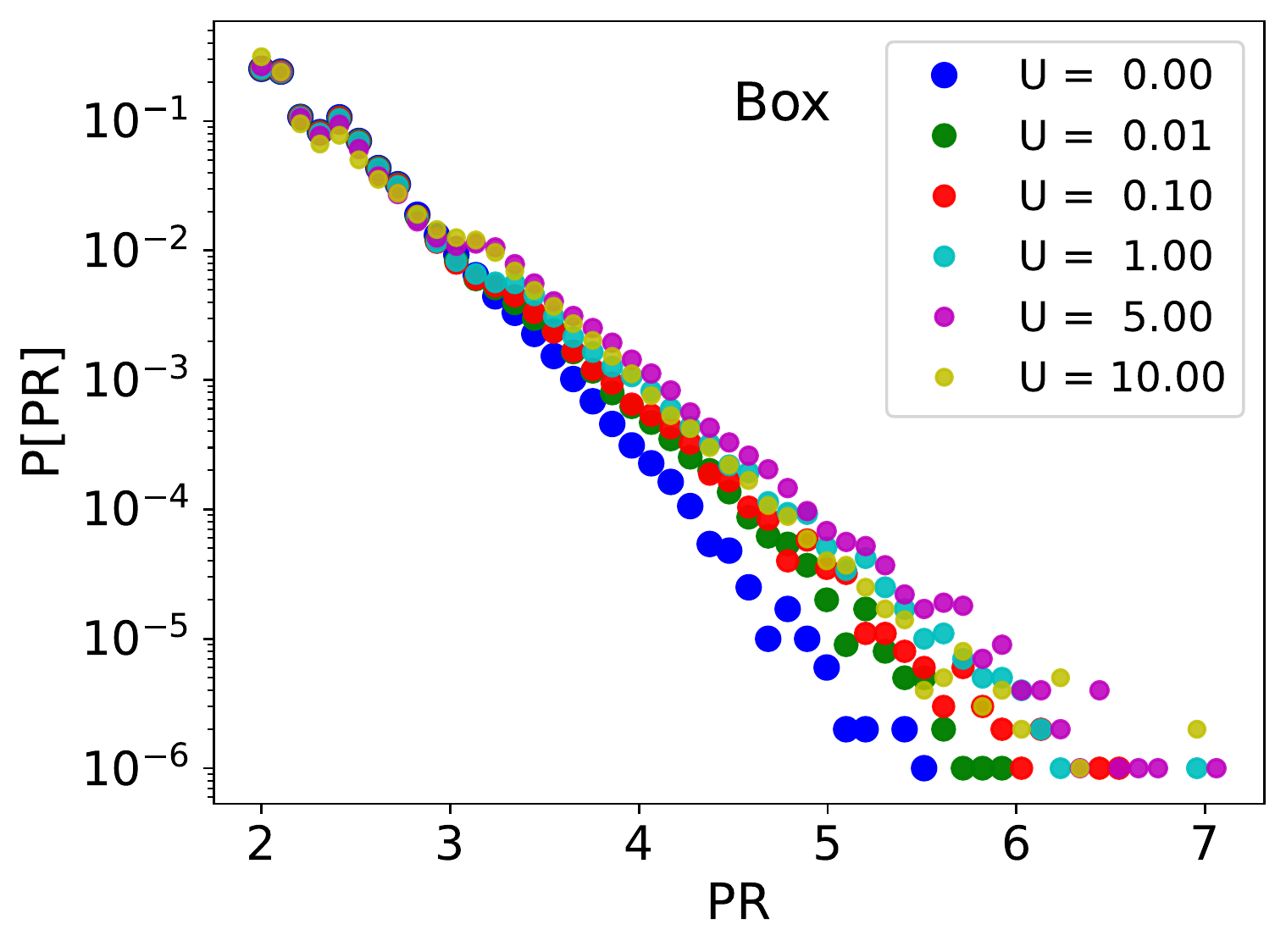} \\
\includegraphics[width=0.5\linewidth]{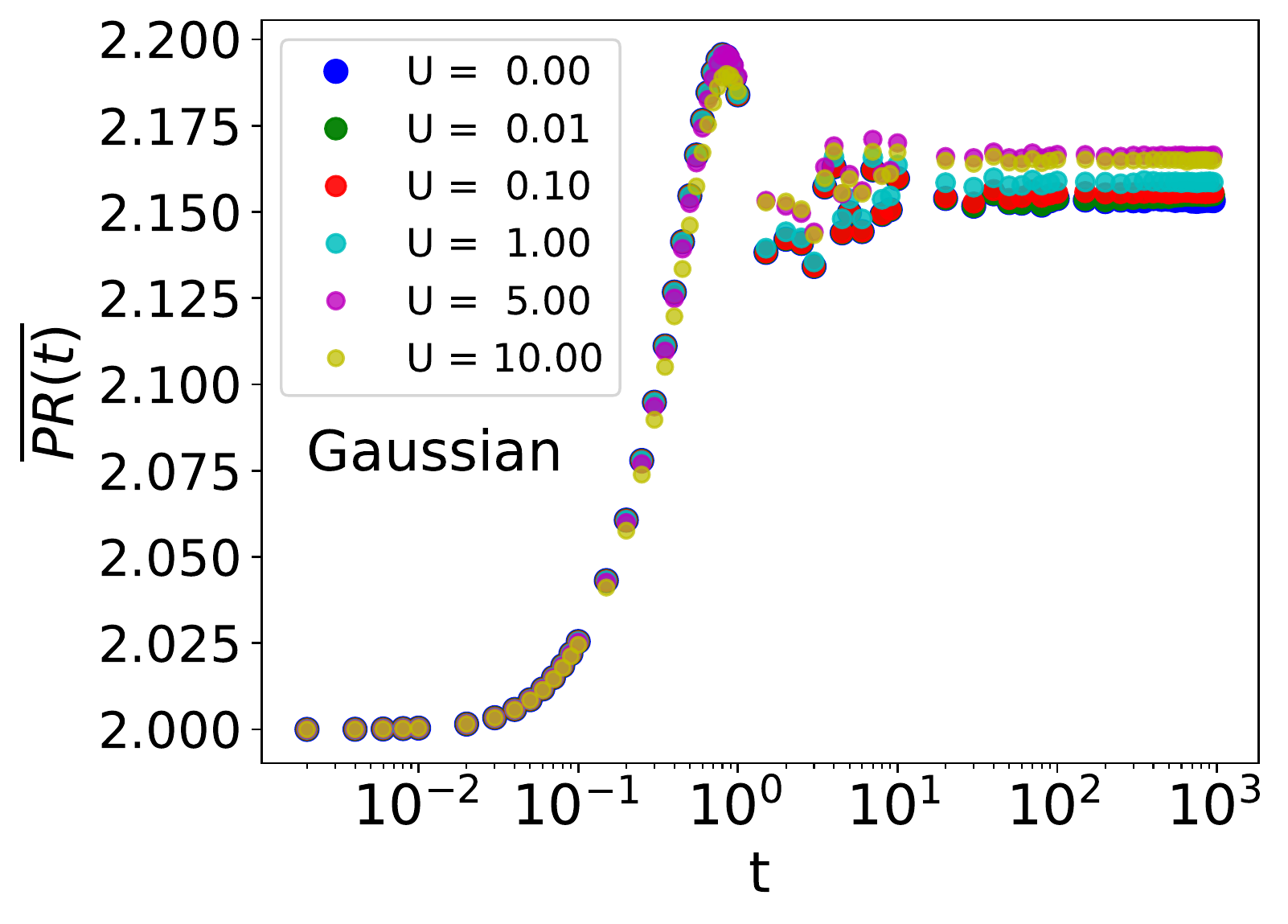} &
\includegraphics[width=0.5\linewidth]{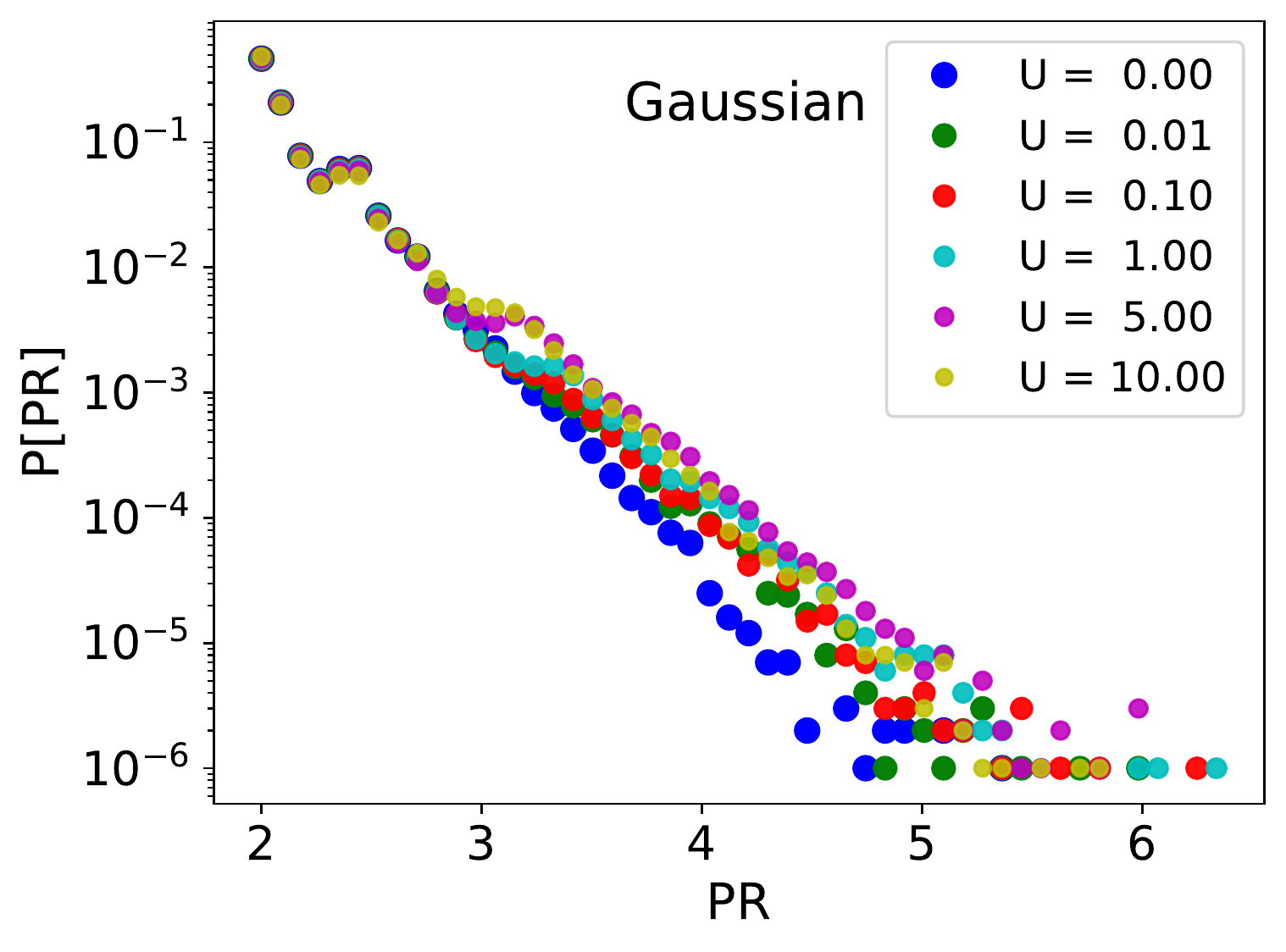} \\
\includegraphics[width=0.5\linewidth]{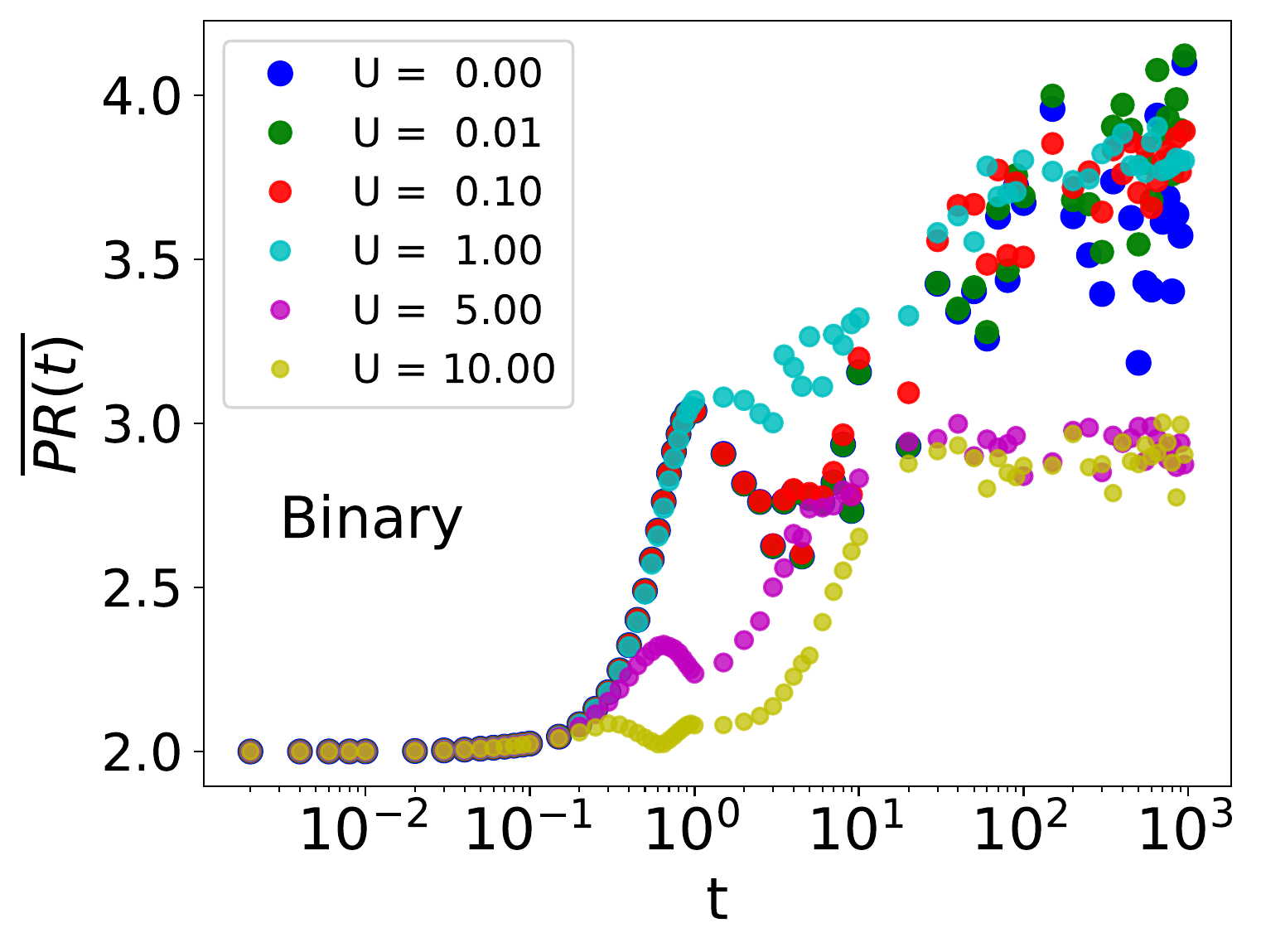} &
\includegraphics[width=0.5\linewidth]{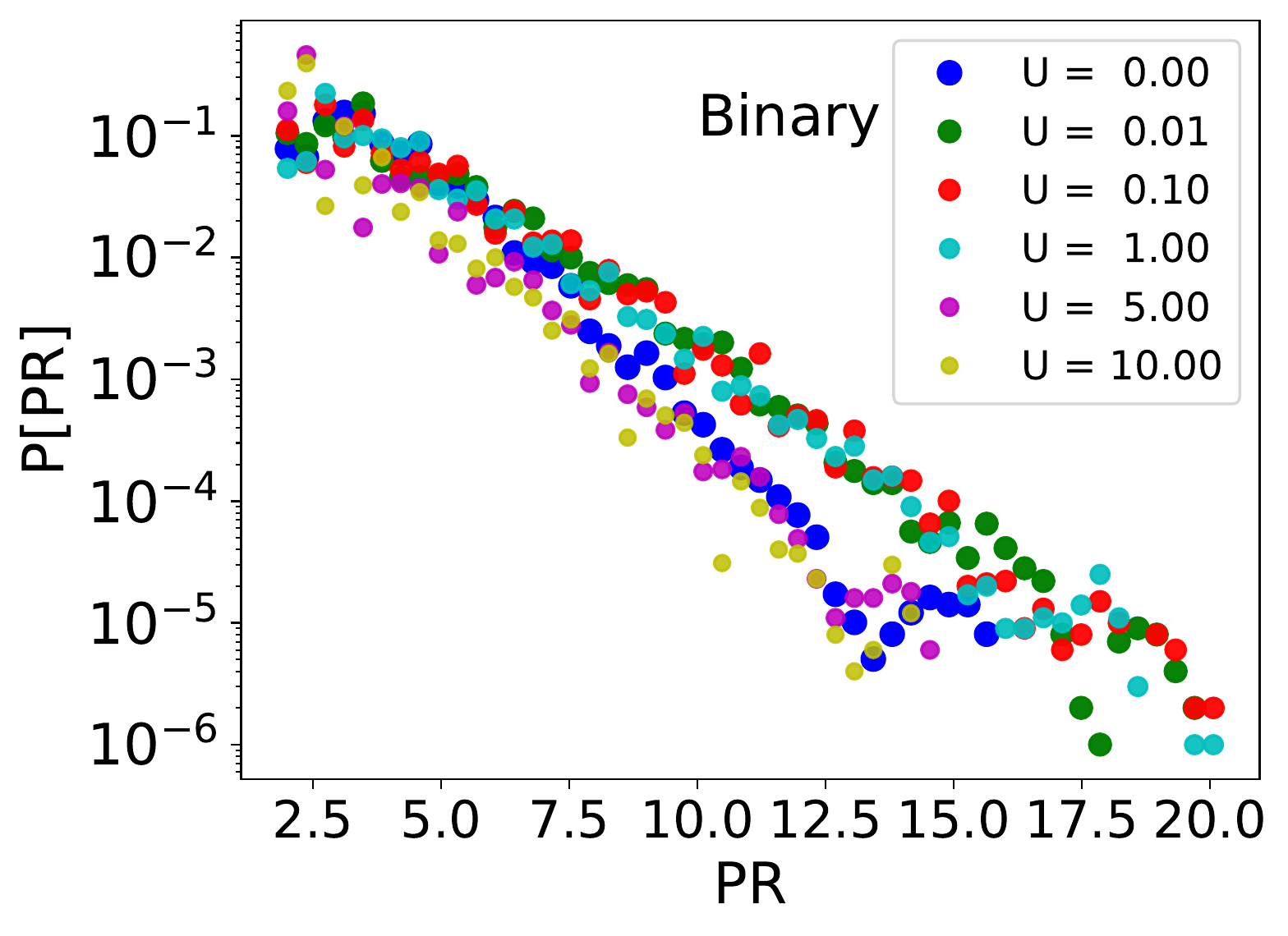} \\
\includegraphics[width=0.5\linewidth]{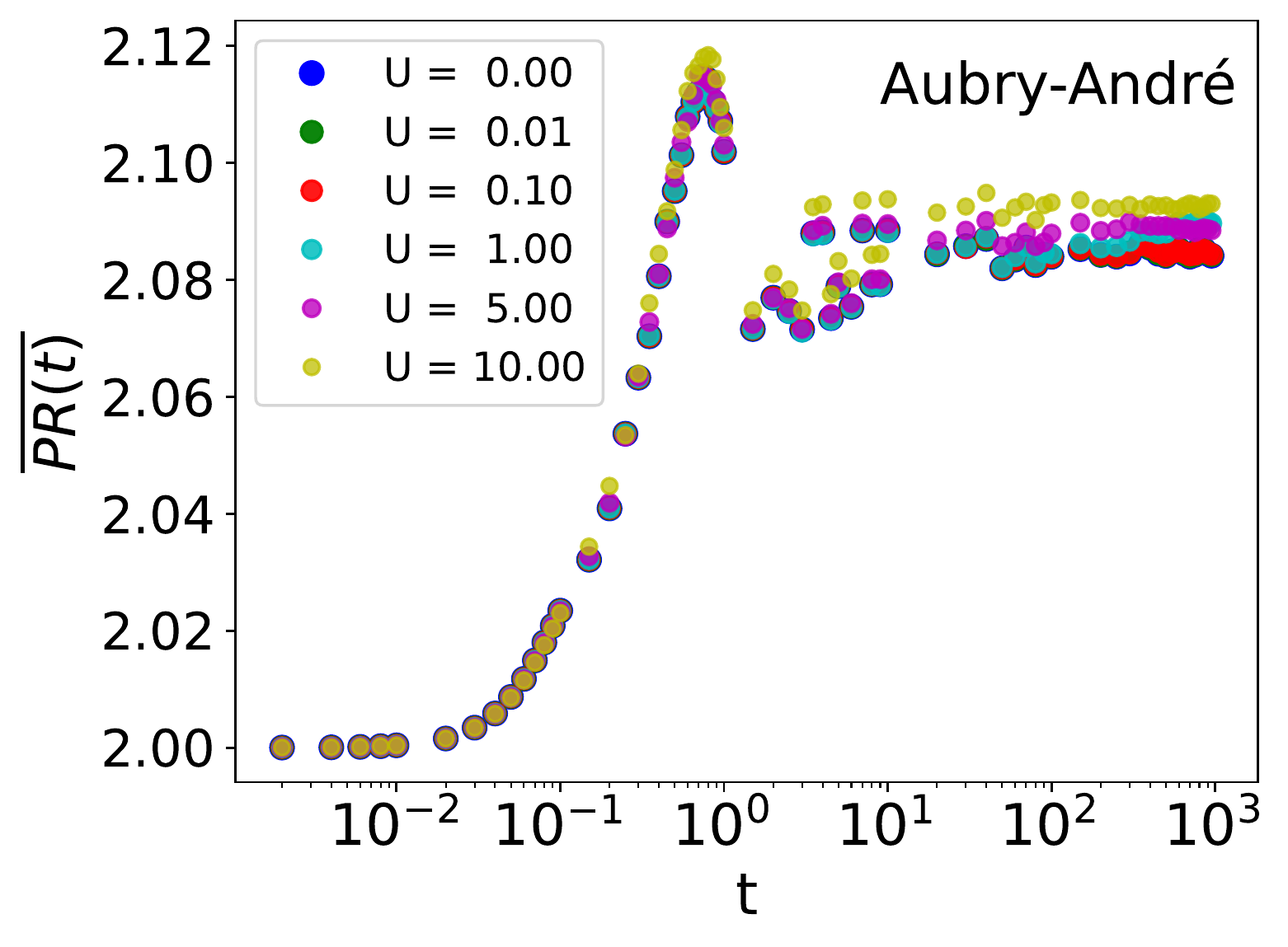} &
\includegraphics[width=0.5\linewidth]{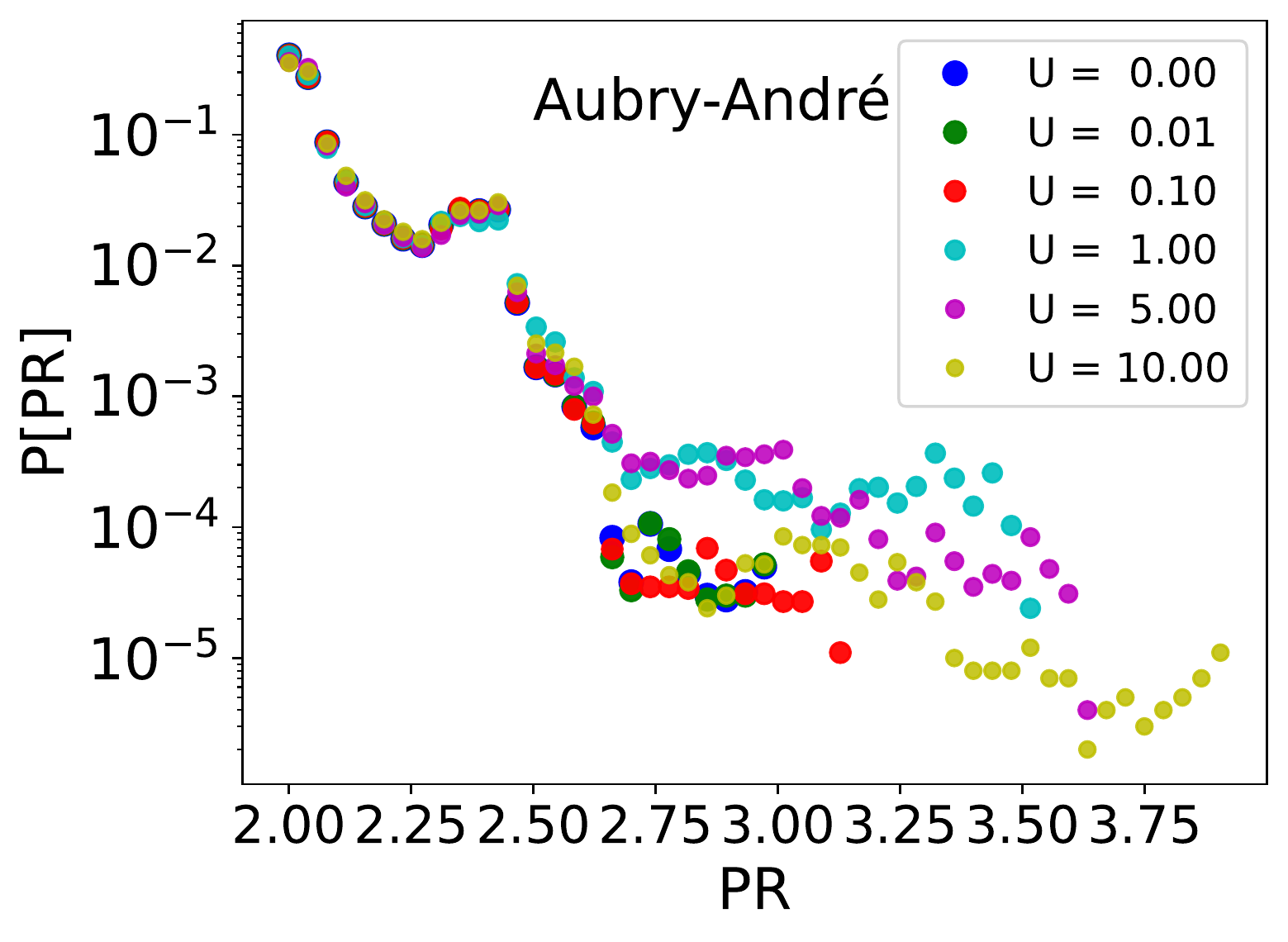} \\
\end{tabular}
\caption{Effect of the interaction strength $U$ on dynamics of the PR. From top to bottom, the disorder follows a box, Gaussian, binary and Aubry-Andr\'{e} distribution, with a disorder strength $W_{\text{And}} = 40$ ($W_{\text{MBL}} = 10$). On the left plots, it is displayed the mean value of the PR for different times. On the right plots, there is the distribution of the PR in the long-time limit. The system has $L=10^6$ sites.}
\label{fig:effect_Jz_dyn_Wlarge}

\end{figure}

We have also studied the effect of the interaction in the time evolution of the following initial states, formed by two particles localized on two consecutive sites ($i,i+1$):
\begin{equation}
\label{initial_states_N2}
    \ket{\psi_i(t=0)} = a_i^\dag a_{i+1}^\dag \ket{0}, \quad i\in[1, L-1], \; L = 10^6.
\end{equation}
In the left plots of Fig.~\ref{fig:effect_Jz_dyn_Wlarge}, we can see how the mean PR, over the $L-1$ wavefunctions, evolves in time. When the disorder follows a binary distribution, the time evolution is completely different from the other three distributions. Even at values of the time around $t\sim 10^3$, in units of the hopping, the mean PR continues increasing in the binary case, for $U = 0, 0.01, 0.1, 1$, while in the other distributions, the value of the mean PR have already converged when $t\sim 10^2$. The values of the mean PR for the binary disorder are much larger compared with the other distributions.

The effect of the interaction is more pronounced in the binary disorder than in the other cases. Only in the binary distribution, there are large regions where all the sites have the same potential. On that region, the physics follows the disorder-free scenario, where the particles can freely hop. Large values of the interaction, compared with the hopping, lead towards the formation of bound states, where the two-particles are always in consecutive sites and therefore affecting their dynamics.

In the right plots of Fig.~\ref{fig:effect_Jz_dyn_Wlarge}, we show the distribution of the PR in the long-time limit. For random disorder, long tails are present in the distribution of the PR, while being absent in the Aubry-Andr\'{e} model. Such long tails have an impact in the mean value of the PR, separating the mean value from the value of the median. The maximum value of the PR for the binary distribution is much larger than in the other cases.

\subsection{Weak disorder physics: Effect of the interaction on the Eigenfunctions, when random potentials are considered}

\begin{figure}
\centering
\begin{center}
\includegraphics[width=\linewidth]{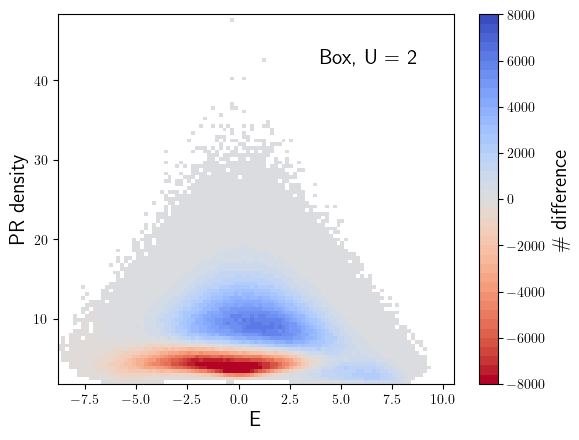} \\
\includegraphics[width=\linewidth]{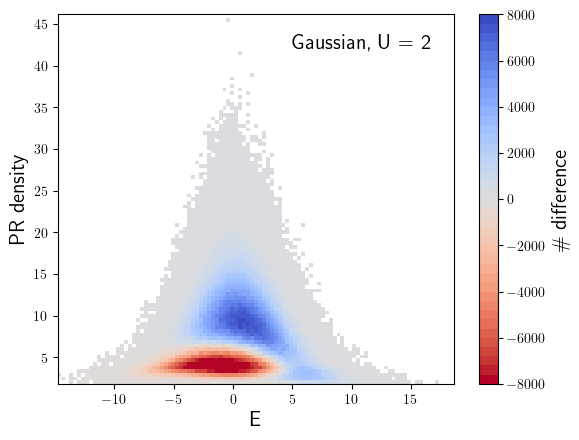} \\
\includegraphics[width=\linewidth]{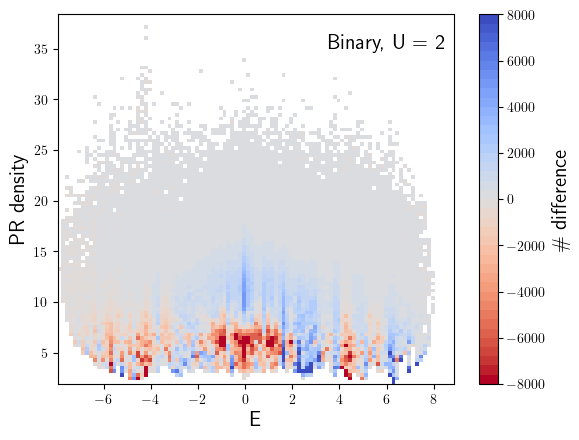} 
\end{center}
\caption{ Histograms showing the effect of the interaction strength $U = 2$, on the distribution of the Energy and density PR for systems of $L=10^6$ sites. From top to bottom, the disorder follows a box, Gaussian and binary distributions, with disorder strength $W_{\text{And}} = 6$ ($W_{\text{MBL}} = 1.5$) for the box distribution and $W_{\text{And}} = 4$ ($W_{\text{MBL}} = 1$), for the Gaussian and binary disorders. The interaction moves the data from red to blue areas.}
\label{fig::int_effect_histo_E_PR_all}
\end{figure}

The effect of the interaction on the eigenstates for low values of the disorder is remarkable, as we can see in the histograms of the Energy and PR in Fig.~\ref{fig::int_effect_histo_E_PR_all}, for $U = 2$. In this type of histograms, the effect of the interaction consists in depleting the red area and move eigenfunctions into the blue areas. In the gray areas, there is not enough data to suppress the statistical noise.

For the box and Gaussian distributions, there are some similarities in their respective histograms. In both cases, there is a blue region in the middle of the spectrum on top of the red region, a clear indication of interaction-induced delocalization phenomena. For large values of the energy and low values of the PR, there is another blue region, where the data for interacting systems accumulate for strong values of the interaction, see Fig.~s~\ref{fig::int_effect_histo_E_PR} and \ref{fig::int_effect_histo_E_PR_G} in Appendix~\ref{appendix} to see the effect of increasing the values of the interaction strength. Such phenomena originates from the creation of bound states, where the two particles are always next to each other. The formation of such bound states is clear in the limit of large interaction strength. For the binary distribution, the red and blue areas are fragmented, due to the structure of the distribution, however, a blue area on top of a red region around energy zero is still visible.

\begin{figure}
\centering
\begin{center}
\includegraphics[width=\linewidth]{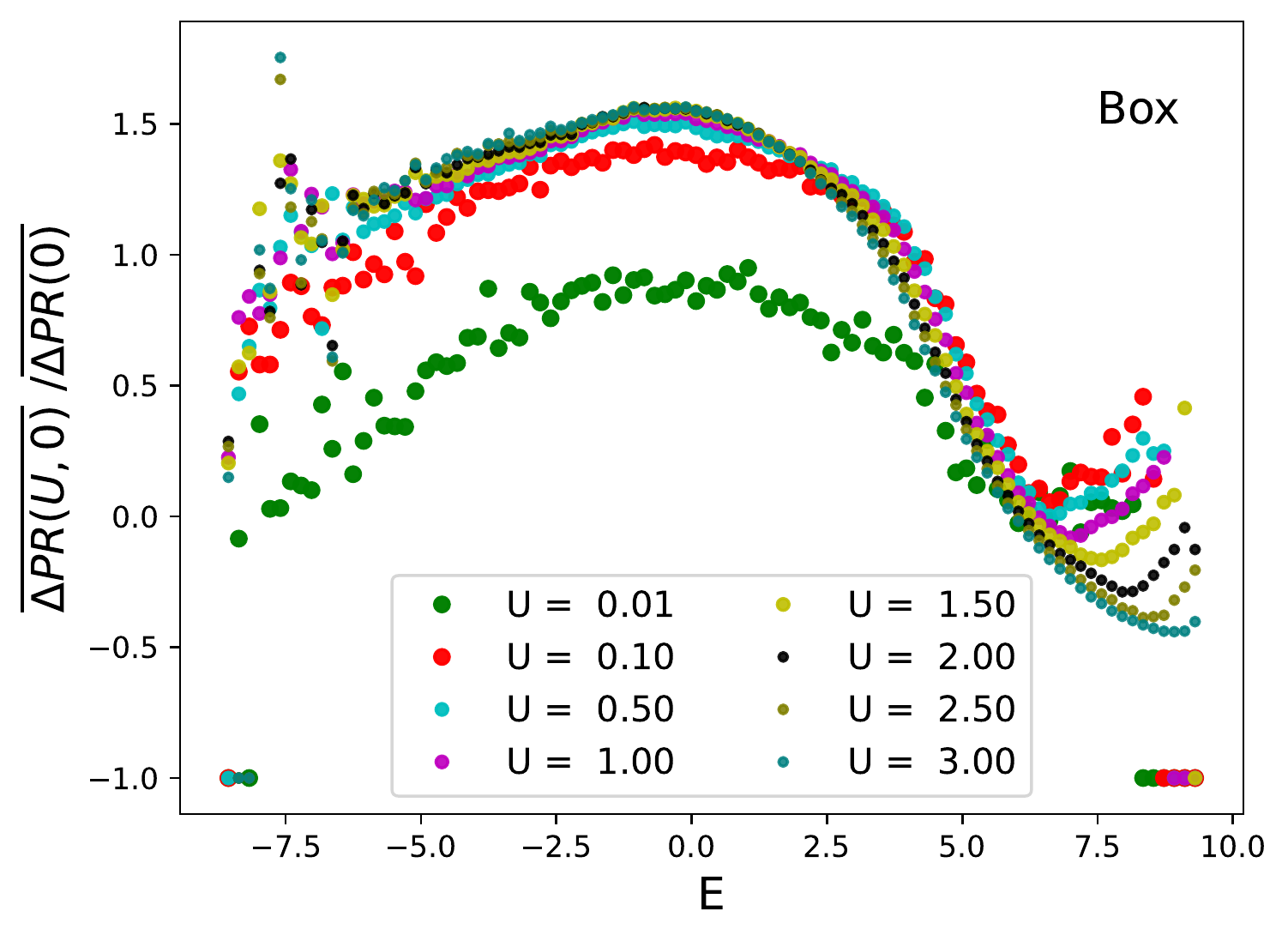} \\
\includegraphics[width=\linewidth]{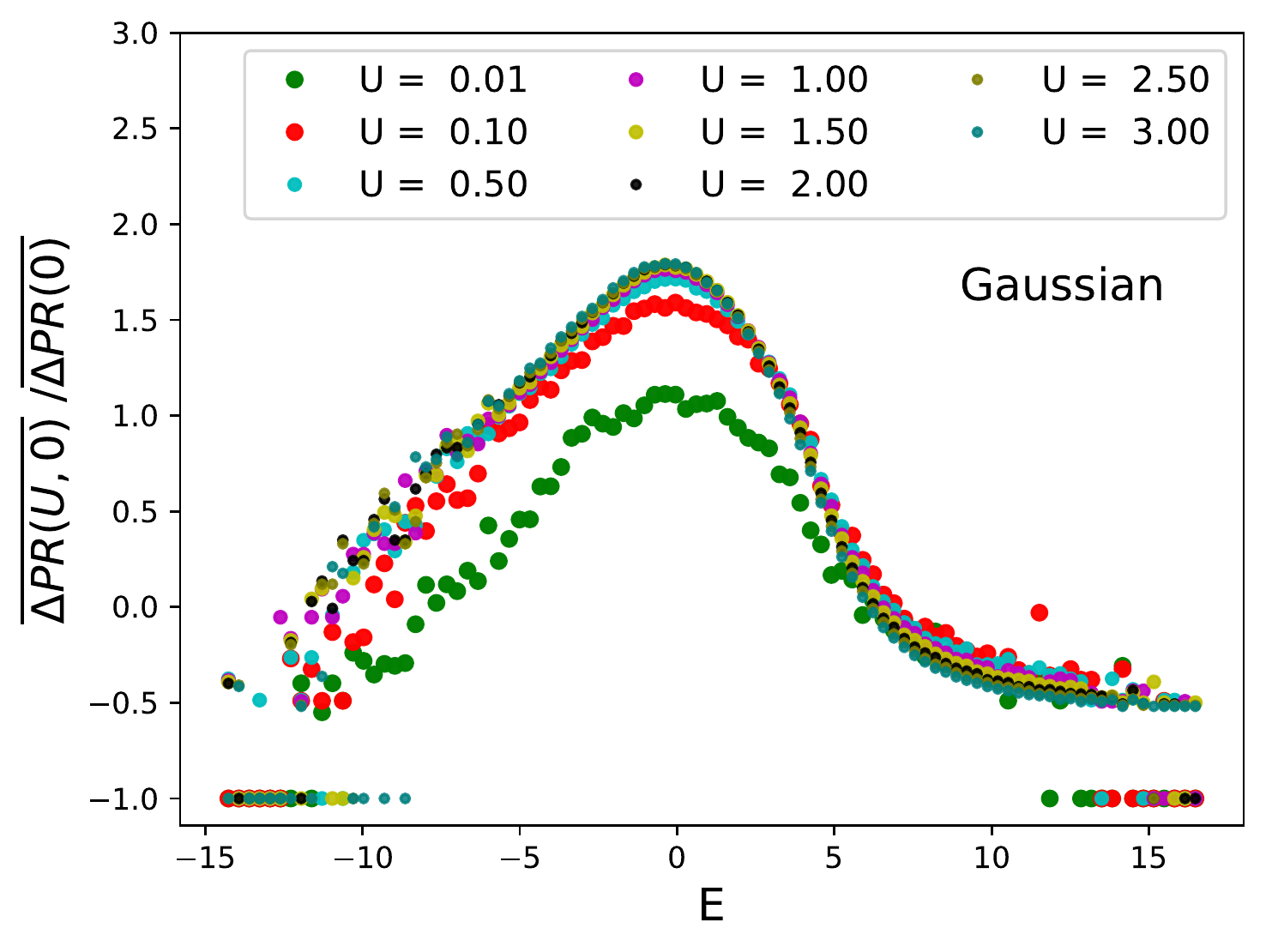} \\
\includegraphics[width=\linewidth]{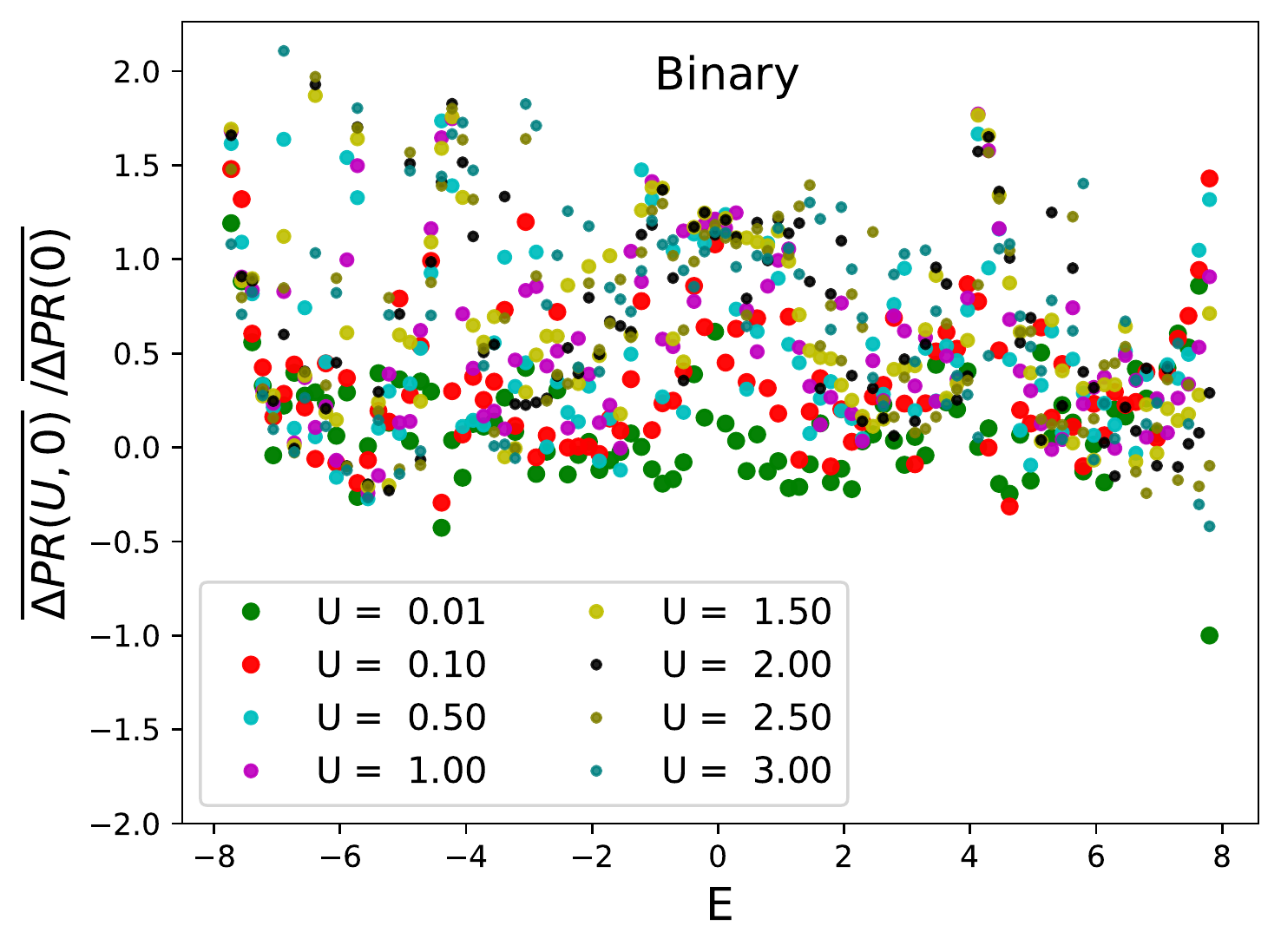}
\end{center}
\caption{ Effect of the interaction strength $U$ on the mean values of the PR, as a function of the energy window. The system has $L=10^6$ sites. From top to bottom, the potential follows a box, Gaussian and binary distribution. The disorder strength is $W_{\text{And}} = 6$ ($W_{\text{MBL}} = 1.5$) for the box distribution and $W_{\text{And}} = 4$ ($W_{\text{MBL}} = 1$), for the Gaussian and binary disorders. }
\label{fig::int_effect_mean_all}
\end{figure}

In Fig.~\ref{fig::int_effect_mean_all}, we can see the effect of the interaction on the mean value of the PR as a function of the energy window. In detail, the difference of the mean PR, $\overline{\Delta \text{PR(U, 0)}}$, is calculated from the data in the histograms of Fig.~\ref{fig::int_effect_histo_E_PR_all}. We group the data in two categories, the one where the histogram has positive values (blue regions) and the one with negative entries (red regions). We calculate the mean PR for the two categories, obtaining the values $\overline{\Delta \text{PR(U)}}$ (blue) and $\overline{\Delta \text{PR(0)}}$ (red). The mean difference is:
\begin{equation}
\overline{\Delta \text{PR(U, 0)}} = \overline{\Delta \text{PR(U)}} - \overline{\Delta \text{PR(0)}}.
\end{equation}

For the box and Gaussian distributions, we can see a smooth dependence of the change of the mean PR on both the interaction strength and energy window. As expected from their respective histograms of Fig.~\ref{fig::int_effect_histo_E_PR_all}, the eigenstates with energy zero are the ones most delocalized due to the interaction. For large values of the energy, we can also see that the interaction increases the localization of the eigenstates, which is the contrary effect that the interaction has on the eigenstates in the middle of the spectrum. For the binary distribution, there is no clear pattern, since the mean PR depends too strongly on the energy window. It seems that the majority of the data is above the value of zero, indicating an overall increase of the delocalization due to the interaction.

\subsection{Weak disorder physics: Effect of the interaction in the dynamics, when random potentials are considered}

\begin{figure}
\centering
\begin{tabular}{c c}
\includegraphics[width=0.5\linewidth]{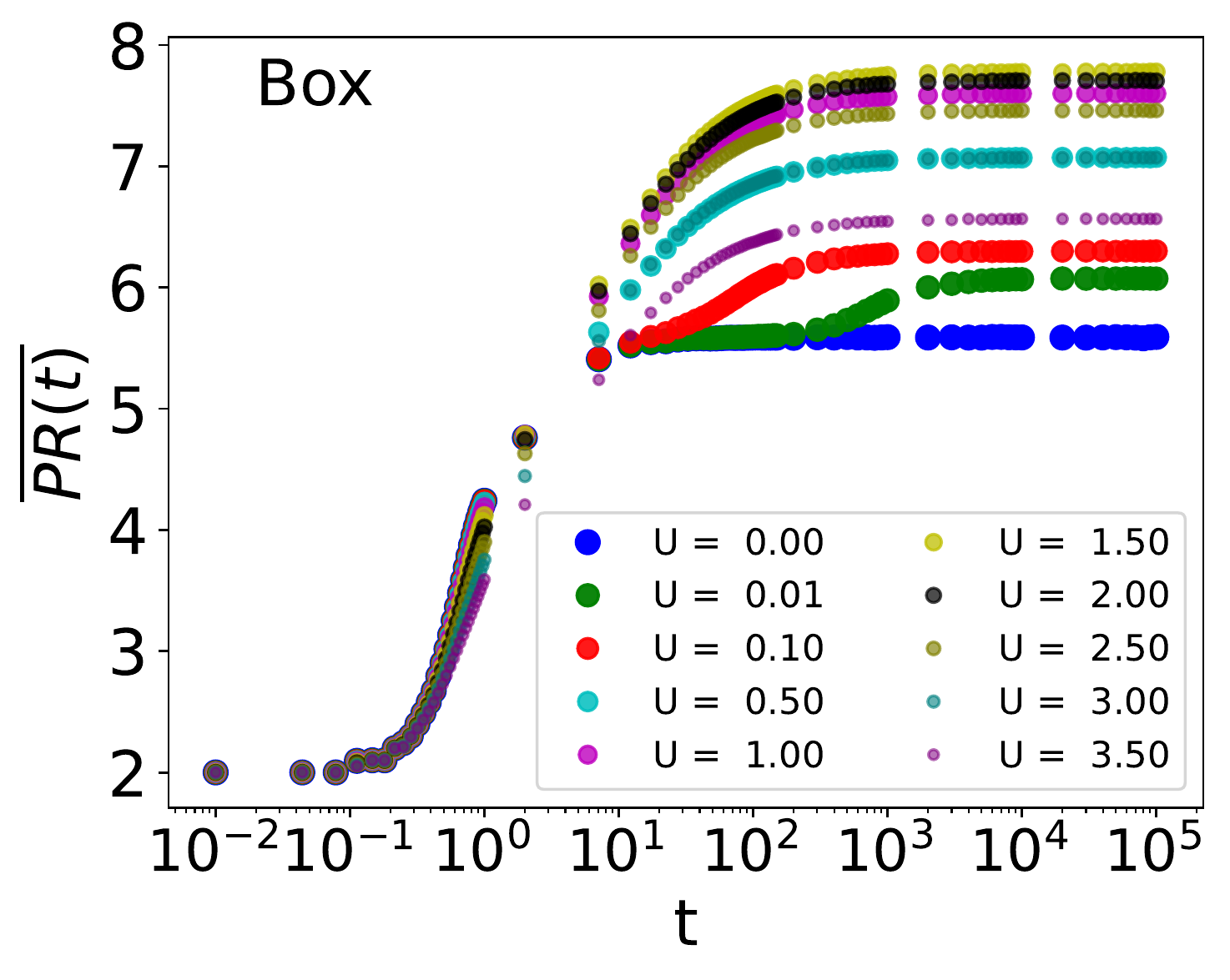} &
\includegraphics[width=0.5\linewidth]{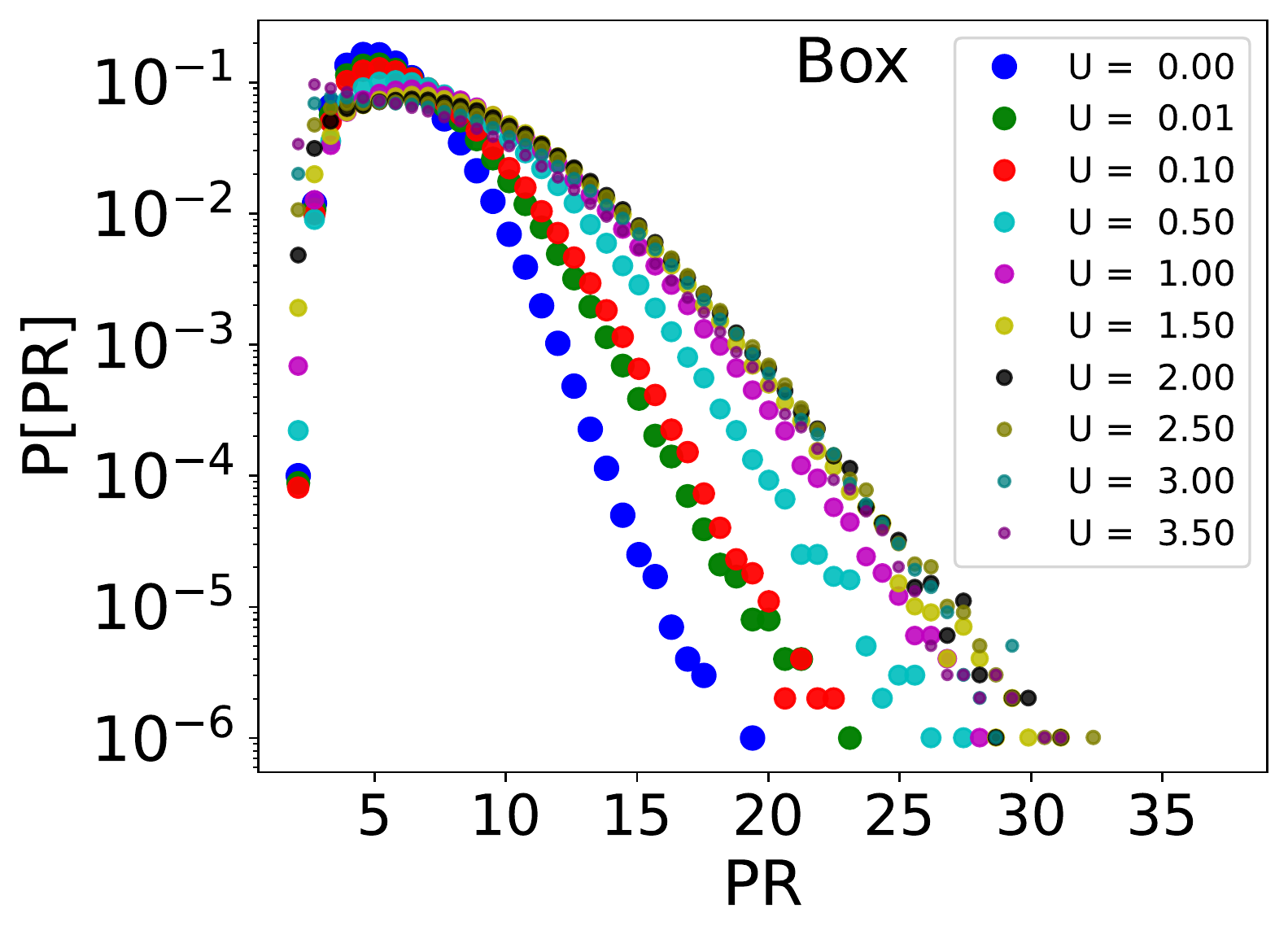} \\
\includegraphics[width=0.5\linewidth]{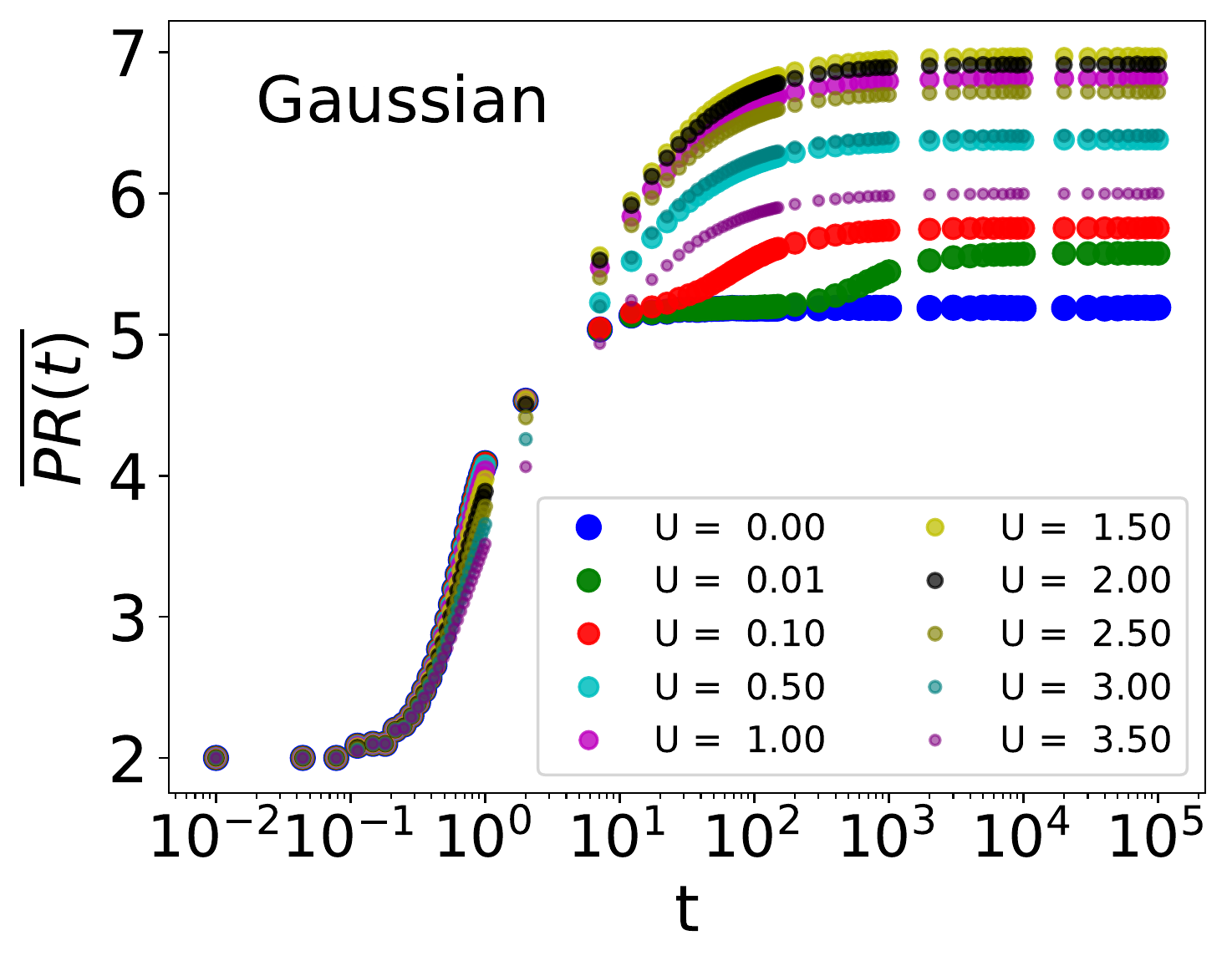} &
\includegraphics[width=0.5\linewidth]{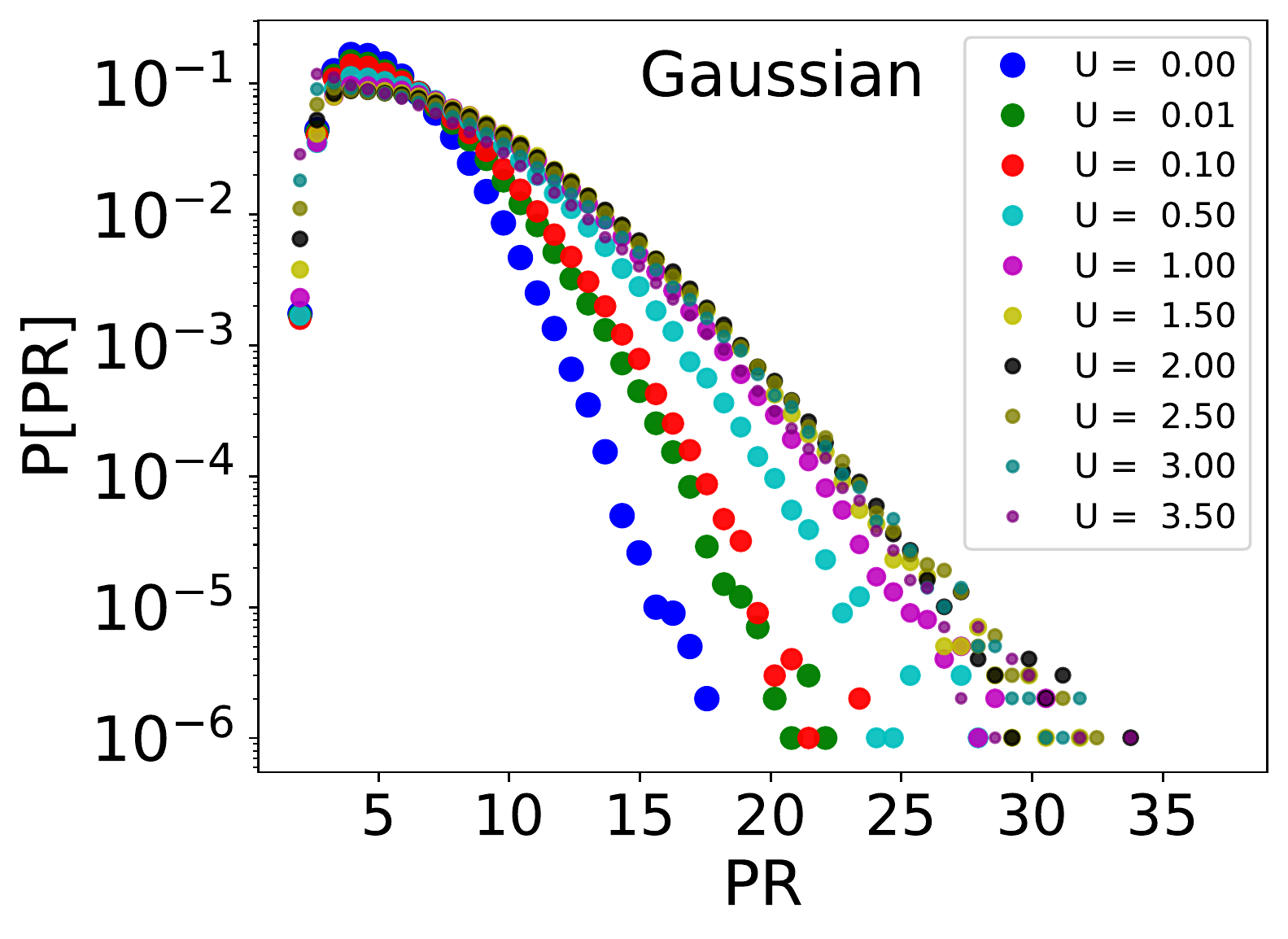} \\
\includegraphics[width=0.5\linewidth]{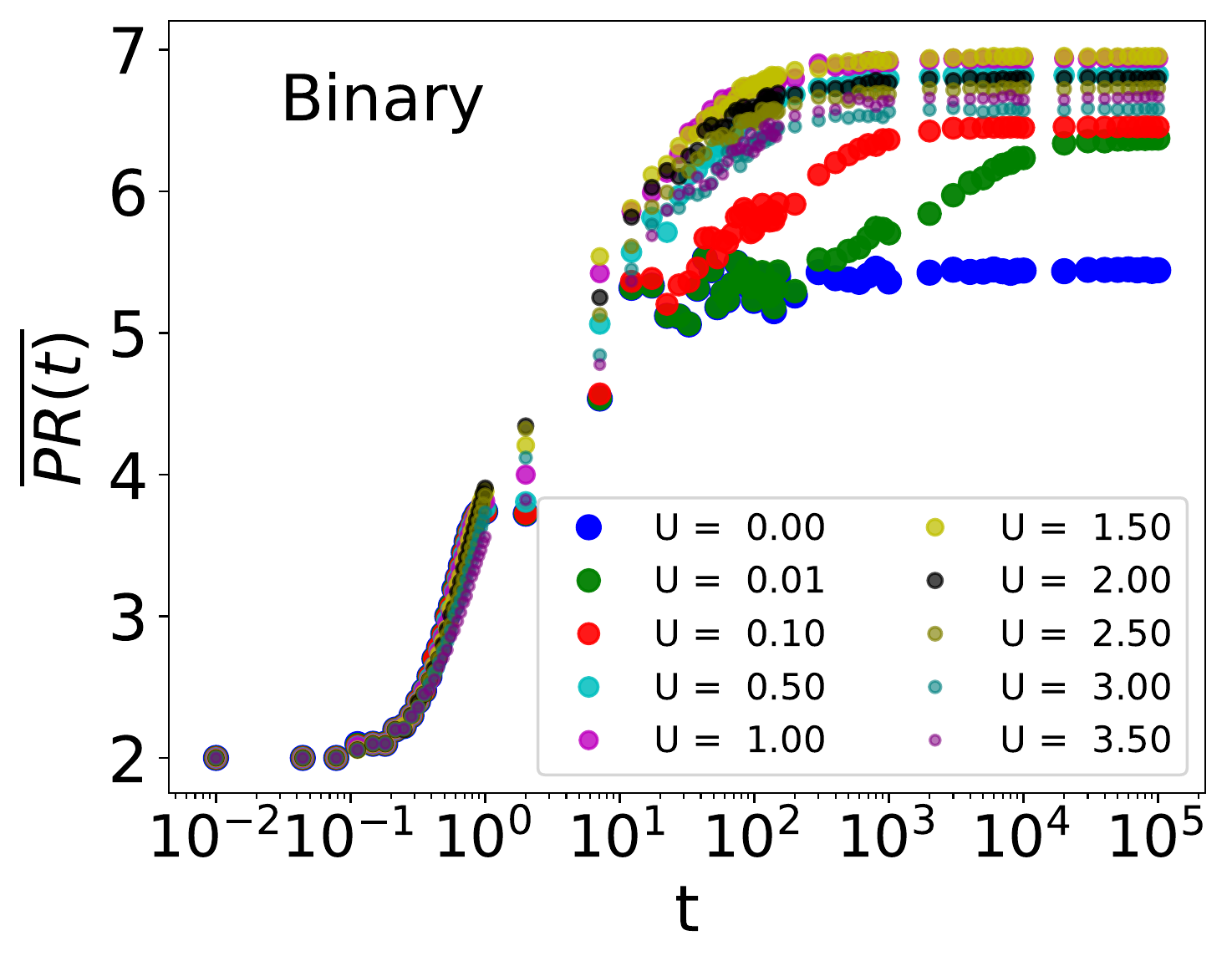} &
\includegraphics[width=0.5\linewidth]{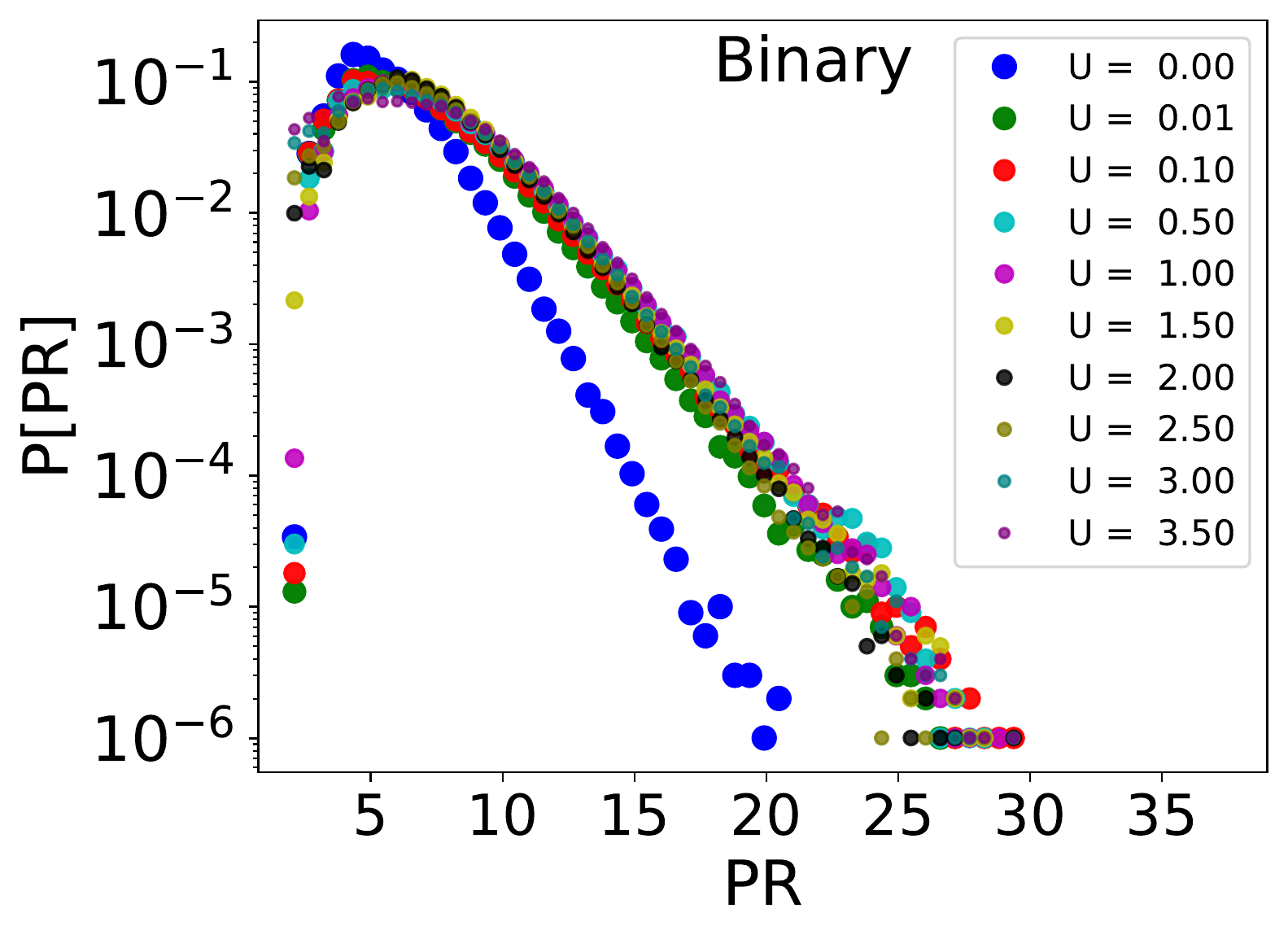}
\end{tabular}
\caption{Effect of the interaction strength $U$ on dynamics of the PR. From top to bottom, the disorder follows a box, Gaussian and binary distribution. The disorder strength is $W_{\text{And}} = 6$ ($W_{\text{MBL}} = 1.5$) for the box distribution and $W_{\text{And}} = 4$ ($W_{\text{MBL}} = 1$) for the others. On the left plots, it is displayed the mean value of the PR for different times. On the right plots, there is the distribution of the converged PR.}
\label{fig:effect_Jz_dyn_Wsmall}

\end{figure}

In this Section, we show the influence of the interaction in the dynamical properties. The initial state considered are given in Eqn.~\ref{initial_states_N2} and the results are displayed in Fig.~\ref{fig:effect_Jz_dyn_Wsmall}. 

In the left plots of Fig.~\ref{fig:effect_Jz_dyn_Wsmall}, the mean value of the PR at several values of the time $t$ is displayed. For weak values of the interaction, $U = 0.01, 0.1$, and for the three random distribution considered, the minimum amount of time that the system must evolve in order to observe a deviation in the mean PR, between interacting and non-interacting systems, is $t > 1/U$. For each distribution, there is an optimal value of the interaction $U\in [1.5,2] $ such that the long time limit of the PR is maximized. Further increasing the interaction results in a reduced mean PR in the long time limit, due to the formation of the bound state, which reduces the two-particle problem into a single-particle problem with an effective hopping $t^2/U$, obtained from second-order perturbation theory.

In the right plots of Fig.~\ref{fig:effect_Jz_dyn_Wsmall}, the distribution of the PR in the long-time limit is displayed. Strong values of the interaction strength do not have a large impact of the long tails of the distribution, instead they increase the proportion of PR with low values, effectively reducing the mean PR in the long-time limit.

\section{Conclusion}

We have developed a real-space divide and conquer algorithm which is based
on the localized nature of eigenfunctions in few-particle Anderson problems. 
Our approach allows to solve systems with a linear scaling in system size. 
This enabled us to obtain the entire spectrum of disorder instances of $10^9$ sites for the single particle problem. This is an interesting achievement, because it allows to investigate the localization properties of {\em all} eigenfunctions, including the tail of the distributions. These tails might easily get overlooked in a similar investigation of a series of finite size systems of a fixed system size $M\ll L$. The exquisite level of statistics with $10^9$ eigenfunctions allowed us to reveal intriguing structures in the energy-PR histograms for box distribution, where the sharp boundaries of the disorder distribution leave distinct traces in the histogram. Similar features are not found for smooth disorder distributions such as the Gaussian one. We have also explored cases of disorder distributions with partial or complete sets of delocalized eigenstates. Our method is reliable here in that it diagnoses when the algorithm is unable to find part or all of the eigenfunctions.

For the two-interacting-particles problem we are restricted to smaller subsystems $M$ compared to the single particle case, but otherwise we can also
scale to very large system sizes because of $M^5$. Here the goal of the algorithm is not to find all $L \choose 2$ eigenfunctions, but only those which are actually affected by the interaction. The number of such eigenstates increases linearly with $L$. Using this scheme we have been able to identify the interaction induced localization in an energy and PR resolved way, giving insights which energies and PR regions are depleted or enhanced by the interactions. 

As further directions one could extend the scheme to two-dimensional problems on the one hand, or try to scale up to larger particle number, or even finite density on the other hand. For the latter, complexity is expected to scale exponentially in particle number, leading to quick reduction of the affordable subsystem sizes, but the method would still scale linearly with system size. A more fruitful approach might be to formulate a finite size embedding of the information about the two particle problem into a Matrix Product State (MPS) or Operator (MPO). Such MPS or MPO could then be used as an initial state for further optimisation with the existing DMRG-like algorithms~\cite{PhysRevB.94.041116, PhysRevLett.116.247204, PhysRevLett.118.017201, PhysRevB.99.104201, PhysRevX.7.021018}.

\begin{acknowledgments}
We thank F.~Evers for discussions. We acknowledge support by the Austrian Science Fund FWF under projects (I 4548 and F-4018). The code used in this work is available on \href{https://github.com/lluisher/Divide-and-Conquer-solves-localization}{GitHub}. The data and plot scripts to produce the figures of this paper are available on \href{https://doi.org/10.5281/zenodo.7349693}{Zenodo}.

\end{acknowledgments}

\bibliography{mbl_biblio}

\newpage

\section*{Appendix}
\appendix
\section{Additional data}
\label{appendix}


In this Section of the Appendix, we want to include some extra plots that we consider interesting, but for the sake of space, they were not included in the main text, both for the single-particle and TIP problem.

Regarding the single-particle, in the main text, we have studied the density of states, the localization length and the Participation Ratio, as three relevant physical quantities, but there are other physically relevant observables, like the distribution of the gap ratio, $r_n$, and its mean value. This quantity is interesting because it can be used to discriminate between MBL and ergodic phases \cite{PhysRevB.75.155111}.

\begin{figure}[h]
\centering
\begin{tabular}{c}
\includegraphics[width=1\linewidth]{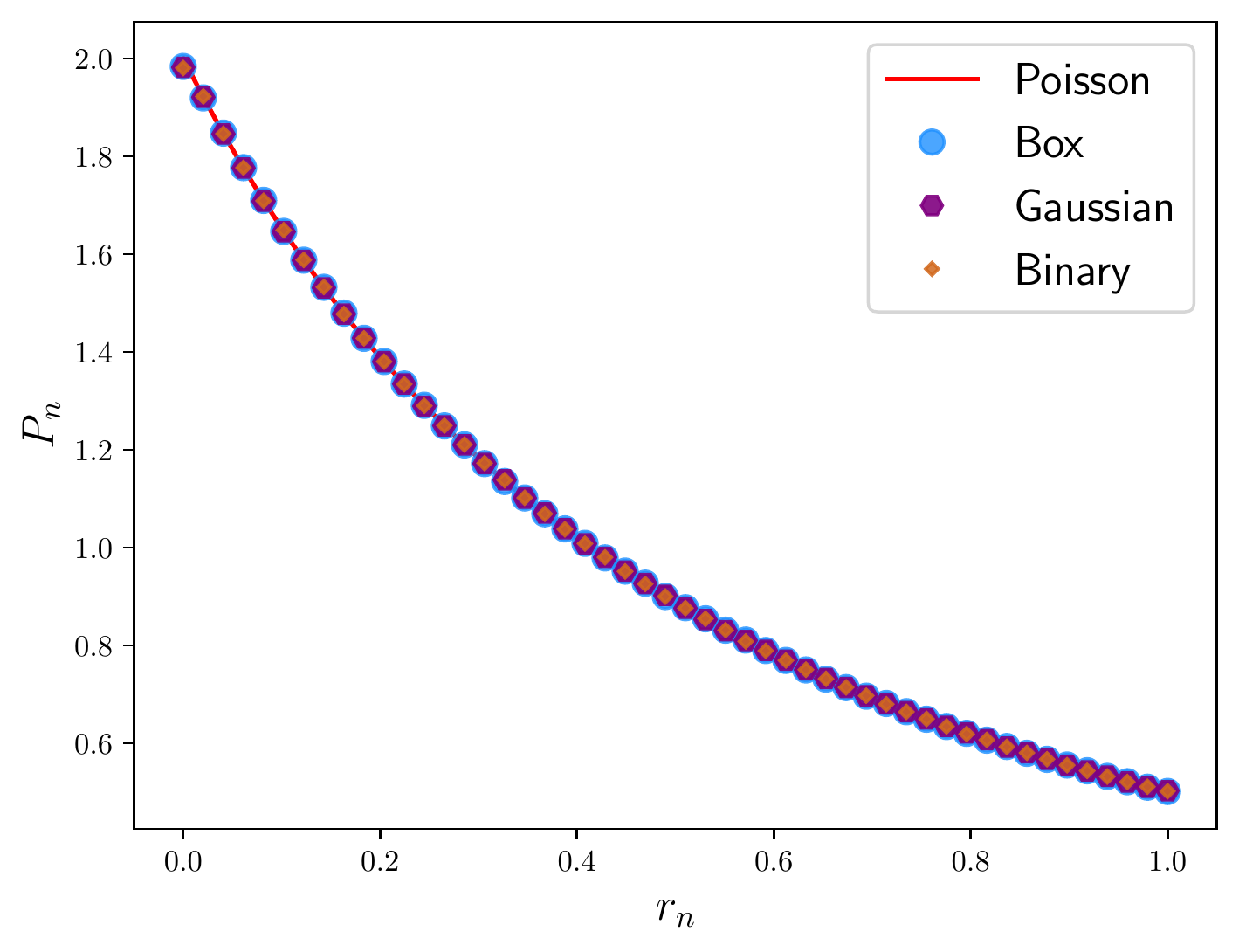}
\end{tabular}
\caption{ Distribution of the energy gap, obtained from the full spectrum of a system of $L=10^8$ sites. Three different types of random distributions are used to represent the on-site disorder, box, Gaussian and binary. The variance of the potential is always $V=0.05$. For comparison, in red, the Poisson distribution is shown.}
\label{fig:extra_check_gap}

\end{figure}

Given the set of ordered energies, $E_i$, we can calculate the energy gap $\delta_n = E_{n+1} - E_n$ and the distribution of gap ratio, $r_n = \min(\delta_n, \delta_{n+1})/\max(\delta_n, \delta_{n+1})$. In the localized physics regime, the distribution of $r$-values follows a Poissonian distribution $2/(1+r^2)$, while in the diffusive regime, $r_n$ follows a Gaussian-orthogonal ensemble (GOE) \cite{PhysRevB.82.174411}. In Fig.~\ref{fig:extra_check_gap}, we show the distribution of the gap ratio in the weak disorder regime. We can see that the data follows a Poissonian distribution.

Regarding the TIP problem, in order to visualize the effect of the interaction, we have generated some histograms that display how the distribution of the energy and PR change for different values of the interaction. In the main text, we have displayed such plots in Fig.~\ref{fig::int_effect_histo_E_PR_all}, for one value of the interaction strength. Here, we want to include the histograms for all the values of the interaction strength that we have considered, in order to see the progression from weak to strong values of the interaction. 

\begin{figure}
\centering
\begin{tabular}{c c}
\includegraphics[width=0.5\linewidth]{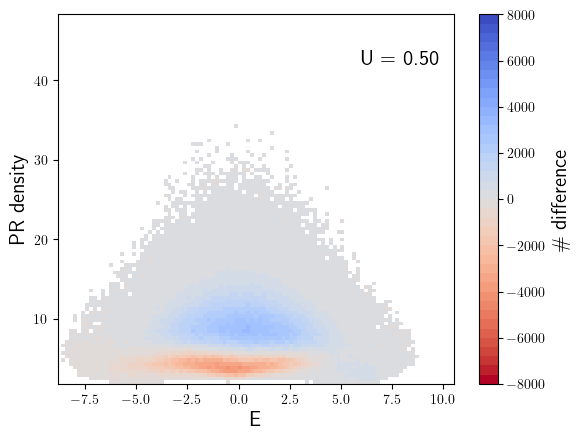} &
\includegraphics[width=0.5\linewidth]{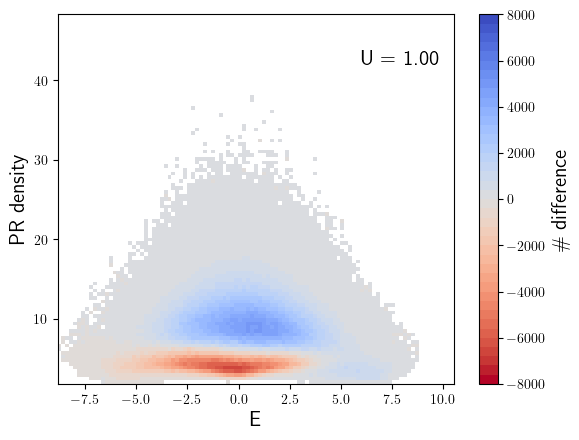} \\
\includegraphics[width=0.5\linewidth]{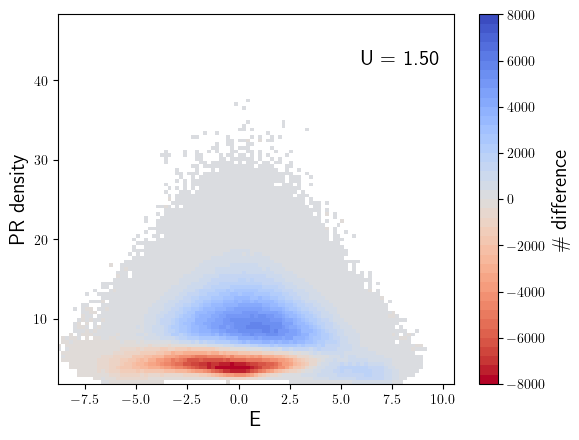} &
\includegraphics[width=0.5\linewidth]{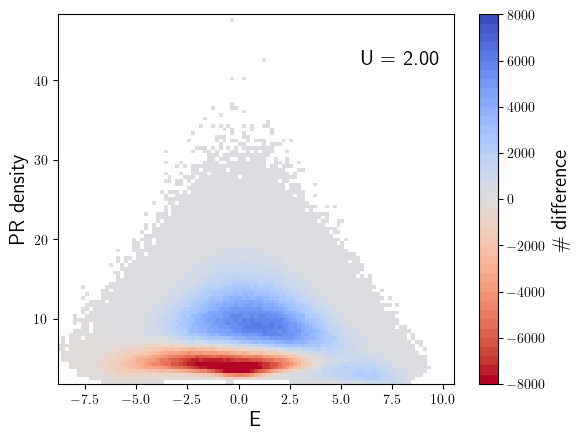} \\
\includegraphics[width=0.5\linewidth]{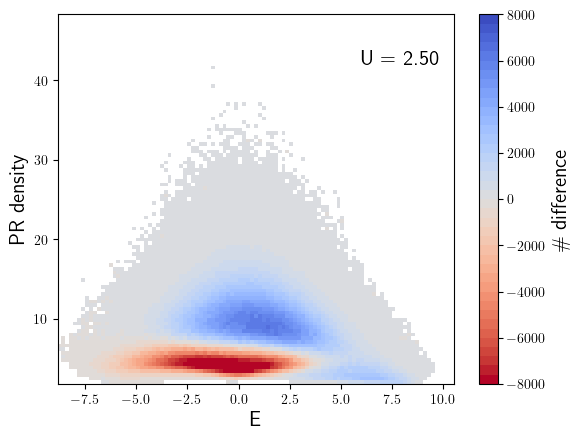} &
\includegraphics[width=0.5\linewidth]{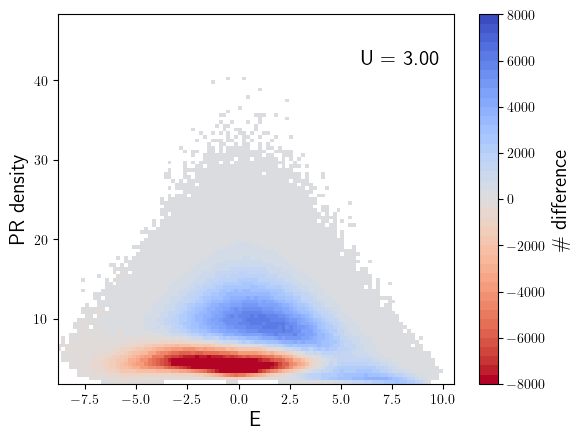} \\
\end{tabular}
\caption{ Histograms showing the effect of the interaction strength $U$, which is different in each plot, on the distribution of the energy and PR for systems of $L=10^6$ sites with a disorder that follows a box distribution and strength $W_{\text{And}} = 6$ ($W_{\text{MBL}} = 1.5$). The interaction moves the data from red to blue areas.}
\label{fig::int_effect_histo_E_PR}
\end{figure}

\begin{figure}[h!]
\centering
\begin{tabular}{c c}
\includegraphics[width=0.5\linewidth]{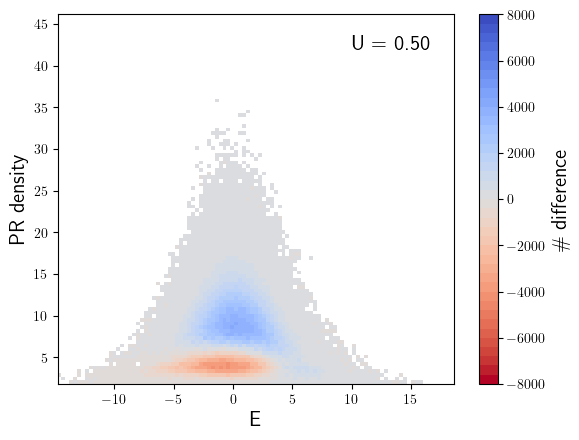} &
\includegraphics[width=0.5\linewidth]{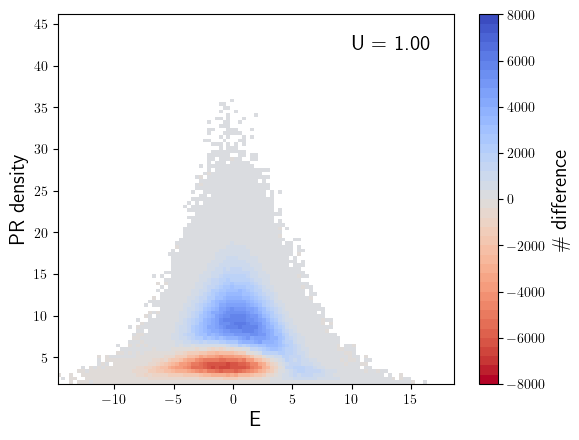} \\
\includegraphics[width=0.5\linewidth]{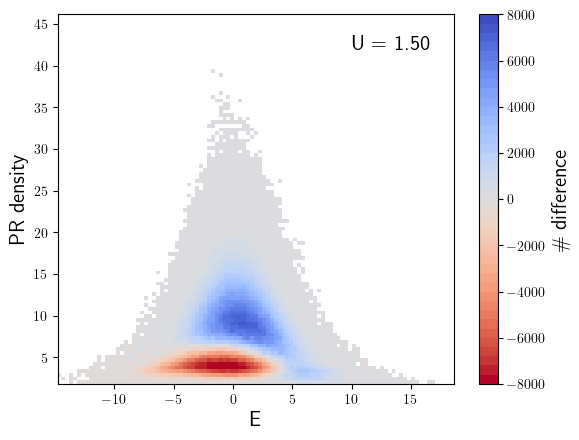} &
\includegraphics[width=0.5\linewidth]{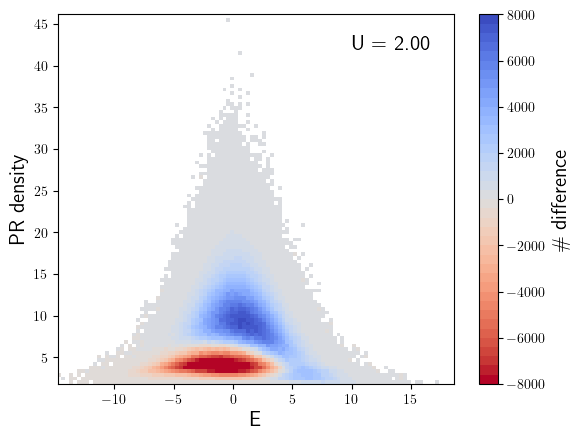} \\
\includegraphics[width=0.5\linewidth]{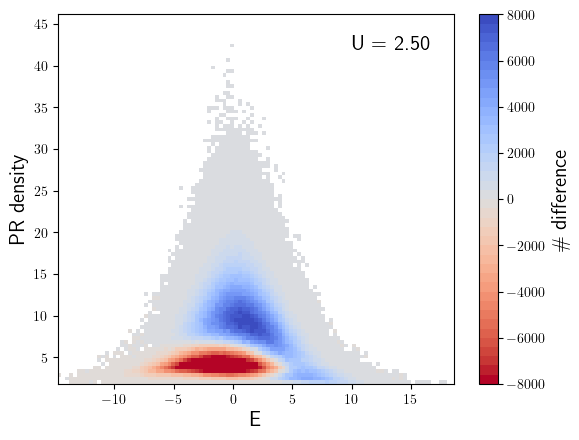} &
\includegraphics[width=0.5\linewidth]{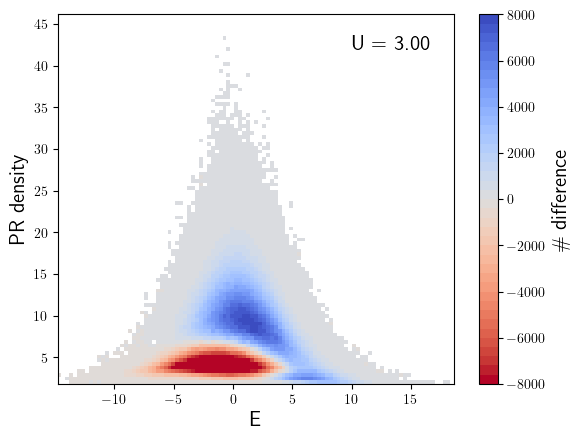} \\
\end{tabular}
\caption{ Histograms showing the effect of the interaction strength $U$, which it is different in each plot, on the distribution of the energy and PR for systems of $L=10^6$ sites with a disorder that follows a Gaussian distribution and strength $W_{\text{And}} = 4$ ($W_{\text{MBL}} = 1$). The interaction moves the data from red to blue areas. On the left column, from top to bottom, the values of $U$ are 0.5, 1.5 and 2.5, respectively. On the right column, from top to bottom, the values of $U$ are 1, 2 and 3, respectively.}
\label{fig::int_effect_histo_E_PR_G}
\end{figure}

\begin{figure}[h!]
\centering
\begin{tabular}{c c}
\includegraphics[width=0.5\linewidth]{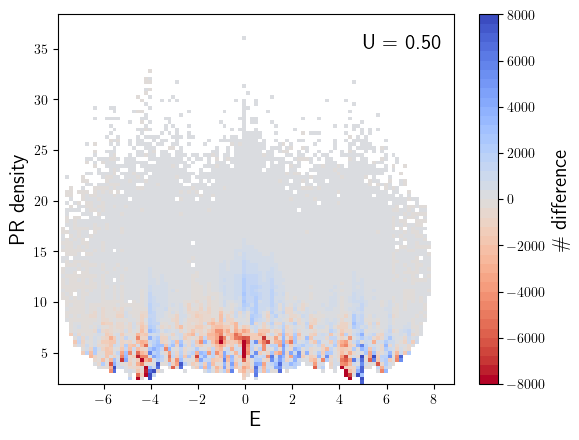} &
\includegraphics[width=0.5\linewidth]{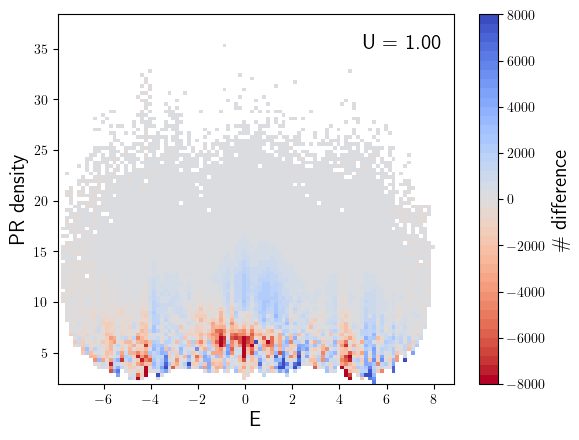} \\
\includegraphics[width=0.5\linewidth]{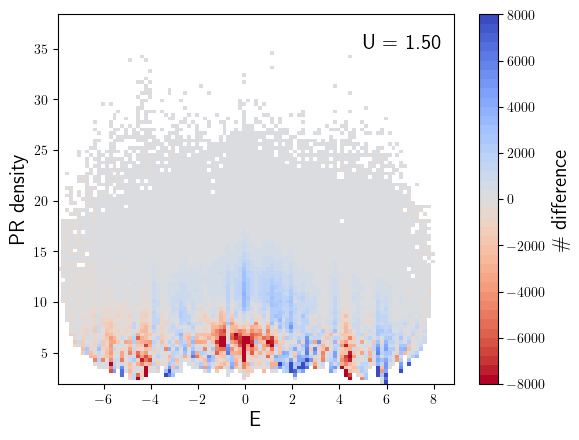} &
\includegraphics[width=0.5\linewidth]{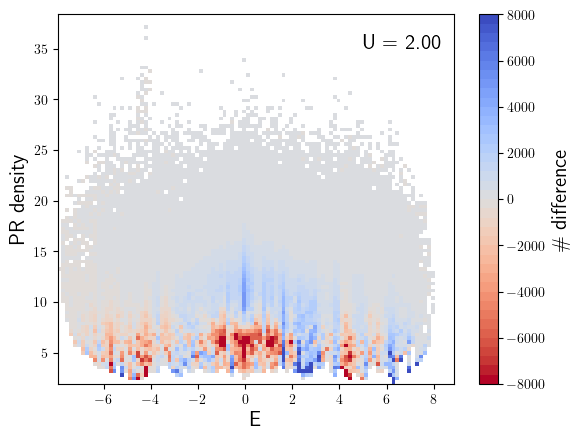} \\
\includegraphics[width=0.5\linewidth]{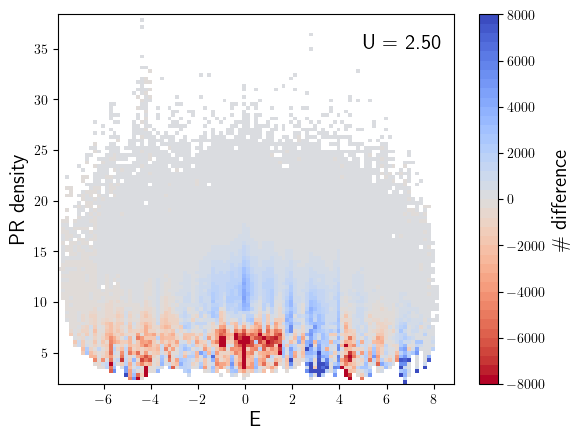} &
\includegraphics[width=0.5\linewidth]{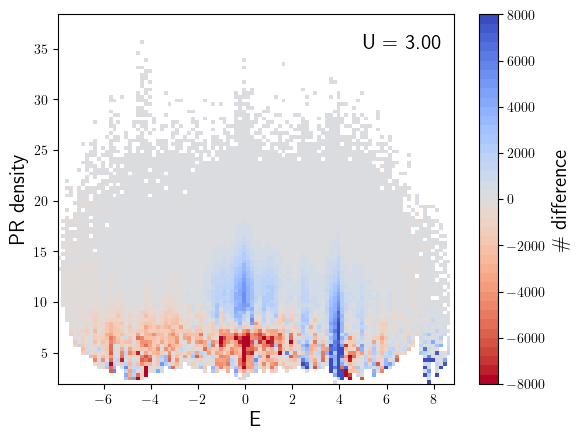}
\end{tabular}
\caption{ Histograms showing the effect of the interaction strength $U$, which it is different in each plot, on the distribution of the energy and PR for systems of $L=10^6$ sites with a disorder that follows a binary distribution and strength $W_{\text{And}} = 4$ ($W_{\text{MBL}} = 1$). The interaction moves the data from red to blue areas. On the left column, from top to bottom, the values of $U$ are 0.5, 1.5 and 2.5, respectively. On the right column, from top to bottom, the values of $U$ are 1 and 2, respectively.}
\label{fig::int_effect_histo_E_PR_bi}
\end{figure}

The data for box, Gaussian and binary distribution is presented in Fig.~\ref{fig::int_effect_histo_E_PR}, Fig.~\ref{fig::int_effect_histo_E_PR_G} and Fig.~\ref{fig::int_effect_histo_E_PR_bi}, respectively. The system has $L=10^6$ sites.

\section{Difference between box and Gaussian distribution in the limit of strong disorder for the single-particle physics}
\label{appendix_box_G}

In this Section of the Appendix, we would like to provide evidences showing the difference between a box and a Gaussian distributions in the single-particle physics. The structures that appear in the density of states when the box distribution is considered, but not present in the Gaussian distribution, have been already explained in the main text, here we just want to present the data from the box distribution next to the data from the Gaussian distribution, for an easy comparison between the two.

\begin{figure*}[h]
\centering
\begin{tabular}{c c}
\includegraphics[width=0.5\linewidth]{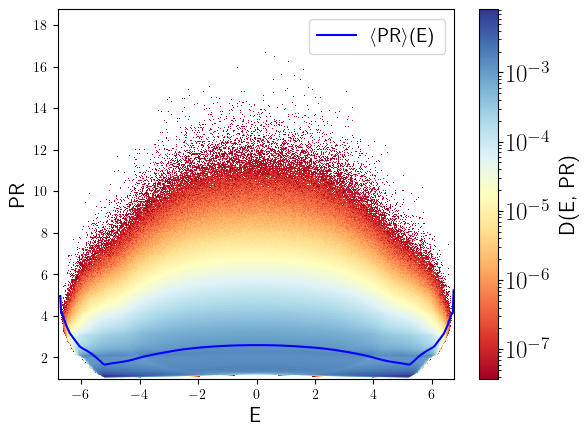} &
\includegraphics[width=0.5\linewidth]{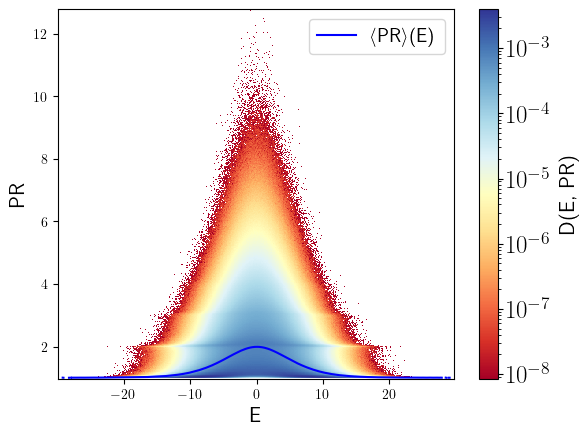} \\
\includegraphics[width=0.5\linewidth]{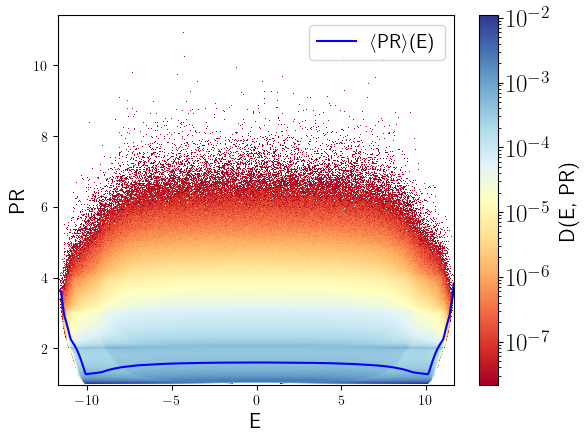} &
\includegraphics[width=0.5\linewidth]{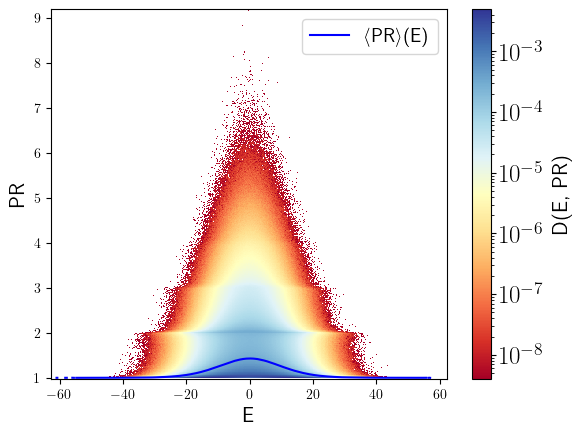} \\
\includegraphics[width=0.5\linewidth]{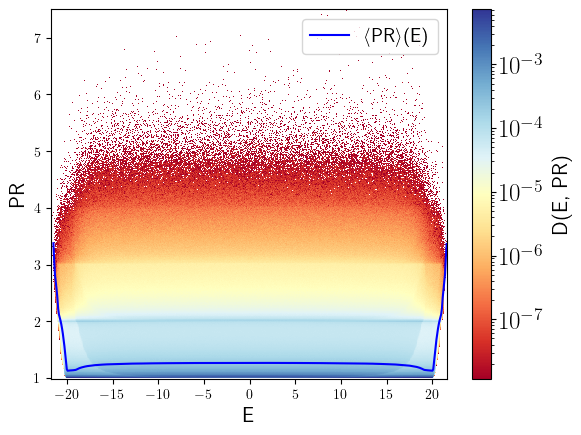} &
\includegraphics[width=0.5\linewidth]{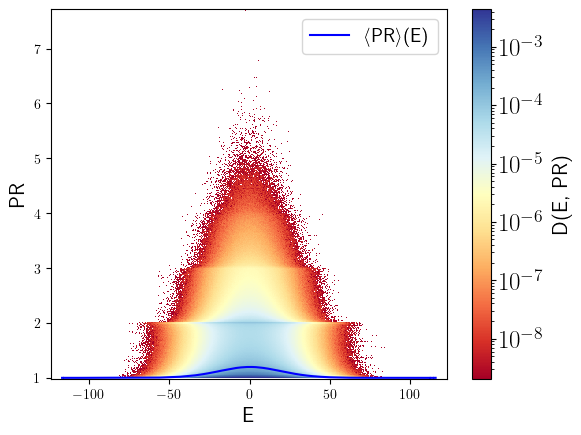}
\end{tabular}
\caption{ Histograms of the energy and PR of the eigenstates of a system of $L=10^9$ sites. The color is determined from the value of the density of states and its scale is a logarithmic one. The solid blue line is the mean value of the PR. On the left column, the disorder follows a box distribution and on the right column, the disorder is sampled via a Gaussian distribution. From top to bottom, the values of the disorder strength are respectively, $W_{\text{And}} = 10, 20, 40$ ($W_{\text{MBL}}=2.5, 5, 10$). }
\label{fig:E_PR_several_W_strong}

\end{figure*}

In Fig.~\ref{fig:E_PR_several_W_strong}, we can see the histogram of the energy and PR, and the value the density of states can be inferred from the color-scale. 

\begin{figure*}[h]
\centering
\begin{tabular}{c c}
\includegraphics[width=0.5\linewidth]{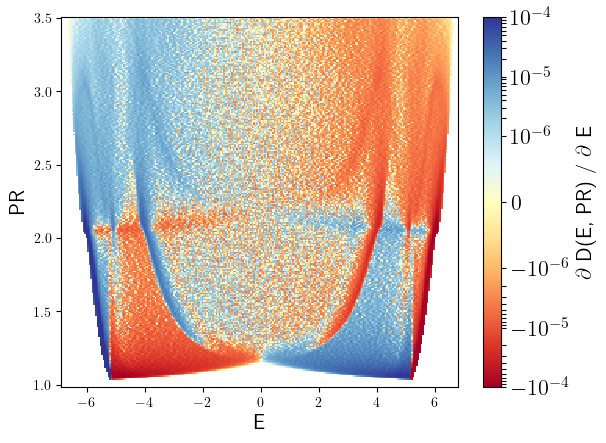} &
\includegraphics[width=0.5\linewidth]{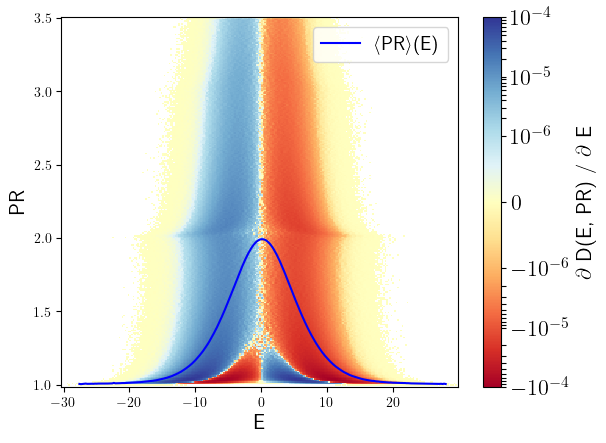} \\
\includegraphics[width=0.5\linewidth]{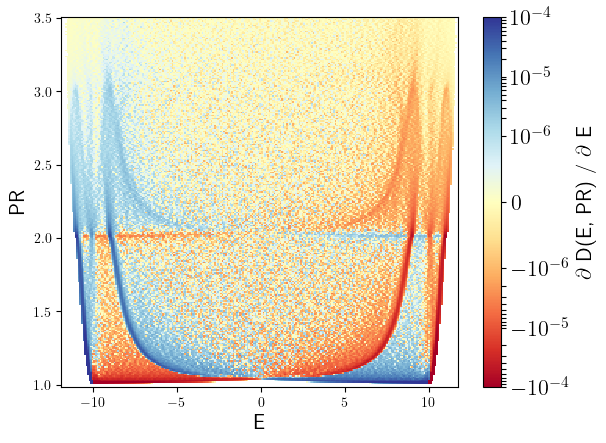} &
\includegraphics[width=0.5\linewidth]{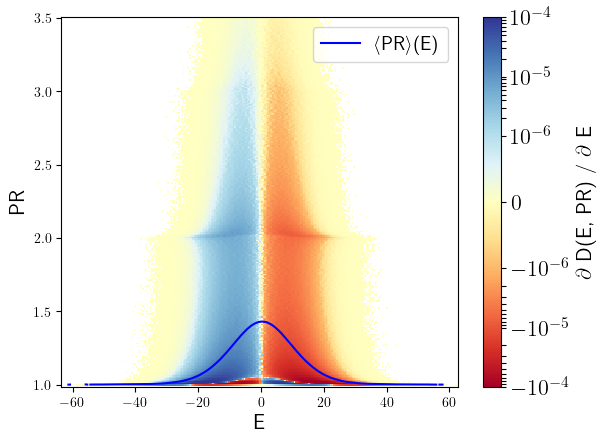} \\
\includegraphics[width=0.5\linewidth]{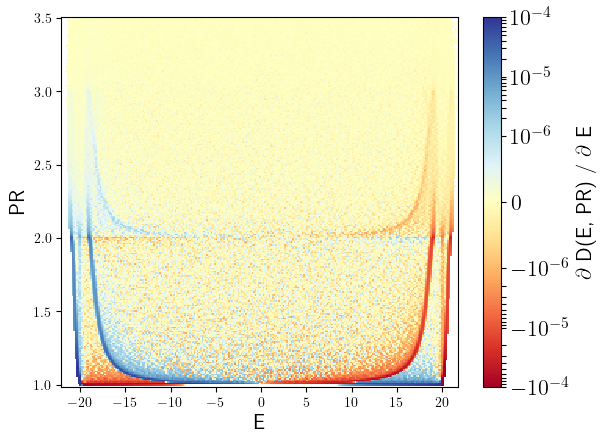} &
\includegraphics[width=0.5\linewidth]{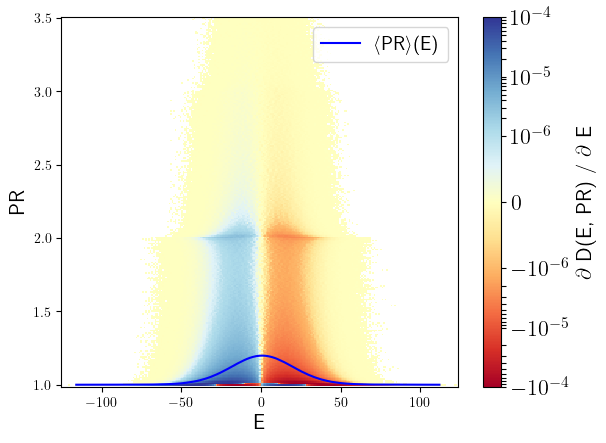}
\end{tabular}
\caption{ Histograms of the energy and PR of the eigenstates of a system of $L=10^9$ sites. The color is determined from the value of the derivative of the density of states with respect to the energy. The colorscale is logarithmic. On the left column, the disorder follows a box distribution and on the right column, the disorder is sampled via a Gaussian distribution and we add the mean value of the PR represented by the solid blue line. From top to bottom, the values of the disorder strength are respectively, $W_{\text{And}} = 10, 20, 40$ ($W_{\text{MBL}}=2.5, 5, 10$). }
\label{fig:derivative_E_PR_several_W_strong}

\end{figure*}

In Fig.~\ref{fig:derivative_E_PR_several_W_strong}, there is the histogram of the energy and PR, but now the color-scale is related with the value of the derivative of the density of states with respect to the energy. In those plots, we can clearly see a parabolic structure in the box distribution, which it is absent for the Gaussian distribution.

\begin{figure*}[h]
\centering
\begin{tabular}{c c}
\includegraphics[width=0.5\linewidth]{4D_box_W5d0.png} &
\includegraphics[width=0.5\linewidth]{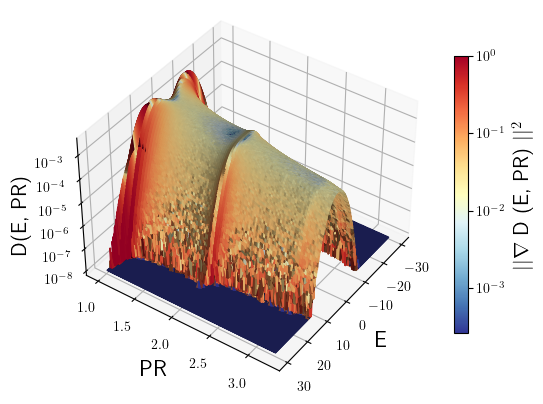} \\
\includegraphics[width=0.5\linewidth]{4D_box_W10d0.png} &
\includegraphics[width=0.5\linewidth]{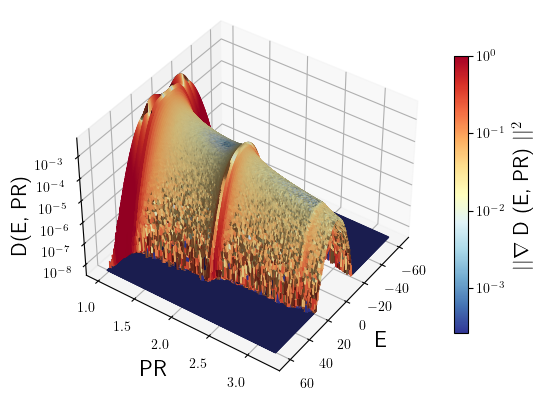} \\
\includegraphics[width=0.5\linewidth]{4D_box_W20d0.png} &
\includegraphics[width=0.5\linewidth]{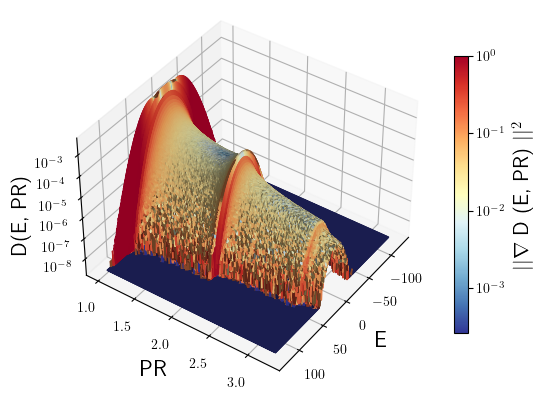}
\end{tabular}
\caption{ Histograms of energy, PR and density of states, from a system of  $L=10^9$ sites. The color is obtained from the norm of the gradient of the density of states. On the left column, the disorder follows a box distribution and on the right column, the disorder is sampled via a Gaussian distribution. From top to bottom, the disorder strength is $W_{\text{And}}=10, 20, 40$ ($W_{\text{MBL}} = 2.5, 5, 10$).  }
\label{fig:4D_box_G}

\end{figure*}

In Fig.~\ref{fig:4D_box_G}, we combine the previous histograms to form a 4D-histogram, where the energy, PR and density of states conform the axis and the color scale is determined by the norm of the gradient of the density of states.

\begin{figure}[h]
\centering
\begin{tabular}{c c}
\includegraphics[width=0.5\linewidth]{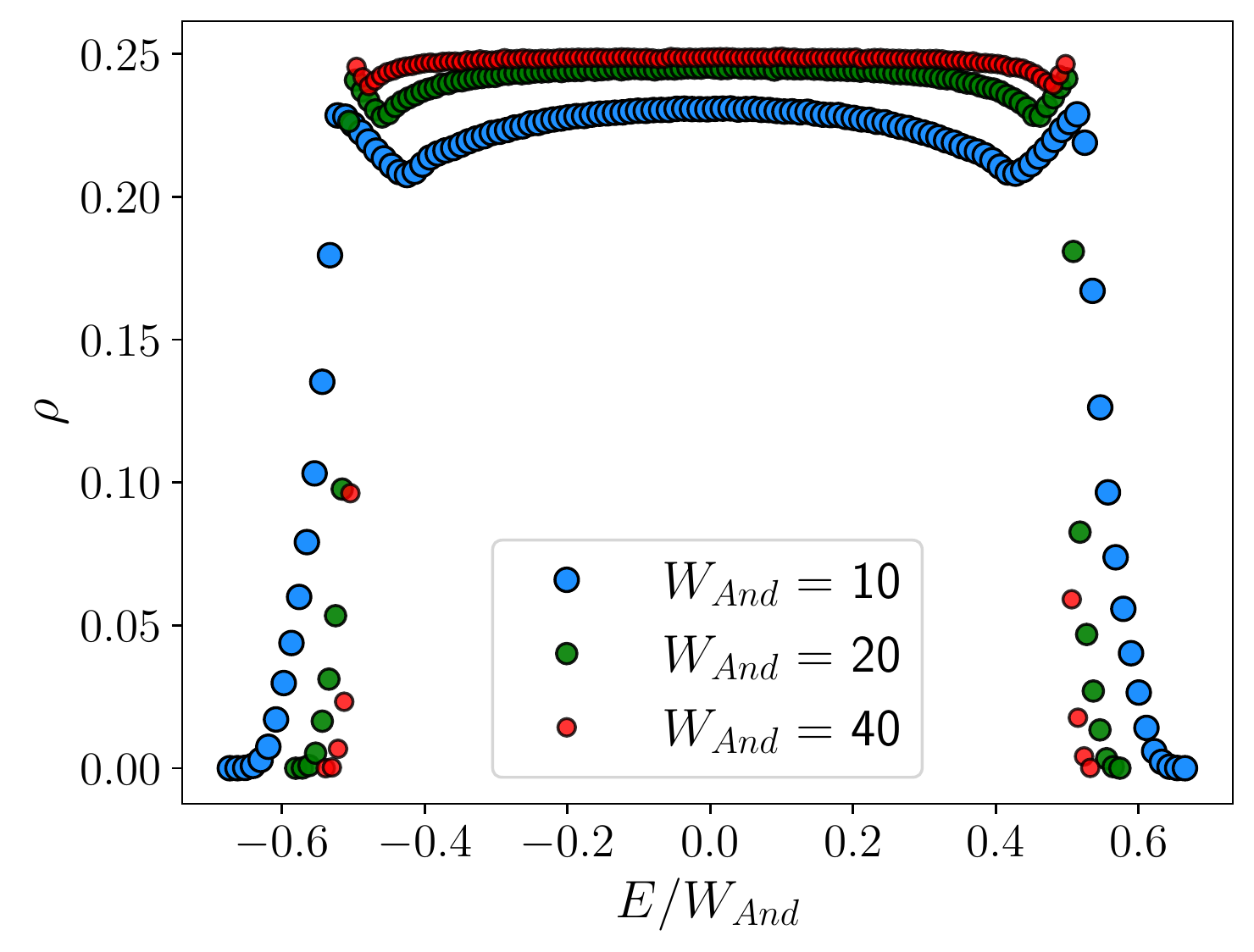} &
\includegraphics[width=0.5\linewidth]{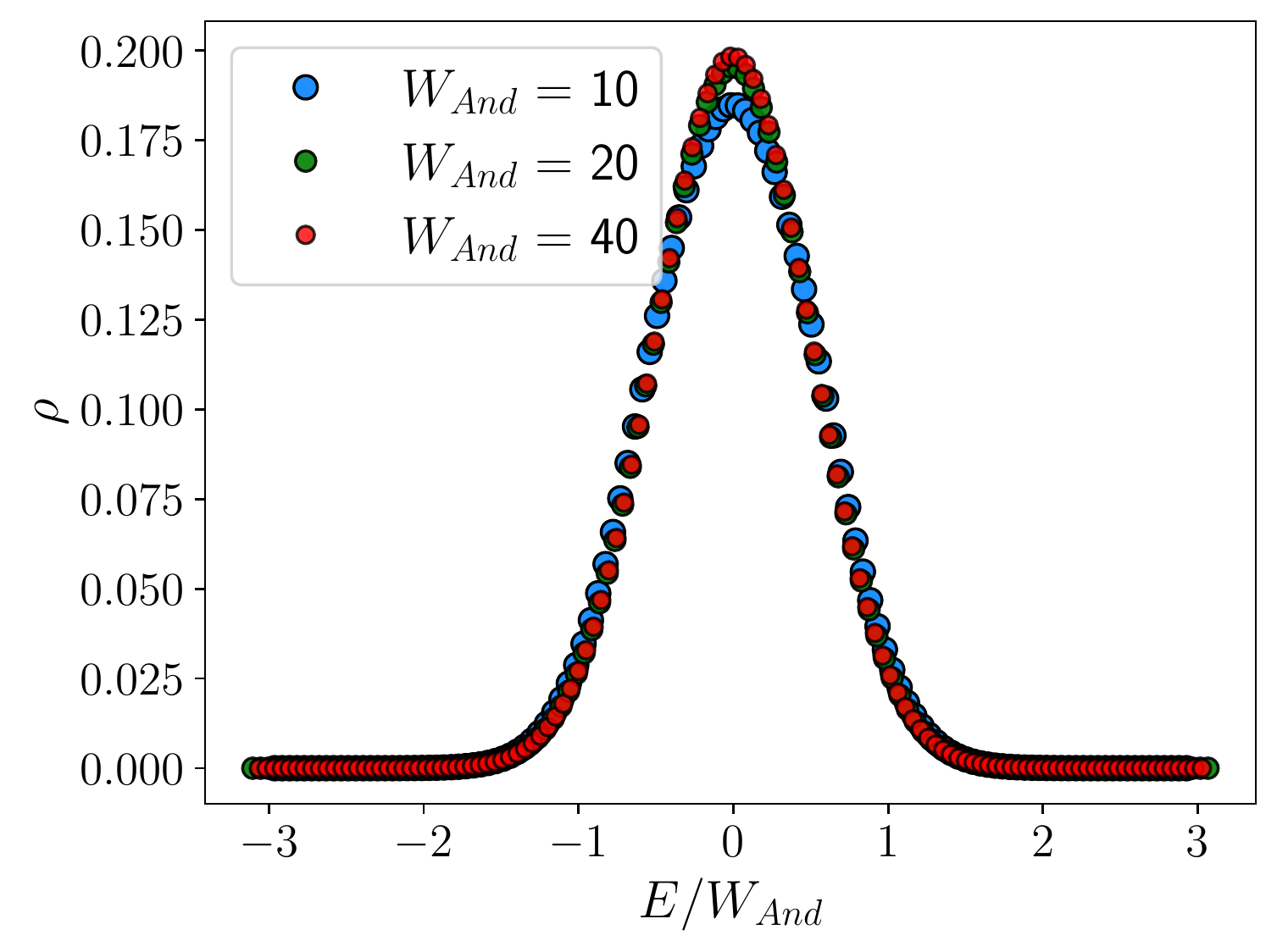} \\
\includegraphics[width=0.5\linewidth]{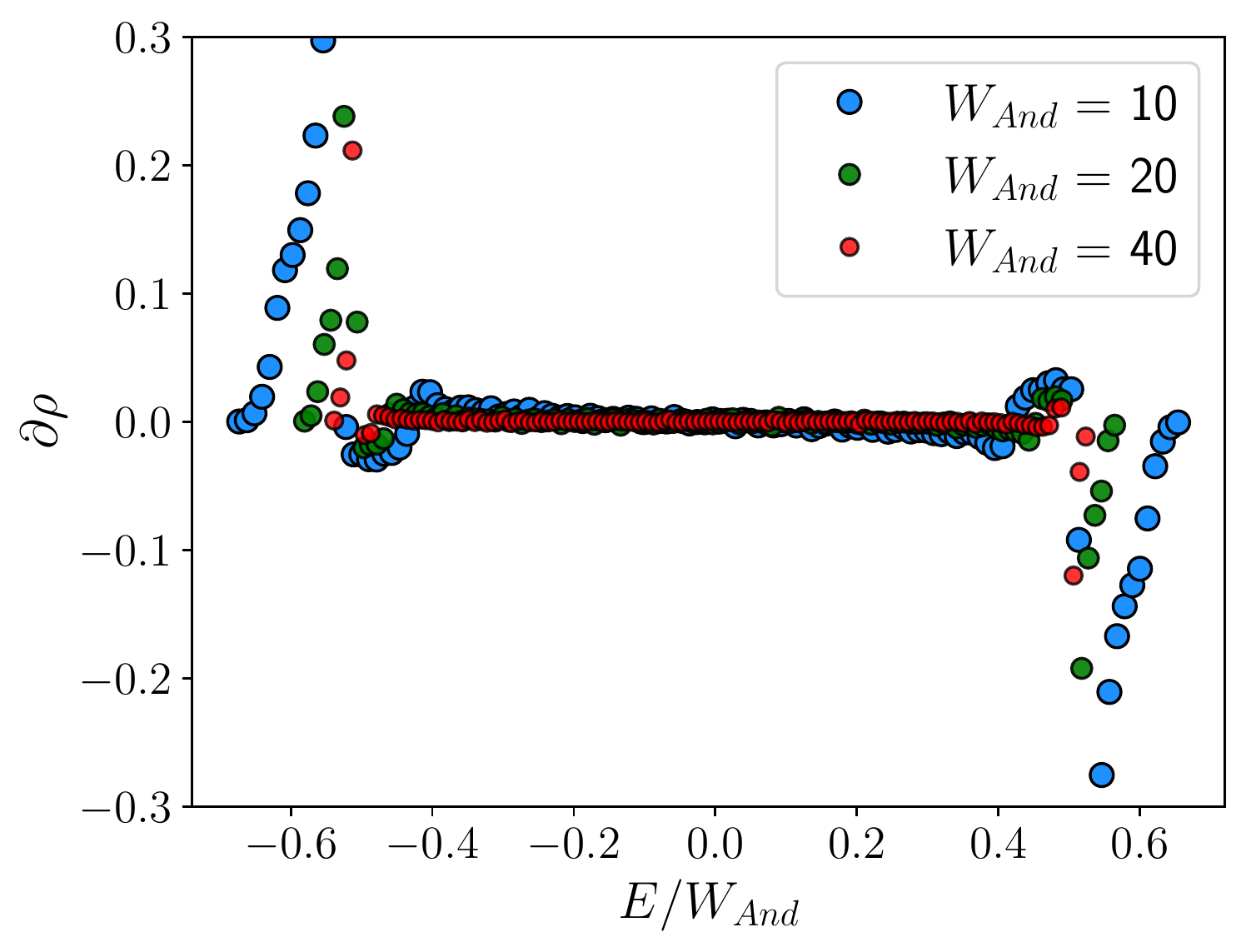} &
\includegraphics[width=0.5\linewidth]{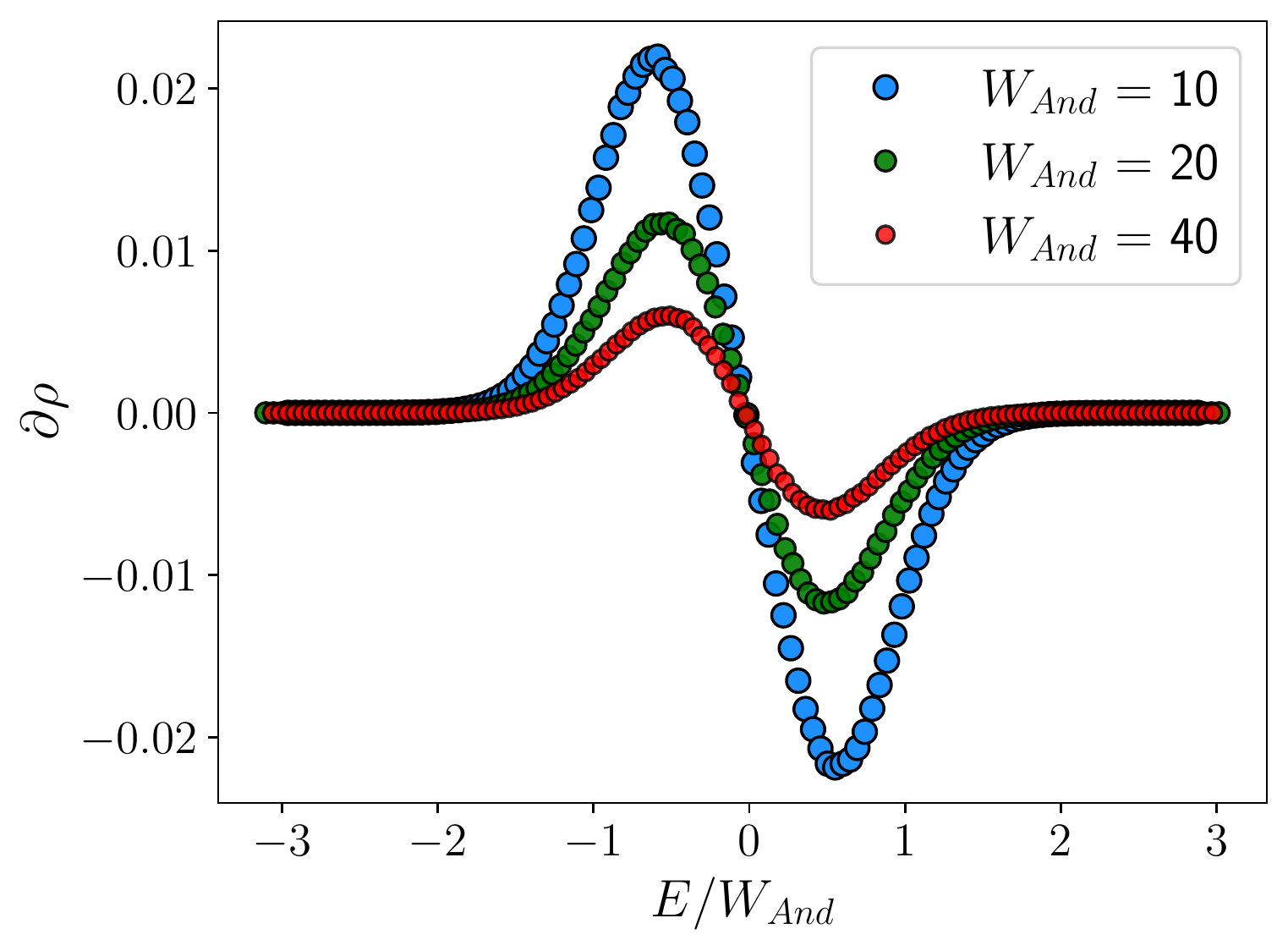} 
\end{tabular}
\caption{ Top, density of states, bottom, its derivative over the energy, from a system of  $L=10^9$ sites. On the left column, the disorder follows a box distribution and on the right column, the disorder is sampled via a Gaussian distribution. }
\label{fig:density_box_vs_G}

\end{figure}

\begin{figure}[h]
\centering
\begin{tabular}{c c}
\includegraphics[width=0.5\linewidth]{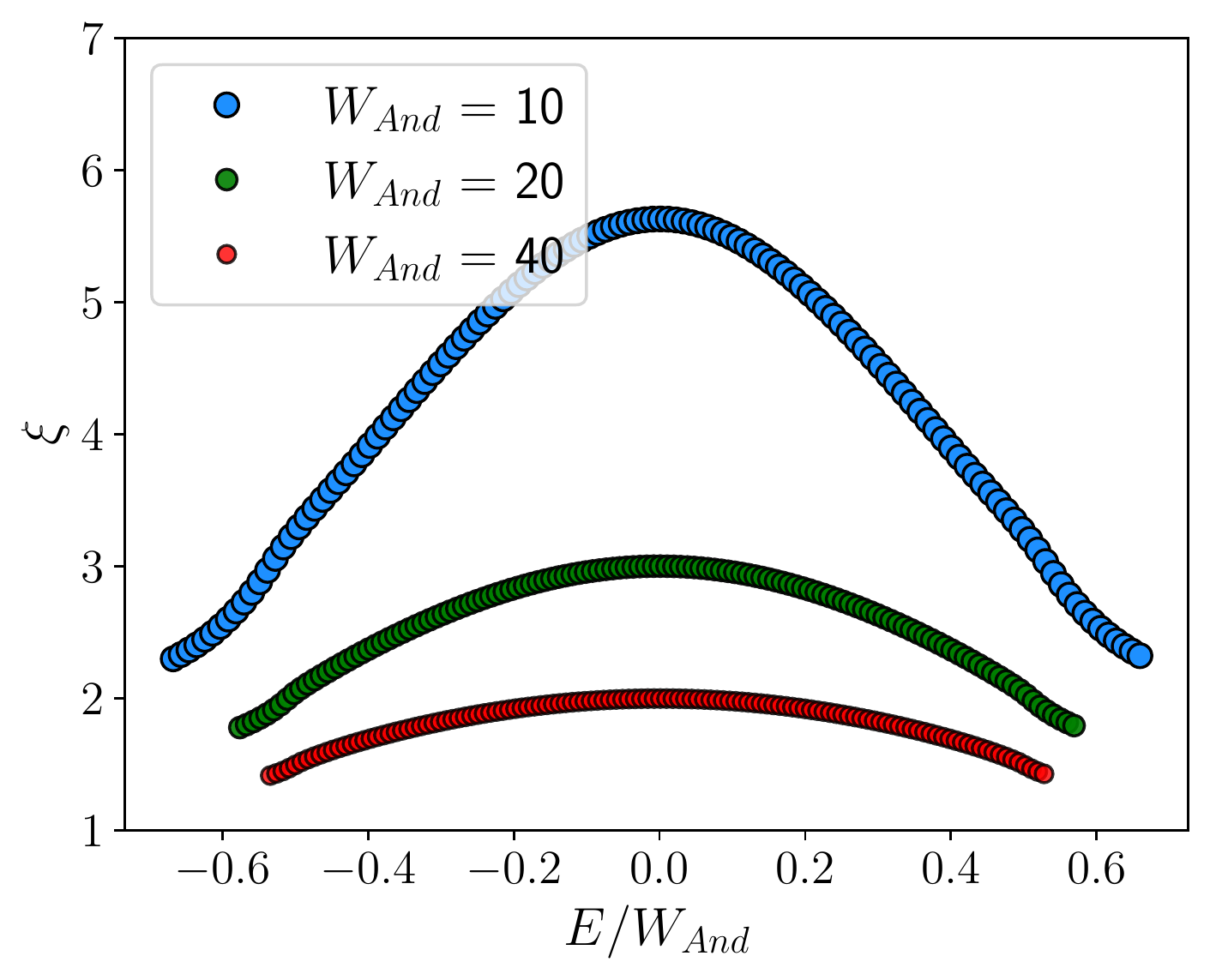} &
\includegraphics[width=0.5\linewidth]{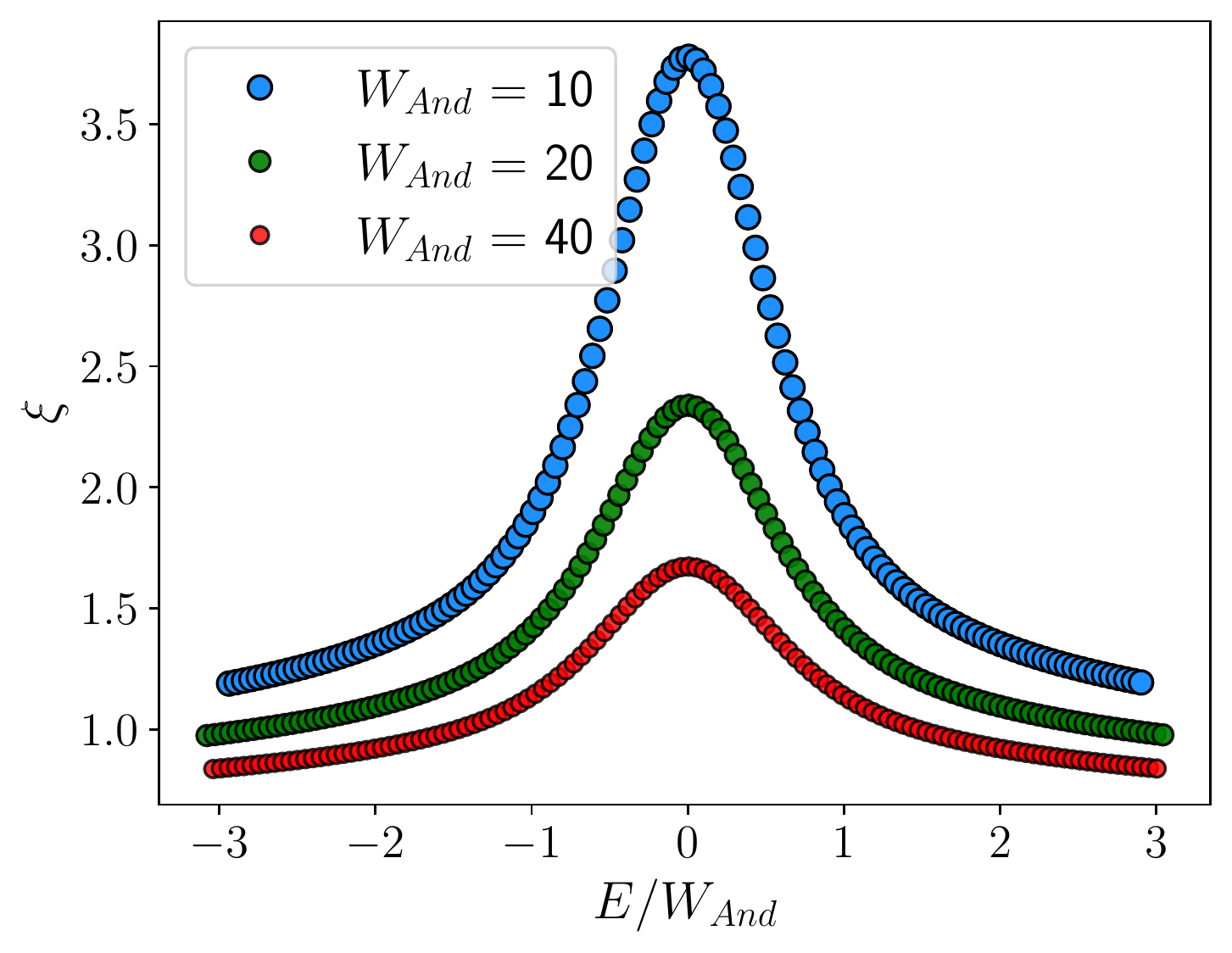} \\
\includegraphics[width=0.5\linewidth]{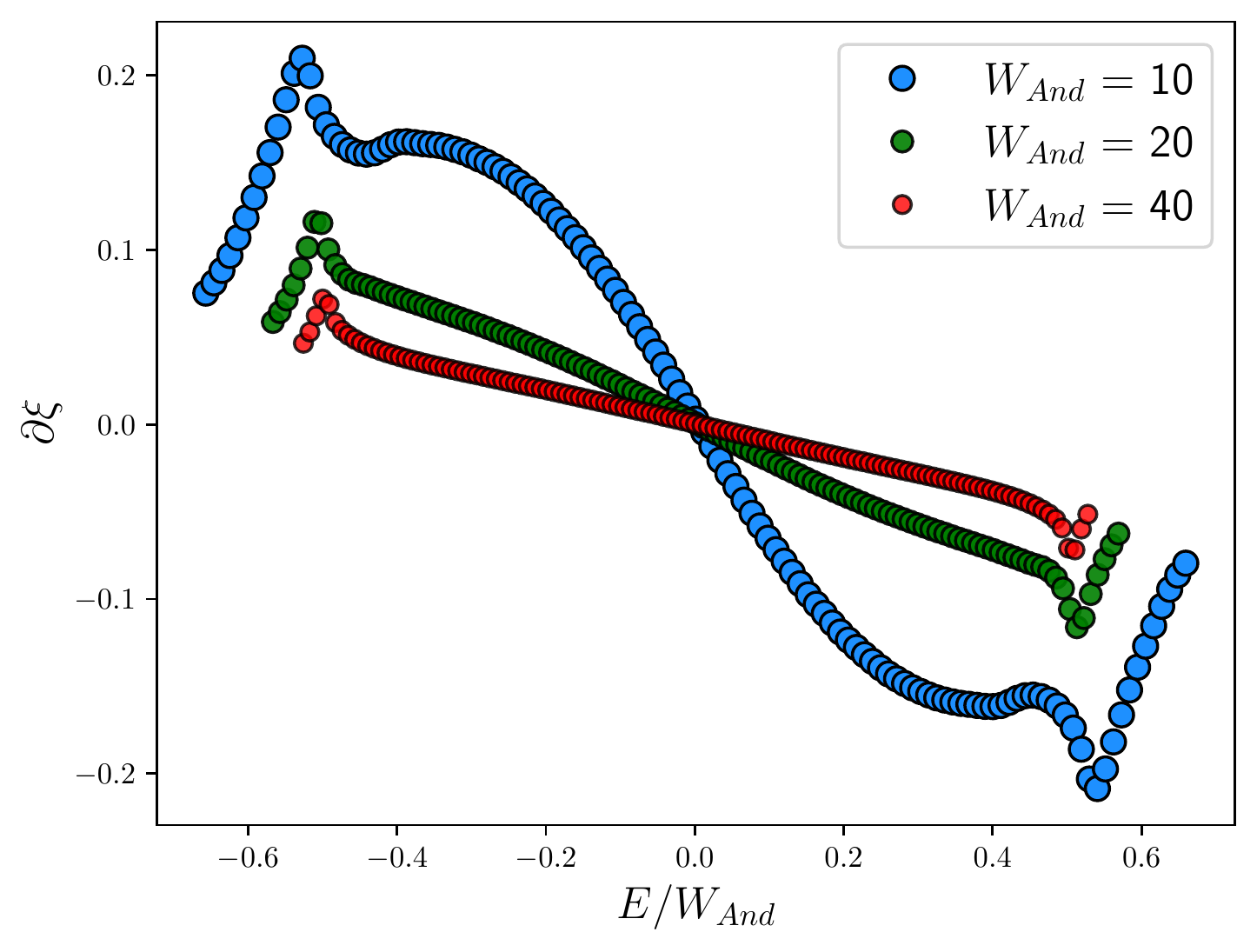} &
\includegraphics[width=0.5\linewidth]{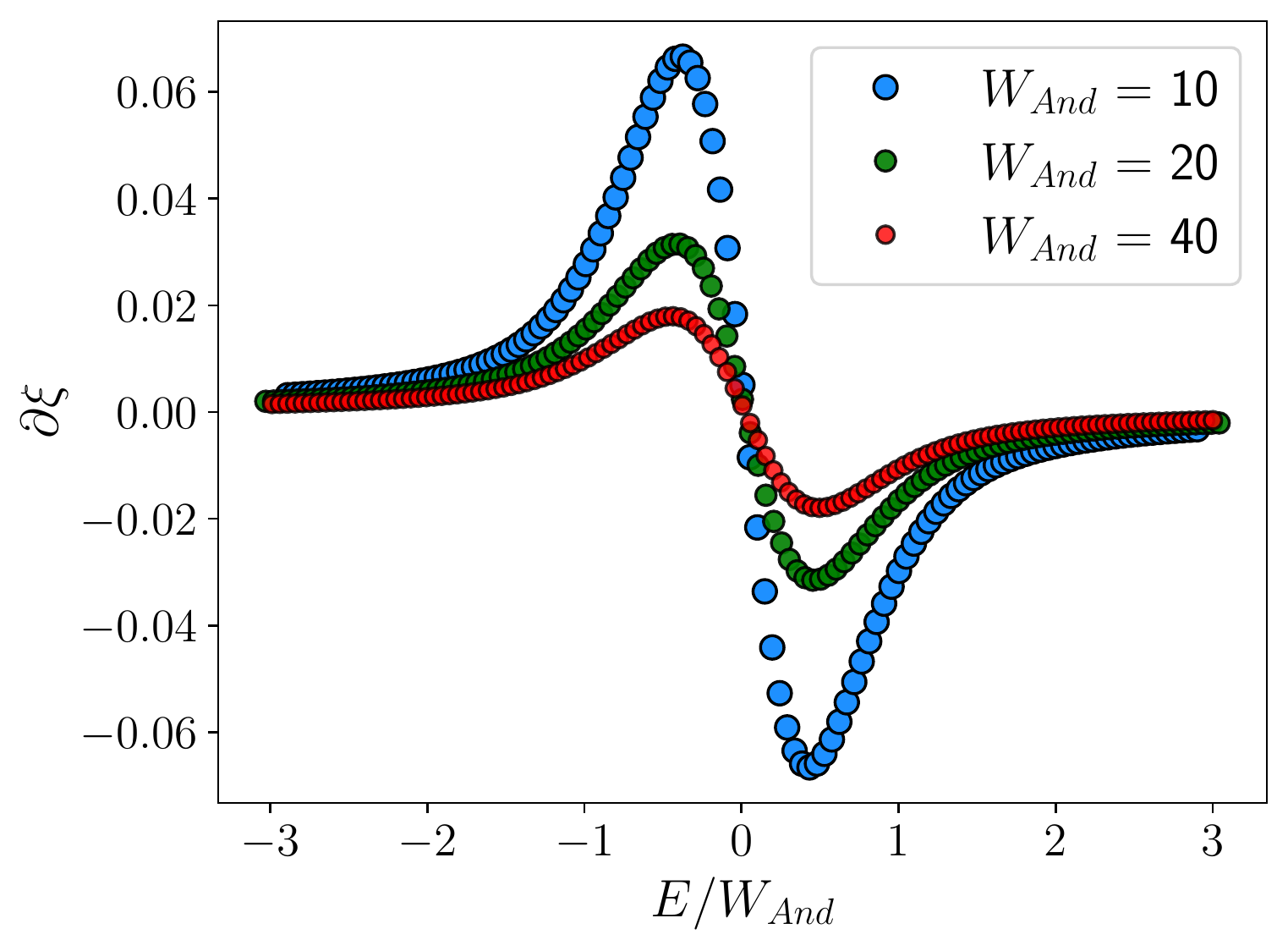} 
\end{tabular}
\caption{ Top, localization length, bottom, its derivative over the energy, from a system of  $L=10^9$ sites. On the left column, the disorder follows a box distribution and on the right column, the disorder is sampled via a Gaussian distribution. }
\label{fig:localization_box_vs_G}

\end{figure}

Up to now, we have considered that the density of states depend on the PR, but this is not the standard choice, generally, the density of states depends only on the energy. In order to make sure that the previous structures are not an artifact of imposing the PR-dependency in the density of states, in Fig.~\ref{fig:density_box_vs_G} and Fig.~\ref{fig:localization_box_vs_G}, we show the density of states, its derivative, the localization length and its derivative. There are clear differences between the box and Gaussian distributions. For instance, in the box distribution, the density of states has kinks leading to divergences when we derive the density of states over the energy. In the Gaussian distribution, there are no such kinks in the density of states and its derivative is a continuous function.

\section{Justification of the termination criterion for the TIP problem}
\label{appendix_justify_stop_N2}

In this section of the Appendix we provide a justification for the termination criterion adopted in the main text when dealing with the TIP problem.

Let us remember that in the TIP problem, we do not want to obtain all the eigenstates, only those that are modified when the nearest-neighbor interaction is switch on. This implies that we need a different way to determine if the size of the subsystem is large enough to fit all the ``eigenstates of interest". The first step towards finding the convergence criteria is to define properly what ``eigenstates of interest" means. Let us start by the intuitive answer. The ``eigenstates of interest" of those eigenstate of the interacting Hamiltonian that are different from the corresponding eigenstates of the non-interacting Hamiltonian. In others words, the expectation value of the occupancy of the orbitals is not always either zero or one for the ``eigenstates of interest", the interaction allows the eigenstates to have overlap with more than two orbitals. Let us define the following quantity $\kappa$:
\begin{equation}
\kappa(\ket{\psi}) = \prod_{x=1}^L f_x(1-f_x),
\end{equation}
where $f_x$ is the occupancy of the $x$-th orbital of the Anderson model by the wavefunction $\ket{\psi}$. For the eigenstates of the non-degenerate two non-interacting particles, $\{\ket{\Phi_n}\}$, we have $\kappa(\ket{\Phi_n})$ is equal to zero, while for eigenstates of the interacting problem, $\{\ket{\Psi_n}\}$, we have $\kappa(\ket{\Psi_n}) \geq 0$. We want to emphasize that if a wavefunction $\ket{\psi}$ fulfills $\kappa(\ket{\psi}) = 0$, then $\ket{\psi}$ is an eigenstate of the non-interacting problem. 

The ``eigenstates of interest" are the subset of $\{\ket{\Psi_n}\}$, refer as $A$, such that $\kappa(\ket{\Psi_n} \in A) > \delta$, for a finite $\delta$, used to take care of round-off errors. A tentative criteria to stop, which also helps us understand the properties of the problem,  is following: Apply the DaC method with a subsystem size equal to $M_1$ and $M_2 > M_1$ and solve the system keeping the random disorder the same with both $M_1$ and $M_2$. We will find more eigenstates when using a subsystem of size $M_2$ than when the subsystems have size $M_1$, the key point is to determine if all the ``extra" eigenstates when using $M_2$, $\ket{\phi_n}$, fulfill $\kappa(\ket{\phi_n}) \leq \delta$. If so, it means that all the new eigenstates obtained using $M_2$, compared from the found eigenstates when using $M_1$, are eigenstates of the non-interacting problem, up to machine precision. Due to the choice of the subsystems, in order to make sure that all the eigenstates obtained when using a subsystem of size $M_1$ are also obtained when considering $M_2$ as the size of the subsystem, we need to impose that $M_2 = 2 M_1$.

There is a problem in this approach, namely, we need to be able to use the DaC method when the subsystem size is $2M_1$ in order to check if the subsystem $M_1$ is large enough for the given random potential. The maximum subsystem size we can deal with is $M_{\max} = 200$, therefore, with this approach, we can only solve systems with a disorder strength strong enough to fit the eigenstates on subsystems of size $M_1 = 100$, instead of working with subsystems of size $M = 200$. We need to find another criteria to determine if we have obtained all the ``eigenstates of interest" and that allows us to use the data obtained for subsystems of size $M = 200$. With larger subsystems, we can deal with smaller disorder strength, where the effect of the interaction is expected to be more significant.

Our way to determine if the subsystem is large enough is based on the mean missing population of the sites $(i, i+1)$, for $i \in [1, L-1]$, denoted by $\overline{\epsilon}$. We proceed to justify why $\overline{\epsilon}$ determines the convergence of our results. Let $H$ be the Hamiltonian of the TIP problem, $H_0$ is the non-interacting part, consisting on the hopping term and the on-site disorder, and $H_I$ in the term encoding the nearest-neighbor interaction, with $H = H_0 + H_I$. If $\ket{\psi_\alpha}$ is an eigenstate of $H_0$ then the energy variance of $\ket{\psi_\alpha}$ with respect to $H$ is given by $\sigma^2_H(\ket{\psi_\alpha})$:
\begin{equation}
\label{var_H_TIP}
\sigma^2_H(\ket{\psi_\alpha}) = \braket{\psi_\alpha | H_I^2 | \psi_\alpha} - \braket{\psi_\alpha | H_I | \psi_\alpha}^2.
\end{equation}
Similarly, if $\ket{\phi_\beta}$ is an eigenstate of $H$ then the energy variance of $\ket{\phi_\beta}$ with respect to $H$ is given by $\sigma^2_{H_0}(\ket{\phi_\beta})$:
\begin{equation}
\label{var_H0}
\sigma^2_{H_0}(\ket{\phi_\beta}) = \braket{\phi_\beta | H_I^2 | \phi_\beta} - \braket{\phi_\beta | H_I | \phi_\beta}^2.
\end{equation}

We want to provide an example to grasp what exactly Eqn.~\ref{var_H_TIP} is telling us. Let us assume that exists one eigenstate, $\ket{\psi}$, of $H_0$ that fulfills the following equation
\begin{equation}
\label{cond_eigen}
\braket{\psi|H_i|\psi} = \braket{\psi|H_i^2|\psi} = 0.
\end{equation}
Let us emphasize that the existence of such eigenstate of $H_0$, in general, is not guaranteed to exist. But if it exists, then it is an eigenstate of $H$ too. For the very specific case where Eqn.~\ref{cond_eigen} holds, the interaction term does not affect at all the state $\ket{\psi}$, and therefore, it is also an eigenstate of $H$. Let us focus again in our particular case, where the interacting term $H_I$ is:
\begin{equation}
H_I = U \sum_{i=1}^{L-1} \ket{i, i+1}\bra{i, i+1}.
\end{equation}
Let us remark that our $H_I$ fulfills the following equation:
\begin{equation}
\label{HI_projector}
H_I^2 = U H_I.
\end{equation}
Defining $P_\alpha = \braket{\psi_\alpha|H_I|\psi_\alpha}/U$, assuming that $\ket{\psi_\alpha}$ is an eigenstate of $H_0$ and using the Eqn.~\ref{HI_projector}, we can calculate the energy variance of $\ket{\psi_\alpha}$ with respect to $H$ via:
\begin{equation}
\label{error_H}
\begin{split}
&\sigma^2_H(\ket{\psi_\alpha}) = U^2P_\alpha (1-P_\alpha), \\
&P_\alpha = \sum_{i=1}^{L-1} \abs{\braket{i, i+1 | \psi_\alpha}}^2 = \sum_{i=1}^{L-1} \abs{c_{i, i+1}}^2.
\end{split}
\end{equation}
Eqn.~\ref{error_H} fits with the intuition that far away particles, whose mean position differs more than their localization length, do not feel each other. The same idea extends to eigenstates representing those stationary and far away particles. Given a set of eigenstates of $H$, $\{\phi_\alpha\}$, let us define the mean missing population, $\overline{\epsilon}$, as:
\begin{equation}
\label{def_mean}
\overline{\epsilon}(\{\phi_\alpha\}) = 1 - \frac{1}{L-1}\sum_{i=1}^{L-1}\sum_\alpha \abs{\braket{\phi_\alpha |i,i+1}}^2.
\end{equation}
Eqn.~\ref{def_mean} calculates which is the average missing population in all the consecutive sites $(i, i+1)$, which are precisely the sites where $H_I$ acts.
If $\{\phi_\alpha\}$ is the full set of eigenstates of $H$, then $\overline{\epsilon} = 0$, since for the full set of eigenstates, all sites $(i,j)$ are fully populated. If $\{\phi_\alpha\}$ is a subset of the eigenstates and the subset $\{\psi_\beta\}$ completes the orthonormal basis, then we know:
\begin{equation}
\label{sum_missing}
\begin{split}
\sum_\beta \sum_{i=1}^{L-1} \abs{\braket{\psi_\beta |i,i+1}}^2 &= \sum_\beta P_\beta = \\
& = (L-1)\overline{\epsilon}(\{\phi_\alpha\}).
\end{split}
\end{equation}
The combination of Eqn.~\ref{error_H} and Eqn.~\ref{sum_missing} provides the justification to use the mean missing population as a criteria to stop. Let us remember that the goal is to determine when the subsystem is large to fit all the ``eigenstates of interest", the ones that are different from the eigenstates of the non-interacting problem. All the eigenstates which are not obtained using the DaC algorithm are considered to the same as the corresponding ones of the non-interacting problem and Eqn.~\ref{error_H} tells us how to bound the error when doing such association and Eqn.~\ref{sum_missing} can be used to provide an expectation value of the error, as we show next. After applying the DaC algorithm with a subsystem size $M$ to find eigenstates of a systems of $L$ sites, we obtain the subset of eigenstates $\{\phi_\alpha\}$. The number of element of the subset, $N_\alpha$ ,  scales linearly with $L$, $N_\alpha = C_{M, h} L$, where $C_{M, h}$ is the average number of eigenstates obtained per subsystem and it depends mainly in the size of the subsystem $M$ and the random potential $h$. We can calculate the mean missing population of the subset $\{\phi_\alpha\}$ via Eqn.~\ref{def_mean}, obtaining a value of $\overline{\epsilon}$. We start assuming that the total missing population, $(L-1)\overline{\epsilon}$ is shared among all the missing eigenstates, $N_\beta = N_T - N_\alpha$, where $N_T$ is the total number of eigenstates, $N_T = L(L-1)/2$. If that would be the case, then the average value of $P_\beta$, which is the norm of the states after applying $H_I$, and the variance with respect to $H$, $\overline{\sigma}^2_H$, would be:
\begin{equation}
\overline{P}_\beta = \frac{(L-1)\overline{\epsilon}}{N_T - N_\alpha} \sim \frac{\overline{\epsilon}}{L}, \qquad \overline{\sigma}^2_H \sim U^2 \frac{\overline{\epsilon}}{L}.
\end{equation}
The localization of the eigenstates is in contradiction with the assumption that all missing eigenstates have an similar overlap with the consecutive sites. One better assumption would be that the mean missing population is shared fairly among the subset of eigenstates with the mean distance between the two particles smaller than the localization length. To determine how many eigenstates are missing, we can extrapolate from the number of obtained eigenstates, $N_\alpha$. If $N_\alpha = C_{M, h} L$ eigenstates occupy up to $1-\overline{\epsilon}$ of the total population per sites $(i, i+1)$, then we can expect that the number of missing eigenstates with a finite overlap with the sites $(i,i+1)$, $N_\beta$, to be:
\begin{equation}
N_\beta \sim N_\alpha \overline{\epsilon} \sim C_{M, h} L\overline{\epsilon}.
\end{equation}
The expected values of $P_\beta$ and the variance with respect to $H$ are now:
\begin{equation}
\overline{P}_\beta \sim	\frac{\overline{\epsilon}}{C_{M, h}}, \qquad \overline{\sigma}^2_H \sim U^2 \frac{\overline{\epsilon}}{C_{M, h}}.
\end{equation}
To summarize, if the assumption of distributing evenly the missing population holds, the mean missing population $\overline{\epsilon}$ is an indicator of the proportion of the ``eigenstates of interest" missing and also an upper bound of the average error when assuming that the missing eigenstates are like the corresponding non-interacting eigenstates. The criteria to stop is when $\overline{\epsilon} \ll 1$. Of course, the previous assumption does not hold in the general case, only in the scenario of fully localization of the eigenstates of the TIP problem, which it is guaranteed for a truly random potential~\cite{Shepelyansky1994_4}. If we consider instead a quasi-periodic potential, we know that for certain interaction strength, metallic states appear~\cite{Flach2012_4}, $\{\Psi_\beta\}$. The probability to find the two particles in consecutive sites for those metallic eigenstates is quite large. Therefore, the mean missing population is not shared fairly among the missing eigenstates, only a few instances, the number of metallic states $\{\Psi_\beta\}$, share the total missing population $(L-1)\overline{\epsilon}$. If we want to be completely sure that none of the missing eigenstates is different from one of eigenstates of the non-interacting problem, then we need to increase $M$ until we achieve that the total missing population is small, $(L-1)\overline{\epsilon} \ll 1$. As a summary of the previous reasoning, the mean missing population, $\overline{\epsilon}$, can be used as a criteria to determine if the subsystem size $M$ is large enough. Under certain assumptions, we only need that $\overline{\epsilon} \ll 1$ in order to stop the algorithm. If we do not want to take any assumption, then we require that $(L-1)\overline{\epsilon} \ll 1$.

\end{document}